%Paper: hep-th/9505150
%From: lerche@nxth04.cern.ch
%Date: Wed, 24 May 95 18:37:11 -0500

%%%%%%%%%%%%%%%%%%%%%%%%%%%%%%%%%%%%%%%%%%%%%%%%%%%%%%%%%%
%%% this paper has 9 uuencoded compressed
%%% postscript figures appended, and also
%%% contains hypertex-links if texed with lanlmac.
%%%%%%%%%%%%%%%%%%%%%%%%%  %%%%%%%%%%%%%%%%%%%%%%%%%%%%%%%
% this set of macros is an extension of harvmac
% (9/91 or later) and its hypertex version, lanlmac.
% for info on hypertex, see
% http://xxx.lanl.gov/hypertex/
%%%%%%%%%%%%%%%%%%% %%%%%%%%%%%%%%%%%%%%%%%%%%%
% determine hypertex mode
\newif\iflanl
\openin 1 lanlmac
\ifeof 1 \lanlfalse \else \lanltrue \fi
\closein 1
\iflanl
    \input lanlmac
\else
    \message{[lanlmac not found - use harvmac instead}
    \input harvmac
\fi
\newif\ifhypertex
\ifx\hyperdef\UnDeFiNeD
    \hypertexfalse
    \message{[HYPERTEX MODE OFF}
    \def\hyperref#1#2#3#4{#4}
    \def\hyperdef#1#2#3#4{#4}
    \def\hypernoname{}
    \def\e@tf@ur#1{}
    \def\hth/#1#2#3#4#5#6#7{{\tt hep-th/#1#2#3#4#5#6#7}}
    \def\CERN{\address{CERN, Geneva, Switzerland}}
\else
    \hypertextrue
    \message{[HYPERTEX MODE ON}
%hypertex links to xxx.lanl.gov:
  \def\hth/#1#2#3#4#5#6#7{
  {\tt hep-th/#1#2#3#4#5#6#7}}
\def\CERN{\address{

Theory Division, CERN, Geneva, Switzerland}}
\fi
%%%%%%%%%%%%%%%%%%%%%%% %%%%%%%%%%%%%%%%%%%%%%%
\newif\ifdraft

\noblackbox
\catcode`\@=11
\newif\iffrontpage
%%%%%%%%%%%%%%%%%%% %%%%%%%%%%%%%%%%%%%%%%%%%%%%%%%%%%%%%%%%%%%%%%
%%%%% sizes, offsets etc
%%%%%%%%%%%%%%%%%%% %%%%%%%%%%%%%%%%%%%%%%%%%%%%%%%%%%%%%%%%%%%%%%
\ifx\answ\bigans
\def\titleft{\titsm}
\magnification=1200%\baselineskip=14pt plus 2pt minus 1pt
%
%%%%% unreduced mode: %%%%
%\voffset=0.35truein\hoffset=0.250truein
\advance\hoffset by-0.075truein
\advance\voffset by1.truecm
\hsize=6.15truein\vsize=600.truept\hsbody=\hsize\hstitle=\hsize
\else\let\lr=L
\def\titleft{\titla}
\magnification=1000%\baselineskip=14pt plus 2pt minus 1pt
%
%%%%% reduced mode: %%%%%%%
\hoffset=-0.75truein\voffset=-.0truein
%?\hoffset=-.25truein\voffset=-.0truein
\vsize=6.5truein
\hstitle=8.truein\hsbody=4.75truein
\fullhsize=10truein\hsize=\hsbody
\fi
\parskip=4pt plus 15pt minus 1pt
%
%%%%%%%%%%%%%%%%%%% %%%%%%%%%%%%%%%%%%%%%%%%%%%%%%%%%%%%%%%%%%%%%%
%%%%% figures
%%%%%%%%%%%%%%%%%%% %%%%%%%%%%%%%%%%%%%%%%%%%%%%%%%%%%%%%%%%%%%%%%
\newif\iffigureexists
\newif\ifepsfloaded
\def\epsfcheck{
\ifdraft% to speed up
\input epsf\epsfloadedtrue
\else
  \openin 1 epsf
  \ifeof 1 \epsfloadedfalse \else \epsfloadedtrue \fi
  \closein 1
  \ifepsfloaded
    \input epsf
  \else
\immediate\write20{NO EPSF FILE --- FIGURES WILL BE IGNORED}
  \fi
\fi
\def\epsfcheck{}}
\def\checkex#1{
%\ifdraft
%\figureexistsfalse\immediate%
%\write20{Draftmode: figure #1 not included}
\figureexiststrue
%\else
\relax
    \ifepsfloaded \openin 1 #1
        \ifeof 1
           \figureexistsfalse
  \immediate\write20{FIGURE FILE #1 NOT FOUND}
        \else \figureexiststrue
        \fi \closein 1
    \else \figureexistsfalse
    \fi
%\fi
}
\def\missbox#1#2{$\vcenter{\hrule
\hbox{\vrule height#1\kern1.truein
\raise.5truein\hbox{#2} \kern1.truein \vrule} \hrule}$}
\def\lfig#1{%  this is to call the figure in the text
\let\labelflag=#1%
\def\numb@rone{#1}%
\ifx\labelflag\UnDeFiNeD%
{\xdef#1{\the\figno}%
\writedef{#1\leftbracket{\the\figno}}%
\global\advance\figno by1%
}\fi{\hyperref{}{figure}{{\numb@rone}}{Fig.{\numb@rone}}}}
\def\figinsert#1#2#3#4{%  this inserts the figure
\epsfcheck\checkex{#4}%
\def\figsize{#3}%
\let\flag=#1\ifx\flag\UnDeFiNeD
{\xdef#1{\the\figno}%
\writedef{#1\leftbracket{\the\figno}}%
\global\advance\figno by1%
}\fi
\goodbreak\midinsert%
\iffigureexists
\centerline{\epsfysize\figsize\epsfbox{#4}}%
\else%
\vskip.05truein
  \ifepsfloaded
  \ifdraft
  \centerline{\missbox\figsize{Draftmode: #4 not included}}%
  \else
  \centerline{\missbox\figsize{#4 not found}}
  \fi
  \else
  \centerline{\missbox\figsize{epsf.tex not found}}
  \fi
\vskip.05truein
\fi%
{\smallskip%
\leftskip 4pc \rightskip 4pc%
\noindent\ninepoint\sl \baselineskip=11pt%
{\bf{\hyperdef\hypernoname{figure}{{#1}}{Fig.{#1}}}:~}#2%
\smallskip}\bigskip\endinsert%
}
%
%%%%%%%%%%%%%%%%%%% %%%%%%%%%%%%%%%%%%%%%%%%%%%%%%%%%%%%%%%%%%%%%%
%%%%%  fonts
%%%%%%%%%%%%%%%%%%% %%%%%%%%%%%%%%%%%%%%%%%%%%%%%%%%%%%%%%%%%%%%%%
%%%%%%%%%%%%%%%%%%% %%%%%%%%%%%%%%%%%%%%%%%%%%%%%%%%%%%%%%%%%%%%%%
\font\bigit=cmti10 scaled \magstep1

\font\titla=cmr10 scaled\magstep3
\font\tenmss=cmss10
\font\absmss=cmss10 scaled\magstep1

\font\twelvebf=cmbx10 scaled\magstep1

\newfam\mssfam
\font\footrm=cmr8  \font\footrms=cmr5
\font\footrmss=cmr5   \font\footi=cmmi8
\font\footis=cmmi5   \font\footiss=cmmi5
\font\footsy=cmsy8   \font\footsys=cmsy5
\font\footsyss=cmsy5   \font\footbf=cmbx8
\font\footmss=cmss8
\def\footfont{\def\rm{\fam0\footrm}
\textfont0=\footrm \scriptfont0=\footrms
\scriptscriptfont0=\footrmss
\textfont1=\footi \scriptfont1=\footis
\scriptscriptfont1=\footiss
\textfont2=\footsy \scriptfont2=\footsys
\scriptscriptfont2=\footsyss
\textfont\itfam=\footi \def\it{\fam\itfam\footi}
\textfont\mssfam=\footmss \def\mss{\fam\mssfam\footmss}
\textfont\bffam=\footbf \def\bf{\fam\bffam\footbf} \rm}
\def\tenpoint{\def\rm{\fam0\tenrm}
\textfont0=\tenrm \scriptfont0=\sevenrm
\scriptscriptfont0=\fiverm
\textfont1=\teni  \scriptfont1=\seveni
\scriptscriptfont1=\fivei
\textfont2=\tensy \scriptfont2=\sevensy
\scriptscriptfont2=\fivesy
\textfont\itfam=\tenit \def\it{\fam\itfam\tenit}
\textfont\mssfam=\tenmss \def\mss{\fam\mssfam\tenmss}
\textfont\bffam=\tenbf \def\bf{\fam\bffam\tenbf} \rm}
\ifx\answ\bigans\def\abstractfont{\tenpoint}\else
\def\abstractfont{\def\rm{\fam0\absrm}
\textfont0=\absrm \scriptfont0=\absrms
\scriptscriptfont0=\absrmss
\textfont1=\absi \scriptfont1=\absis
\scriptscriptfont1=\absiss
\textfont2=\abssy \scriptfont2=\abssys
\scriptscriptfont2=\abssyss
\textfont\itfam=\bigit \def\it{\fam\itfam\bigit}
\textfont\mssfam=\absmss \def\mss{\fam\mssfam\absmss}
\textfont\bffam=\absbf \def\bf{\fam\bffam\absbf}\rm}\fi
%
%%%%%%%%%%%%%%%%%%%%%%%%%%%%% %%%%%%%%%%%%%%%%%%%%%%%%%%%%%%%%%
%%%%% footnotes   (adapted from PHYZZX)
%%%%%%%%%%%%%%%%%%%%%%%%%%%%% %%%%%%%%%%%%%%%%%%%%%%%%%%%%%%%%%
\def\f@@t{\baselineskip10pt\lineskip0pt\lineskiplimit0pt
\bgroup\aftergroup\@foot\let\next}
\setbox\strutbox=\hbox{\vrule height 8.pt depth 3.5pt width\z@}
\def\vfootnote#1{\insert\footins\bgroup
\baselineskip10pt\footfont
\interlinepenalty=\interfootnotelinepenalty
\floatingpenalty=20000
\splittopskip=\ht\strutbox \boxmaxdepth=\dp\strutbox
\leftskip=24pt \rightskip=\z@skip
\parindent=12pt \parfillskip=0pt plus 1fil
\spaceskip=\z@skip \xspaceskip=\z@skip
\Textindent{$#1$}\footstrut\futurelet\next\fo@t}
\def\Textindent#1{\noindent\llap{#1\enspace}\ignorespaces}
\def\foot{\global\advance\ftno by1%
\attach{\hyperref{}{footnote}{\the\ftno}{\footsymbolgen}}%
\vfootnote{\hyperdef\hypernoname{footnote}{\the\ftno}{\footsymbol}}}%
%   this is for custom footnote marks:
\def\footnote#1{\global\advance\ftno by1%
\attach{\hyperref{}{footnote}{\the\ftno}{#1}}%
\vfootnote{\hyperdef\hypernoname{footnote}{\the\ftno}{#1}}}%
\newcount\lastf@@t           \lastf@@t=-1
\newcount\footsymbolcount    \footsymbolcount=0
\global\newcount\ftno \global\ftno=0
\def\footsymbolgen{\relax\footsym
\global\lastf@@t=\pageno\footsymbol}
\def\footsym{\ifnum\footsymbolcount<0
\global\footsymbolcount=0\fi
{\iffrontpage \else \advance\lastf@@t by 1 \fi
\ifnum\lastf@@t<\pageno \global\footsymbolcount=0
\else \global\advance\footsymbolcount by 1 \fi }
\ifcase\footsymbolcount
\fd@f\dagger\or \fd@f\diamond\or \fd@f\ddagger\or
\fd@f\natural\or \fd@f\ast\or \fd@f\bullet\or
\fd@f\star\or \fd@f\nabla\else \fd@f\dagger
\global\footsymbolcount=0 \fi }
\def\fd@f#1{\xdef\footsymbol{#1}}
\def\space@ver#1{\let\@sf=\empty \ifmmode #1\else \ifhmode
\edef\@sf{\spacefactor=\the\spacefactor}
\unskip${}#1$\relax\fi\fi}
\def\attach#1{\space@ver{\strut^{\mkern 2mu #1}}\@sf}
%
%%%%%%%%%%%%%%%%%%% %%%%%%%%%%%%%%%%%%%%%%%%%%%%%%%%%%%%%%%%%%%%%%
%%%%% References
%%%%%%%%%%%%%%%%%%% %%%%%%%%%%%%%%%%%%%%%%%%%%%%%%%%%%%%%%%%%%%%%%
\newif\ifnref
\def\rrr#1#2{\relax\ifnref\nref#1{#2}\else\ref#1{#2}\fi}
\def\ldf#1#2{\begingroup\obeylines
\gdef#1{\rrr{#1}{#2}}\endgroup\unskip}

\def\doubref#1#2{\refs{{#1},{#2}}}

\nreffalse
\def\refout{\listrefs}
%
%%%%%%%%%%%%%%%%%%% %%%%%%%%%%%%%%%%%%%%%%%%%%%%%%%%%%%%%%%%%%%%%%
%%%%%%% eq numbering
%%%%%%%%%%%%%%%%%%% %%%%%%%%%%%%%%%%%%%%%%%%%%%%%%%%%%%%%%%%%%%%%%
\def\eqn#1{\xdef #1{(\noexpand\hyperref{}%
{equation}{\secsym\the\meqno}%
{\secsym\the\meqno})}\eqno(\hyperdef\hypernoname{equation}%
{\secsym\the\meqno}{\secsym\the\meqno})\eqlabeL#1%
\writedef{#1\leftbracket#1}\global\advance\meqno by1}
\def\eqnalign#1{\xdef #1{\noexpand\hyperref{}{equation}%
{\secsym\the\meqno}{(\secsym\the\meqno)}}%
\writedef{#1\leftbracket#1}%
\hyperdef\hypernoname{equation}%
{\secsym\the\meqno}{\e@tf@ur#1}\eqlabeL{#1}%
\global\advance\meqno by1}
%old:
\def\eqnalign#1{\xdef #1{(\secsym\the\meqno)}
\writedef{#1\leftbracket#1}%
\global\advance\meqno by1 #1\eqlabeL{#1}}
%
%%%%%%%%%%%%%%%%%%% %%%%%%%%%%%%%%%%%%%%%%%%%%%%%%%%%%%%%%%%%%%%%%
%%%%%%  macros for titlepage, marginnotes, etc
%%%%%%%%%%%%%%%%%%% %%%%%%%%%%%%%%%%%%%%%%%%%%%%%%%%%%%%%%%%%%%%%%
\def\hsect#1{\hyperref{}{section}{#1}{section~#1}}
\def\hsubsect#1{\hyperref{}{subsection}{#1}{section~#1}}
\def\chap#1{\newsec{#1}}
\def\chapter#1{\chap{#1}}
\def\sect#1{\subsec{#1}}
\def\section#1{\sect{#1}}
\def\\{\ifnum\lastpenalty=-10000\relax
\else\hfil\penalty-10000\fi\ignorespaces}
\def\note#1{\leavevmode%
\edef\@@marginsf{\spacefactor=\the\spacefactor\relax}%
\ifdraft\strut\vadjust{%
\hbox to0pt{\hskip\hsize%
\ifx\answ\bigans\hskip.1in\else\hskip .1in\fi%
\vbox to0pt{\vskip-\dp
%\vskip4pt
\strutbox\sevenbf\baselineskip=8pt plus 1pt minus 1pt%
\ifx\answ\bigans\hsize=.7in\else\hsize=.35in\fi%
\tolerance=5000 \hbadness=5000%
\leftskip=0pt \rightskip=0pt \everypar={}%
\raggedright\parskip=0pt \parindent=0pt%
\vskip-\ht\strutbox\noindent\strut#1\par%
\vss}\hss}}\fi\@@marginsf\kern-.01cm}
\def\titlepage{%
\frontpagetrue\nopagenumbers\abstractfont%
\hsize=\hstitle\rightline{\vbox{\baselineskip=10pt%
{\abstractfont\pubnum}}}\pageno=0}
\frontpagefalse
\def\pubnum{}
\def\pdate{\number\month/\number\yearltd}
\def\makefootline{\iffrontpage\vskip .27truein
\line{\the\footline}
%\vskip -.1truein\line{\pdate\hfil}
\vskip -.1truein\leftline{\vbox{\baselineskip=10pt%
{\abstractfont\pdate}}}
\else\vskip.5cm\line{\hss \tenrm $-$ \folio\ $-$ \hss}\fi}
\def\title#1{\vskip .7truecm\titlestyle{\titleft #1}}
\def\titlestyle#1{\par\begingroup \interlinepenalty=9999
\leftskip=0.02\hsize plus 0.23\hsize minus 0.02\hsize
\rightskip=\leftskip \parfillskip=0pt
\hyphenpenalty=9000 \exhyphenpenalty=9000
\tolerance=9999 \pretolerance=9000
\spaceskip=0.333em \xspaceskip=0.5em
\noindent #1\par\endgroup }
\def\autskip{\ifx\answ\bigans\vskip.5truecm\else\vskip.1cm\fi}
\def\author#1{\vskip .7in \centerline{#1}}

\def\address#1{\ifx\answ\bigans\vskip.2truecm
\else\vskip.1cm\fi{\it \centerline{#1}}}
\def\abstract#1{
\vskip .5in\vfil\centerline
{\bf Abstract}\penalty1000
{{\smallskip\ifx\answ\bigans\leftskip 2pc \rightskip 2pc
\else\leftskip 5pc \rightskip 5pc\fi
\noindent\abstractfont \baselineskip=12pt
{#1} \smallskip}}
\penalty-1000}
\def\endpage{\tenpoint\supereject\global\hsize=\hsbody%
\frontpagefalse\footline={\hss\tenrm\folio\hss}}
\def\ack{\goodbreak\vskip2.cm\centerline{{\bf Acknowledgements}}}
\def\append#1#2{\global\meqno=1\global
\subsecno=0\xdef\secsym{\hbox{#1.}}
\bigbreak\bigskip\noindent{\twelvebf Appendix #1. #2}\message{(#1.
#2)} \writetoca{\twelvebf{Appendix {#1.} {#2}}}\nobreak}
%
%%%%%%%%%%%%%%%%%%%%%%%%%%%%% %%%%%%%%%%%%%%%%%%%%%%%%%%%%%%%%%
\def\bfone{\relax{\rm 1\kern-.35em 1}}
\def\inbar{\vrule height1.5ex width.4pt depth0pt}
\def\IC{\relax\,\hbox{$\inbar\kern-.3em{\mss C}$}}
\def\ID{\relax{\rm I\kern-.18em D}}
\def\IF{\relax{\rm I\kern-.18em F}}
\def\IH{\relax{\rm I\kern-.18em H}}
\def\II{\relax{\rm I\kern-.17em I}}
\def\IN{\relax{\rm I\kern-.18em N}}
\def\IP{\relax{\rm I\kern-.18em P}}
\def\IQ{\relax\,\hbox{$\inbar\kern-.3em{\rm Q}$}}
\def\IR{\relax{\rm I\kern-.18em R}}
\font\cmss=cmss10 \font\cmsss=cmss10 at 7pt
\def\ZZ{\relax\ifmmode\mathchoice
{\hbox{\cmss Z\kern-.4em Z}}{\hbox{\cmss Z\kern-.4em Z}}
{\lower.9pt\hbox{\cmsss Z\kern-.4em Z}}
{\lower1.2pt\hbox{\cmsss Z\kern-.4em Z}}\else{\cmss Z\kern-.4em
Z}\fi}
\def\a{\alpha} \def\b{\beta} \def\d{\delta}
 \def\c{\gamma}
 \def\l{\lambda}
\def\L{\Lambda} 
 
\def\cC{{\cal C}} 
\def\cF{{\cal F}}

\def\cL{{\cal L}} \def\cM{{\cal M}}
 \def\cO{{\cal O}}

\def\nup#1({Nucl.\ Phys.\ $\us {B#1}$\ (}
\def\plt#1({Phys.\ Lett.\ $\us  {#1}$\ (}
\def\cmp#1({Comm.\ Math.\ Phys.\ $\us  {#1}$\ (}
\def\prp#1({Phys.\ Rep.\ $\us  {#1}$\ (}
\def\prl#1({Phys.\ Rev.\ Lett.\ $\us  {#1}$\ (}
\def\prv#1({Phys.\ Rev.\ $\us  {#1}$\ (}
\def\mpl#1({Mod.\ Phys.\ Let.\ $\us  {A#1}$\ (}
\def\ijmp#1({Int.\ J.\ Mod.\ Phys.\ $\us{A#1}$\ (}
\def\tit#1|{{\it #1},\ }
%
%%%%%%%%%%%%%%%%%%%%%%%%%%%%%%%% %%%%%%%%%%%%%%%%%%%%%%%%%%%%%%
%%%%% misc %%%%
%%%%%%%%%%%%%%%%%%%%%%%%%%%%%%%% %%%%%%%%%%%%%%%%%%%%%%%%%%%%%%

%

\def\ni{\noindent}
\def\tilde{\widetilde}
\def\bar{\overline}
\def\us#1{\underline{#1}}

\def\hat{\widehat}
\def\hyp{\vrule height 2.0pt width 2.5pt depth -1.5pt}

\def\Coeff#1#2{{#1\over #2}}
\def\Coe#1.#2.{{#1\over #2}}
\def\coeff#1#2{\relax{\textstyle {#1 \over #2}}\displaystyle}
\def\coe#1.#2.{\relax{\textstyle {#1 \over #2}}\displaystyle}

\def\shalf{\relax{\textstyle {1 \over 2}}\displaystyle}

\def\to{\rightarrow}
\def\notin{\hbox{{$\in$}\kern-.51em\hbox{/}}}
\def\shdot{\!\cdot\!}

\def\del{\partial}

\def\nex#1{$N\!=\!#1$}

\catcode`\@=12
%%%%%%%%% end macros  %%%%%%% %%%%%%%%%%%%%%%%%%%%%%%%%%%%%%
%%%%%%%%%%%specific macros   %%%%%%%%%%%%%%%%%%%%%%%%%%%%%
\def\cW{{\cal W}}
\def\weyA{W_{\!A_{n-1}}}
\def\bifset{\Sigma}

\def\cW{{\cal W}}
\def\g{\gamma}
\def\CM{{\cal M}_0}
\def\cW{{\cal W}}
\def\weyA{W_{\!A_{n-1}}}
\def\simpA{\cW_{\!A_{n-1}}}
\def\bifset{\Sigma}
\def\g{\gamma}
\def\o#1{\omega_{#1}}
\def\O#1{\Omega_{#1}}
\def\ad#1{a_{D#1}}
\def\cF{{\cal F}}
\def\ul#1{^{P_{#1}}}
\def\ql{^{Q_1}}
\def\p{\partial}
\def\a{\alpha}
\def\b{\beta}
\def\c{\gamma}
\def\t{\theta}
\def\ta{\theta_\a}
\def\tb{\theta_\b}

\def\L{\Lambda}
\def\CM{{\cal M}_0}
\def\Class{{\rm class}}
\def\class{{}}
\def\sclass{{}}
\def\onel{{\rm 1\hyp loop}}
\def\vp{\varpi}

\def\Re{{\rm Re\,}}
\def\Im{{\rm Im\,}}
\def\ln{{\,\log\,}}
%%%%%%%%%%%%%%%%%%%%%%%%%%%% %%%%%%%%%%%%%%%%%%%%%%%%%%%%%
%%%%%%%%%%%%%%%%%%%%%%%%%%%%% %%%%%%%%%%%%%%%%%%%%%%%%%%
%
\ldf\SWa{N.\ Seiberg and E.\ Witten, \nup426(1994) 19, \hth/9407087.}
\ldf\SWb{N.\ Seiberg and E.\ Witten, \nup431(1994) 484,
\hth/9408099.}
\ldf\KLTYa{A.\ Klemm, W.\ Lerche, S.\ Theisen and S.\ Yankielowicz,,
\plt B344(1995) 169, \hth/9411048}
\ldf\KLTYb{A.\ Klemm, W.\ Lerche, S.\ Theisen and S.\ Yankielowicz,
{\it On the Monodromies of N=2 Supersymmetric Yang-Mills Theory},
Proceedings of the Workshop on Physics from the Planck Scale to
Electromagnetic Scale, Warsaw, 1994 and of the 28th International
Symposium on Particle Theory, Wendisch-Rietz; preprint
CERN-TH-7538-94, \hth/9412158}
\ldf\AF{P. Argyres and A. Faraggi, \prl 73 (1995) 3931, \hth/9411057}
\ldf\Arn{See e.g., V.\ Arnold, A.\ Gusein-Zade and A.\ Varchenko,
{\it Singularities of Differentiable Maps I, II}, Birkh\"auser 1985.}
\ldf\CDFLLR{A. Ceresole, R. D'Auria and T. Regge, \nup414 (1994) 517;
see also: A.\ Ceresole, R. D'Auria, S. Ferrara, W. Lerche, J. Louis
and T. Regge, {\it Picard-Fuchs Equations, Special Geometry and
Target Space Duality}, preprint CERN-TH.7055/93, POLFIS-TH.09/93, to
be published in ``Essays on Mirror Symmetry, Vol. 2", B. Green and
S.-T. Yau, eds.}
\ldf\thooft{G.\ `t Hooft, \nup190(1981) 455.}
\ldf\HT{C.\ Hull and
P.\ Townsend, {\it Unity of Superstring Dualities}, preprint
QMW-94-30, \hth/9410167.}
\ldf\NS{N.\ Seiberg, \nup303(1988) 286.}
\ldf\NSbeta{N.\ Seiberg, \plt206(1988) 75.}
\ldf\SCHIF{
V.\ Novikov, M.\ Schifman, A.\ Vainstein,
M\ Voloshin and V.\ Zakharov,
\nup229(1983) 394;
V.\ Novikov, M.\ Schifman, A.\ Vainstein and V.\ Zakharov,
\nup229(1983) 381, 407;
M.\ Schifman, A.\ Vainstein and V.\ Zakharov,
\plt166(1986) 329.}
\ldf\FK{For a general reference on Riemann surfaces, see e.g.
H. Farkas and I. Kra, Riemann Surfaces, Springer 1980}
\ldf\VW{See e.g. B. van der Waerden, Algebra, Springer 1994}
\ldf\Appell{P. Appell and J. Kamp\'e de Feriet, Fonctions
Hyperg\'eometriques
and Hyperspheriques - Polynomes d'Hermite,
Gauthier-Villars, Paris (1929); A. Erdelyi et al, Higher
Transcendental
Functions, Vol I, McGraw-Hill (1953)}
\ldf\Yoshida{M. Yoshida, Fuchsian Differential Equations,
Friedr. Vieweg \& Sohn (1987);
T. Sasaki and M. Yoshida, Math. Ann. 282, 69-93 (1988)}
\ldf\Deligne{P. Deligne, Equations differentielles a points
singulieres
r\'eguliers, LNM 163, Springer (1970)}
\ldf\OrTe{P.\ Orlik and H. Terao,
{\it Arrangements of Hyerplanes}, Springer GTM 300 (1992).}
\ldf\MDSS{M.\ Douglas and S.\ Shenker, {\it
Dynamics of SU(N) Supersymmetric Gauge Theory}, preprint RU-95-12,
\hth/9503163.}
\ldf\Stro{A.\ Strominger, {\it Massless Black Holes and Conifolds in
String Theory}, ITP St.\ Barbara preprint, \hth/9504090.}
\ldf\WiVa{E.\ Witten, {\it String Theory Dynamics In Various
Dimensions},
Princeton preprint, \hth/9503124; C.\ Vafa, as cited therein.}
\ldf\BN{R.\ Brandt and F.\ Neri, \nup161 (1979) 253.}
\ldf\DS{U.\ Danielsson and B.\ Sundborg, {\it The Moduli Space and
Monodromies of N=2 Supersymmetric SO(2r+1) Yang-Mills Theory},
preprint USITP-95-06, UUITP-4/95, \hth/9504102.}
\ldf\AnnaEtAl{A.\ Ceresole, R.\ D'Auria and S.\ Ferrara, {\it On the
Geometry of Moduli Space of Vacua in N=2 Supersymmetric Yang-Mills
Theory}, preprint POLFIS-TH.07/94, CERN-TH.7384/94, \hth/9408036; A.\
Ceresole, R.\ D'Auria, S.\ Ferrara and A.\ Van Proeyen, {\it On
Electromagnetic Duality in Locally Supersymmetric N=2 Yang--Mills
Theory}, preprint CERN-TH.7510/94, POLFIS-TH. 08/94, UCLA 94/TEP/45,
KUL-TF-94/44, \hth/9412200; {\it Duality Transformations in
Supersymmetric Yang-Mills Theories coupled to Supergravity}, preprint
CERN-TH 7547/94, POLFIS-TH. 01/95, UCLA 94/TEP/45, KUL-TF-95/4,
\hth/9502072.}
\ldf\DS{U.\ Danielsson and B.\ Sundborg, {\it The Moduli Space and
Monodromies of N=2 Supersymmetric SO(2r+1) Yang-Mills Theory},
preprint USITP-95-06, UUITP-4/95, \hth/9504102.}
\ldf\PAMD{P.\ Argyres and M.\ Douglas, {\it
New Phenomena in SU(3) Supersymmetric Gauge Theory}, preprint
IASSNS-HEP-95/31, RU-95-28, \hth/9505062.}
\ldf\Nf{A.\ Hanany and Y.\ Oz, {\it On the Quantum Moduli Space of
N=2 Supersymmetric SU(Nc) Gauge Theories},
preprint TAUP-2248-95,WIS-95/19/May-PH, \hth/9505075;
P.\ Argyres, M.\ Plesser and A. Shapere, {\it The Coulomb Phase of
N=2 Supersymmetric QCD}, preprint IASSNS-HEP-95/32, UK-HEP/95-06,
\hth/9505100.}
\ldf\KaVa{S.\ Kachru and C.\ Vafa, {\it Exact Results for N=2
Compactifications of Heterotic Strings}, preprint HUTP-95/A016,
\hth/9505105.}

%%%%%%%%%%%%%%%%%%%%%%%%%%%% %%%%%%%%%%%%%%%%%%%%%%%%%%%%%
%\draft
\def\pubnum{
\hbox{CERN-TH/95-104}
\hbox{LMU-TPW 95-7}
\hbox{hep-th/9505150}}
\def\pdate{}
\titlepage
\title
{Nonperturbative Effective Actions of N=2 Supersymmetric Gauge
Theories}
\vskip0.cm\autskip
\author{A.\ Klemm, W.\ Lerche}
\CERN
\vskip .5truecm
\centerline{and}
\vskip-1.7truecm
\author{S.\ Theisen\footnote {a}{Work supported in part by GIF -
the German-Israeli Foundation for Scientific Research}}
\address{Sektion Physik, University of Munich, Germany}
\vskip-1.2truecm
\abstract{
We elaborate on our previous work on \nex2 supersymmetric Yang-Mills
theory. In particular, we show how to explicitly determine the low
energy quantum effective action for $G=SU(3)$ from the underlying
hyperelliptic Riemann surface, and calculate the leading instanton
corrections. This is done by solving Picard-Fuchs equations and
asymptotically evaluating period integrals. We find that the dynamics
of the $SU(3)$ theory is governed by an Appell system of type $F_4$,
and compute the exact quantum gauge coupling explicitly in terms of
Appell functions.}
\vskip.4cm
\vfil
\vskip 1.cm
\ni CERN-TH/95-104\hfill\break
\ni May 1995
\endpage
\baselineskip=14pt plus 2pt minus 1pt
%\sequentialequations

%%%%%%%%%%%%%%%%%%%%%%%%%%%%% %%%%%%%%%%%%%%%%%%%%%%%%%%%%%%%
\chapter{Introduction}
%%%%%%%%%%%%%%%%%%%%%%%%%%%%% %%%%%%%%%%%%%%%%%%%%%%%%%%%%%%%

Seiberg and Witten \doubref\SWa\SWb\ have investigated \nex2
supersymmetric gauge theories, with gauge group $G=SU(2)$, and solved
for their exact non-perturbative low energy effective action. Their
construction has been generalized to arbitrary $SU(n)$ gauge groups
in \doubref\KLTYa\AF. A detailed analysis from the
viewpoint of large $n$ was presented in \MDSS. The relation to
special geometry was discussed in \AnnaEtAl. More recently, the
extension to $SO(2n+1)$ was presented in \DS, as well as an analysis
of the non-local behavior at the cusp points in moduli space \PAMD.
In addition, $SU(n)$ gauge theories with extra matter were considered
in \Nf. There is a considerable overlap between various of these
papers.

The purpose of the present paper is to elaborate on our previous work
\KLTYa\ on $SU(n)$ \nex2 Yang-Mills theory, with particular focus on
$G=SU(3)$; some of the material has already been presented in a short
preview \KLTYb.

For arbitrary gauge group $G$, $N=2$ supersymmetric gauge theories
without matter hypermultiplets are characterized by having flat
directions for the Higgs vacuum expectation values, along which the
gauge group is generically broken to the Cartan sub-algebra. Thus,
the
effective theories contain $r\!=\!{\rm rank}(G)$ abelian \nex2 vector
supermultiplets, which can be decomposed into $r$ \nex1 chiral
multiplets $A_i$ plus $r$ \nex1 $U(1)$ vector multiplets $W_{\alpha
i}$
(we will denote the scalar components of $A_i$ by $a_i$). \nex2
supersymmetry implies that the leading piece (with up to two
derivatives) of the low energy effective lagrangian depends only on a
single holomorphic function, $\cF(A)$:
$$
\cL\ =\ {1\over4\pi}{\rm Im}\,\Big[\, \int \!d^4\theta\,\big(\sum
{\del \cF(A)\over\del A_i}\bar A_i\big) \,+\, \int
\!d^2\theta\,{1\over2} \big(\sum {\del^2 \cF(A)\over\del A_i\del
A_j}W_{\alpha i}W_{\alpha j}\big)\Big]\ . %\eqn\effL
$$
The prepotential $\cF$ describes the geometry of the quantum moduli
space $\cM_\L$, whose metric $\tau$ gives the complexified gauge
coupling constant: $\tau(a) =\del^2_a\cF(a) =
{1\over\pi}\theta_{{\rm eff}}(a)+{8\pi i}(g_{{\rm eff}}(a))^{-2}$.

An important point is that $\cM_\L$ has singularities where the local
effective action description breaks down. This is because certain BPS
monopoles become massless for the corresponding vacuum expectation
values. For example, for $G=SU(2)$ \doubref\SWa\SWb, there are
singularities at $u=\pm\L^2$, where $\L$ is the dynamically generated
scale of the theory, and where $u$ is a gauge invariant coordinate of
$\cM_\L$ (for large $u$, $u\sim\shalf \langle a^2\rangle$). These
singularities correspond to a monopole and a dyon becoming massless,
respectively. (There is also a singularity in the semi-classical
region at $u=\infty$, but there are no massless states associated
with it.)

Loops in $\cM_\L$ around these singularities induce non-trivial
monodromy, which acts on the section
$$
\pi\ =\ \left({a_D\atop a}\right)\ ,\qquad\
a_D\ \equiv\ {\del\over\del a}\cF(a)
$$
via matrix multiplication. A crucial insight of Seiberg and Witten
was to use the global monodromy properties of $\pi$ to essentially
fix it and thus, via integration, to find the prepotential $\cF(a)$.
In practice, this was done by viewing $\pi$ as a vector of period
integrals related to an auxiliary elliptic curve, and $\tau$ as
period matrix of this curve.

In \hsect{2}, we will review some ideas of
Seiberg and Witten about $G=SU(2)$ Yang-Mills theory, with emphasis
on
the techniques that we are going to use later. In \hsect{3}, we will
first describe some properties of the classical $SU(n)$ gauge
theories as well as their relation to simple singularities, and then
discuss their monodromy properties. In \hsect{4} we will consider the
exact $G=SU(n)$ quantum theories, which are defined in terms of
certain
hyperelliptic curves. In particular, we will present some
details about the moduli space $\cM_\L$ and its monodromies. We will
also emphasize the relationship between BPS states and the singular
homology of level surfaces. In \hsect{5} we will derive the
Picard-Fuchs equations for $G=SU(3)$, whose solutions give an
alternative representation of the period integrals.

In \hsect{6} we will use these solutions to compute the series
expansion of the exact quantum effective action, in both
semi-classical and dual magnetic semi-classical coordinate patches of
$\cM_\L$. The explicit expressions for the non-perturbative
corrections represent the main results of the present paper, and may
be viewed as predictions that may --in principle-- be checked by some
other sort of computation.

Finally, we will present some conclusions in \hsect{7}, and in two
Appendices we give some details about the computation of the period
integrals, and present explicit expressions for the periods.

%%%%%%%%%%%%%%%%%%%%%%%%%%%%% %%%%%%%%%%%%%%%%%%%%%%%%%%%%%%%
%%%%%%%%%%\input wlnew

\def\ds{\displaystyle}
\overfullrule=0pt

\def\vp{\varpi}
\def\L{\Lambda}
\def\cL{{\cal L}}
\def\cF{{\cal F}}

\chapter{A quick tour through G=SU(2)}

In this section we review and detail some of the results for
gauge group $SU(2)$ \doubref\SWa\SWb. The main purpose is to
familiarize the reader with the techniques that will be used later
in a more involved context.

By hypothesis, the quantum moduli space
$\cM_\L$ of $SU(2)$ Yang-Mills theory
coincides with the moduli space of the elliptic curve
$$
y^2=W_{A_{1}}^2-\Lambda^{4} = (x^2-u)^2-\Lambda^4\ .
\eqn\SWtwocurve
$$
This curve is equivalent to the curve given in the second
paper of Seiberg and Witten \SWb, in that it has the same
$j$-function. The prime interest is in the periods,
$$
\left({\vp_D\atop\vp}\right)\ =\ {\del\over\del u}\,\pi(u)\ \equiv
{\del\over\del u}\left({a_D\atop a}\right)\ \sim\
\left({\oint_\b\atop\oint_\a }\right)\cdot {dx\over y(x,u)}\ ,
\eqn\perdef
$$
since the prepotential can be obtained directly from them by
integration: $\cF=\int_aa_D(a)$. The periods \perdef\ are largely
fixed by their monodromy properties around the singularities of
$\cM_\L$, which just reflect the monodromy properties of the basis
homology cycles $\a$ and $\b$. Specifically, denoting the the four
zero's of $p(x)=y^2(x)$ by
$e^+_1=-\sqrt{u+\Lambda^2}$,
$e^-_1=-\sqrt{u-\Lambda^2}$,
$e^-_2= \sqrt{u-\Lambda^2}$ and
$e^+_2= \sqrt{u+\Lambda^2}$, we define the basis for the homology
cycles
as in \lfig\figone.
\figinsert\figone{The definition of the cuts and  cycles for the
elliptic
curve \SWtwocurve~in the x-plane. This picture correspond to the
choice
of the basepoint $u_0> \Lambda^2$ real.}{0.8in}{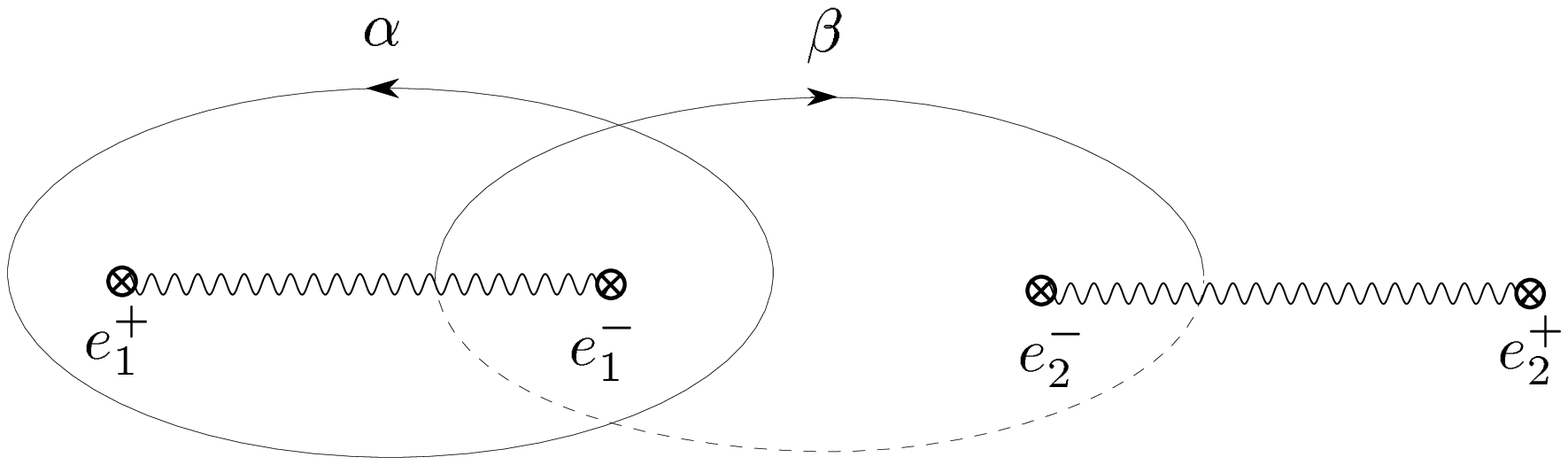}

The singularities in the quantum moduli space are given by the the
zeros of the discriminant of \SWtwocurve, $\Delta_\L=(2
\Lambda)^8(u^2-\Lambda^4)$, and describe the following degenerations
of the elliptic curve:

\noindent
$i_+$) $u\rightarrow +\Lambda^2$, for which $(e^-_1\rightarrow
e^-_2)$, i.e., the cycle $\nu_{+\Lambda^2}=\b$ degenerates,

\noindent
$i_-$) $u\rightarrow -\Lambda^2$, for which $(e^+_1\rightarrow
e^+_2)$, i.e., the cycle $\nu_{-\Lambda^2}=\b- 2\, \a$ degenerates,

\noindent
 $ii$) ${\Lambda^2\over u}\rightarrow 0$, for which
$(e^+_1\rightarrow e^-_1)$ {\sl and } $(e^+_2\rightarrow e^-_2)$. As
two pairs of zero's coincide simultaneously, one has a ``non-stable''
degeneration. With some care (see below) one can conclude:
$\nu_\infty=2\, \a$.

Is is now easy to see that, for example, a loop $\g_{+\L^2}$ around
the singularity at $u=\L^2$ makes $e^-_1$ and $e_2^-$ rotate around
each other, so that the cycle $\a$ gets transformed into $\a-\b$,
c.f., \lfig\figtwo\ (one can investigate the monodromy around
$u=-\L^2$ in an analogous way). Note that the zeros exchange along a
certain path that shrinks as $e^-_1\to e_2^-$. Such paths are
called vanishing cycles, and play an important r\^ole
because they directly determine the monodromies; this will be
explained in \hsect{4.2}.

\figinsert\figtwo{Monodromy paths and cycles that vanishes as one
moves
towards the degeneration point.}{1.8 in}{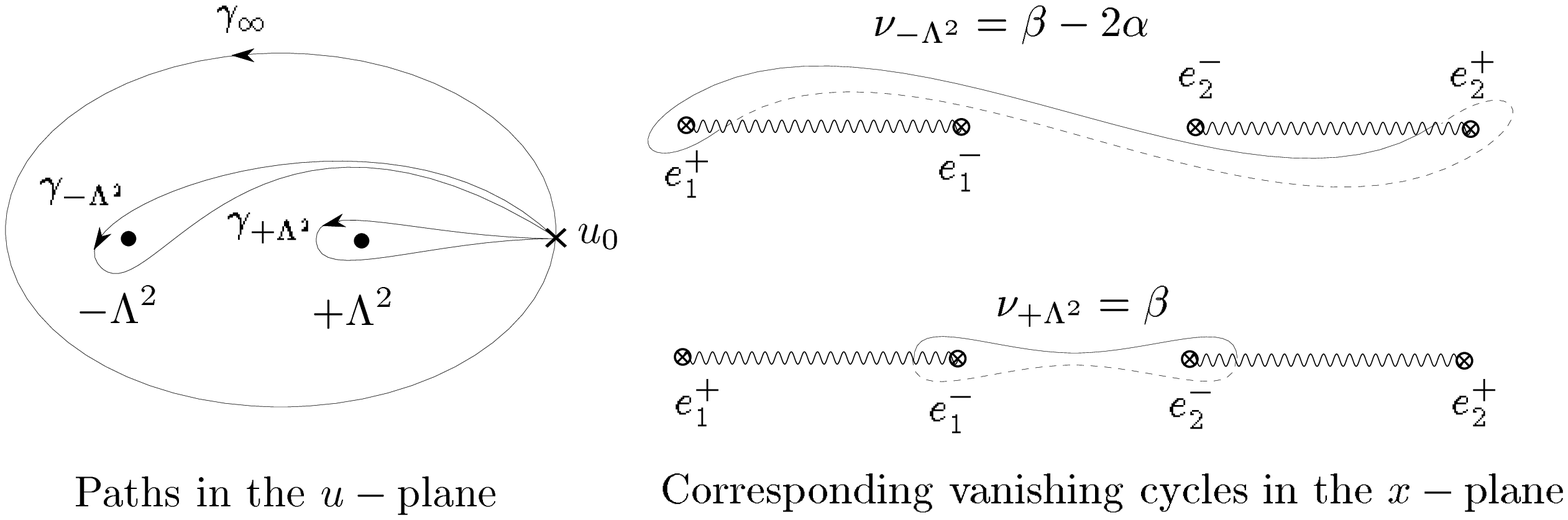}

To obtain the monodromy around ${\Lambda^2\over u}\rightarrow 0$, one
can compactify the $u-$plane to $\IP^1$ with homogeneous coordinates
$(u:\Lambda^2)$, and get the monodromy at infinity from the
consistency condition $M_\infty=M_{+\Lambda^2} M_{-\Lambda^2}$
(\lfig\figtwo). One may also compute the monodromy at
${\Lambda^2\over
u}\rightarrow 0$ directly. To obtain a stable situation with a
single vanishing cycle, one can employ a projective transformation
$x\mapsto {a\,x+b\over c\, x +d}$ that may be fixed by requiring:
$(e_1^+\rightarrow \infty,e_1^-\rightarrow -1, e_2^-\rightarrow 1)$,
plus, for example:
$$
e_2^+\rightarrow \tilde u={ u \Lambda^2\over \sqrt{u^2-\Lambda^4}}\ .
\eqn\uswIuswII
$$
For ${u\over \Lambda^2}\rightarrow \infty$, $\tilde u\rightarrow
1$ i.e. the $\a$-cycle vanishes, but since $(\tilde u-1)\sim
{\Lambda^2\over 2 u^2}+O({\Lambda^8\over u^4})$ this correspond only
to half a loop in the $u-$plane i.e. the vanishing cycle for the
curve \SWtwocurve\ is $\nu_\infty=2\,\a$. One also needs to take into
account a factor $-1$ contributed by the form ${dx\over y}$, as well
as the fact that the monodromy obtained in this way corresponds to
the degeneration point being encircled counter clockwise, ie., the
monodromy will be given by ${M_\infty}^{-1}$.

In summary, one obtains the following monodromies:
%$\left(\matrix{\vp_D(u)\cr \vp(u)}\right) \rightarrow  M_{\gamma}
% %\left(\matrix{\vp_D(u)\cr \vp(u)}\right)$, with
$$
\eqalign{
M_{\infty}=M_{+\Lambda^2} M_{-\Lambda^2}= \left(\matrix{-1& 4\cr 0&
-1}\right),\
M_{+\Lambda^2}=\left(\matrix{1& 0\cr -1& 1}\right),\
M_{-\Lambda^2}=\left(\matrix{-1& 4\cr -1& 3}\right)\ ,
}\eqn\suIImon
$$
which generate $\Gamma_0(4)\subset SL(2\ZZ)$. These monodromies
are consistent with the one-loop $\beta$-function of the weakly
coupled $SU(2)$ theory, with the $\b$-function of the magnetic dual
$U(1)$ theory coupled to a massless monopole of
charges $(g,q)=(1,0)$, and with the $\b$-function corresponding to a
massless dyon of charge $(1,-2)$, respectively \SWb.

In order to obtain the effective action explicitly, one needs to
evaluate the periods \perdef. Instead of directly computing the
integrals,
one may use the fact that the periods form a system of
solutions of the Picard-Fuchs equation associated with \SWtwocurve.
One then has to evaluate the integrals only in leading order, just to
determine the correct linear combinations of the solutions. More
precisely, the PF equations satisfied by the periods
$(\vp_D(u),\vp(u)) \equiv (\partial_u a_D,\partial_u a)$ are given in
terms of the second order differential operator
$
{\cal L}=(\Lambda^4-u^2)\partial_u^2-2 u\partial_u-{1\over4}
$.
In terms of the dimensionless and ${\bf Z}_8$ invariant variable
$\alpha={u^2\over\Lambda^4}$, this turns into
$(\theta_\alpha=\alpha\partial_\alpha)$
$$
{\cal L}=
\theta_\alpha (\theta_\alpha-{1\over2})-
\alpha(\theta_\alpha+{1\over4})^2\ ,
\eqn\PFone
$$
which constitutes the hypergeometric system
$F({1\over4},{1\over4};{1\over2};\alpha)$. It is also possible to
derive a second order differential equation for the section
$\pi\equiv(a_D,a)$ directly. In fact, one easily verifies that ${\cal
L}\partial_u=\partial_u\tilde{\cal L}$ with\foot{The fact that the
operator ${\cal L}\partial_u$ has the alternative factorization,
$\partial_u\tilde{\cal L}$, means that $(a_D,a)$ transform
irreducibly
under monodromy. In the massive case this will no longer be case, and
the three solutions $(a_D,a,{\rm const})$ will mix under monodromy.}
$$
\tilde{\cal L}=\theta_\alpha (\theta_\alpha -{1\over 2}) -\alpha
(\theta_\alpha-{1\over 4})^2\ ,
\eqn\PFtwo
$$
and this form the hypergeometric system
$F(-{1\over4},-{1\over4};{1\over2},\alpha)$. One may also verify
directly that $\oint\lambda$ with $\lambda={i\sqrt{2}\over 4
\pi}2x^2\coeff{dx}y$ satisfies this equation.

The solutions of $\tilde\cL\,\pi=0$ in terms of hypergeometric
functions, and their analytic continuation over the complex plane,
are of course well known. For $|u| > |\Lambda|$ a system of solutions
to the Picard-Fuchs equations is given by $w_0$ and $w_1$ with
$$
w_0(u)={\sqrt{u}\over\L}\sum c(n)({\Lambda^4\over u^2})^n\,,\qquad
c(n)={({1\over4})_n (-{1\over4})_n\over(1)_n^2}
$$
and
$$
w_1(u)=w_0(u) \log({\Lambda^4\over u^2})+
{\sqrt{u}\over\L}\sum d(n) ({\Lambda^4\over u^2})^n\ ,
$$
where
$$
d(n)=c(n)\bigl(2(\psi(1)-\psi(n+1))+\psi(n+{1\over4})-\psi({1\over4})
+\psi(n-{1\over4})-\psi(-{1\over4})\bigr)
$$
and where $(a)_m\equiv \Gamma(a+m)/\Gamma(a)$ is the Pochhammer
symbol. Matching the asymptotic expansions of the period integrals
one finds
$$
a(u)={\Lambda\over \sqrt{2}}w_0(u)\,,
\qquad a_D(u)=-{i\Lambda\over \sqrt{2}\pi}
(w_1(u)+(4-6\log(2)) w_0(u)),$$
which transform under counter-clockwise continuation
of $u$ along $\gamma_\infty$ (c.f., \lfig\figone) precisely as
in \suIImon. These expansions correspond to particular linear
combinations
of hypergeometric functions, the most concise form of which are
$$\eqalign{
a_D(\a)&\ = \oint_\b \lambda  = {i\over 4}\L (\a-1)\,
_2F_1\Big(\Coeff34,\Coeff34, 2 ; 1-\a\Big)\cr
a(\a) &\ = \oint_\a \lambda   =  {1\over 1+i}\L (1- \a)^{1/4}\,
_2F_1\Big(-\Coeff14,\Coeff34,1;\Coeff1{1-a}\Big)\ .  \cr }
\eqn\aaD
$$

{}From these expressions, the prepotential in the semi-classical
regime (near infinity in the moduli space) can readily be obtained to
any given order. Inverting $a(u)$ as series for large $a/\Lambda$
yields for the first few terms ${u(a)\over \Lambda^2}=2 \left(a\over
\Lambda\right)^2 + {1\over 16} \left(\Lambda \over a\right)^2+{5
\over 4096} \left(\Lambda\over a\right)^6 + O(\left(\Lambda\over
a\right)^{10})$. After inserting this into $a_D(u)$, one obtains
${\cal F}$ by integration w.r.t. $a$ as follows:
$$
{\cal F}={i\, a^2\over 2 \pi} \left( 2\log {a^2\over \Lambda^2}- 6 +
8 \log 2 -
\sum_{k=1}^\infty {\cal F}_k \left(\Lambda\over a\right)^{4
k}\right)\ .$$
Specifically, the first few terms of the instanton expansion are:
$$
\vbox{\offinterlineskip\tabskip=0pt
\halign{\strut\vrule#&
\hfil~~$#$~~&
\hfil~~$#$~~&
\hfil~~$#$~~&
\hfil~~$#$~~&
\hfil~~$#$~~&
\hfil~~$#$~~&
\hfil~~$#$~~&
\hfil~~$#$~~&
\hfil~~$#$~~&
\vrule#\cr
\noalign{\hrule}
&k  &1 & 2 &  3& 4   & 5   & 6     & 7     & 8       &\cr
&{\cal F}_k&
\ds{1\over 2^5}&
\ds{5\over 2^{14}}&
\ds{3\over 2^{18}}&
\ds{1469\over 2^{31}}&
\ds{4471\over 2^{34} \cdot 5} &
\ds{40397\over 2^{43}}  &
\ds{441325\over  2^{47}\cdot 7}   &
\ds{866589165\over 2^{64}}&\cr
\noalign{\hrule}}
\hrule}$$

One can treat the dual semi-classical regime is an analogous way.
Near the point $u=\Lambda^2$ where the monopole becomes massless,
we introduce $z=(u-\Lambda^2)/(2 \Lambda^2)$ and rewrite the
Picard-Fuchs operator as
$$
{\cal L}=z (\theta_z-{1\over 2})^2+\theta_z (\theta_z-1)
\eqn\PFdual
$$
At $z=0$, the indices are $0$ and $1$, and we have again one power
series
$$
w_0(z)=\Lambda^2\sum c(n) z^{n+1},
\qquad c(n)=(-1)^n{({1\over2})_n^2\over(1)_n (2)_n}\
\eqn\sersoltwo
$$
and a logarithmic solution
$$
w_1(z)=w_0(z) \log(z)+
\sum d(n) z^{n+1} -4\ ,
\eqn\logsoltwo
$$
with
$$
d(n)=c(n)\Bigl(2(\psi(n+{1\over2})-\psi({1\over2}))
+\psi(1)-\psi(n+1)+\psi(2)-\psi(n+2)\Bigr)\ .
$$
For small $z$ one can easily evaluate
the lowest order expansion for the integrals \aaD\ and
thereby determine the analytic continuation of the
solutions from the weak coupling to the strong coupling domain:
$$
\eqalign{
a_D\ =\ 2 \int_{e_1^-}^{e_2^-}\lambda\ &=\
i\Lambda (z+\ldots)\ =\ i \Lambda w_0(z)\cr
a\ =\ 2 \int_{e_1^+ }^{e_1^- }\lambda\ &=\
{\Lambda\over 2 \pi}(4+ z (1+4 \log(2)) - z\log(z)+\ldots)\cr
&= - {\Lambda\over 2 \pi}(w_1(z)-(1+\log (2)) w_0(z)) \, .}
$$
This exhibits the monodromy of \suIImon\ along the path
$\gamma_{+\Lambda^2}$. Inverting $a_D(z)$ yields $z(a_D)= -2{\tilde
a}_D+\coeff14{{\tilde a}_D}^2+\coeff1{32} {{\tilde a}_D}^3+{\cal
O}({{\tilde a}_D}^4)$, with ${\tilde a}_D\equiv i a_D/\L$. After
inserting
this into $a(z)$ we integrate w.r.t. $a_D$ and obtain the dual
prepotential ${\cal F}_D$ as follows:
$$
{\cal F}_D={i\L^2\over 2 \pi} \left(
{\tilde a}_D^2\ln \Big[-{i\over4}\sqrt{{\tilde a}_D}\Big]
+ \sum_{k=1}^\infty {\cal F}_{D\,k}\, {\tilde a}_D^k
\right),$$
where the lowest threshold corrections ${\cal F}_{D\,k}$ are
$$
\vbox{\offinterlineskip\tabskip=0pt
\halign{\strut\vrule#&
\hfil~~$#$~~&
\hfil~~$#$~~&
\hfil~~$#$~~&
\hfil~~$#$~~&
\hfil~~$#$~~&
\hfil~~$#$~~&
\hfil~~$#$~~&
\hfil~~$#$~~&
\hfil~~$#$~~&
\vrule#\cr
\noalign{\hrule}
&k  &1 & 2 &  3& 4   & 5   & 6     & 7     & 8       &\cr
&{\cal F}_{D\, k}&
\ds{4}&
%-\ds{1\over 2}&
-\ds{3\over 4}&
\ds{1\over 2^4}&
\ds{5\over 2^9}&
\ds{11\over 2^{12}} &
\ds{63\over 2^{16}}  &
\ds{527\over  2^{18}\cdot 5}   &
\ds{3129\over 2^{24}}&\cr
\noalign{\hrule}}
\hrule}$$

Now slightly changing direction, remember that in the first paper of
Seiberg and Witten \SWa, an elliptic curve different from the one in
the second paper \SWb\ was considered. This ``isogenous'' curve has
the form
$$
y^2\ =\ (x-\L^2)(x+\L^2)(x-\tilde u)\ ,
\eqn\SWonecurve
$$
and leads to the monodromy group $\Gamma(2)$.
The motivation for introducing the $\Gamma_0(4)$ curve
\SWtwocurve\ was to have a more convenient electrical charge
normalization, and according to \SWb, the difference between
the curves \SWonecurve\ and \SWtwocurve\ just accounts for this.
This poses, however, a paradox: by comparing the Weierstra\ss\ normal
forms, one finds that the parameters $u$ in \SWtwocurve\ and
$\tilde u$ in \SWonecurve\ are related as follows:\foot{
This is exactly the $SL(2)$ transformation \uswIuswII.}
$$
u\ =\ {\tilde u\over\sqrt{\tilde u^2-\L^4}}\L^2\ .
\eqn\SWmap
$$
This means, however, that the semi-classical regions near infinity
and near the finite points are exchanged for the two curves, which
also means that the electric and magnetic sectors are exchanged. How
can this be reconciled with the statement that the two curves just
differ the normalization of the electric charge ?

The point is that \SWmap\ represents a ($\ZZ_2$-valued) duality
transformation. That is, even though $\pi$ and $\cF(a),\cF_D(a_D)$
transform in a complicated way under
$$
I:\ \qquad \a\ \longrightarrow\ \tilde\a\equiv{\a\over\a-1}\ ,\qquad
\qquad\a\equiv {u^2\over \L^4}\ ,
\eqn\Samap
$$
the physical gauge and dual gauge couplings, $\tau$ and $\tau_D$,
behave in a simple way. More precisely, we find that one can
compactly write the periods $\del_u\pi$ as follows:
$$
\eqalign{
\vp_D(\a)\ &\ = {i \sqrt{2} \over 4 \pi }\oint_\b {dx \over y}
= \ (\Coeff i{2\L}){}_2F_1\Big(\Coeff14,\Coeff14,1;1-\a\Big)\cr
\vp(\a)\ &\ = {i \sqrt{2} \over 4 \pi}\oint_\a {dx \over y}= \
\Coeff1{2\L(1-i)}(1-\a)^{-1/4}
{}_2F_1\Big(\Coeff14,\Coeff14,1;\Coeff1{1-\a}\Big)
\ .\cr}\eqn\Fnicerep
$$
In this particular representation, the arguments of the
hypergeometric
functions simply exchange under the transformation \Samap, and one
finds for expansions around $(1-\a)\in\IR^-$ that
$$
\eqalign{
\tau (\a)\ &\equiv\
{\vp_D(\a)\over\vp(\a)}\ =\ -2{\vp(\tilde \a)\over\vp_D(\tilde\a)}\
\equiv\     2\,\tau_D(\tilde \a)\cr
&=\ \tilde\tau_D(\tilde \a)\ .\cr}
\eqn\taumap
$$
Here, $\tilde\tau_D(\tilde \a)$ is the dual coupling corresponding to
the $\Gamma(2)$ curve \SWonecurve. (For expanding around the fixed
point, $\a=0$, one has to take into account that because of
$(1-\a)\not\in\IR^-$ one picks up an extra overall phase, ie.,
$\tau(u\sim0)= (i+1)+\cO(u) = -2i\tau_D(\tilde u\sim 0)$.) We thus
see that rescaling the electric charges, performing the isogeny map
$I$ (exchanging the curves) and exchanging the electric and magnetic
sectors is the identity map. In fact, \Samap\ can be viewed as the
effect
of the transformation $I:\ \tau\to-\coeff2{\tau}$ on the
$\Gamma_0(4)$ modular function $u$. This is not to be confused with
the $SL(2,\ZZ)$ $S$-duality transformation $S:\
\tau\to-\coeff1{\tau}$, under which the period matrix is invariant:
$$
\tau(\a)\ =\ \tau_D(\hat\a)\ , \ \
\hat\a\ =\
{{{{( 3\,{\sqrt{ \alpha-1 }} + {\sqrt{\alpha }} ) }^2}}\over
     {8\,(  {\sqrt{\a(\alpha-1) }} + \alpha-1 ) }}\ .
\eqn\Staumap
$$
Note that since \taumap, \Staumap\ act non-trivially on the physical
moduli space, these transformations are not a symmetries of the
theory, but rather relate the different semi-classical regimes to
each other.
%%%%%%%%%%%%%%%%%%%%%%%%%%%%% %%%%%%%%%%%%%%%%%%%%%%%%%%%%%%%

%%%%%%%%%%%%%%%%%%%%%%%%%%%%% %%%%%%%%%%%%%%%%%%%%%%%%%%%%%%%
\chapter{Semi-classical Yang-Mills theory}
\section{Classical moduli spaces and Simple Singularities}

Before we jump in the discussion of the exact quantum theory, we
first explain some properties of the classical and semi-classical
(perturbative) theory. Most of these features, especially the
relation to simple singularities, directly generalize to groups other
than $G=SU(n)$.

The classical potential for the scalar superfield component
$\phi$ is given by
$$
V_\class(\phi)\ =\ \Coeff1{g^2}[\phi,\phi^\dagger]^2\ .
\eqn\Vpot
$$
It leads to a continuous family of inequivalent ground states, which
constitutes the classical moduli space $\cM_0$. In order to
characterize $\cM_0$, note that one can always rotate $\phi$ into
the Cartan sub-algebra,
$$
\phi=\sum_{k=1}^{n-1}a_k H_k
\eqn\VEV
$$
with $H_k=E_{k,k}-E_{k+1,k+1},\,
(E_{k,l})_{i,j}=\delta_{ik}\delta_{jl}$. For generic eigenvalues of
$\phi$, the $SU(n)$ gauge symmetry is broken to the maximal torus
$U(1)^{n-1}$, whereas if some eigenvalues coincide, some larger,
non-abelian group $H\subseteq G$ remains unbroken. Precisely which
gauge bosons are massless for a given background $\vec a=\{a_k\}$,
can easily be read off from the central charge formula,
$$
Z_{ q}(a)\  =\ \vec q\cdot \vec a\ ,\ \ \ \
{\rm with\ }\ \  m^2({ q})=2|Z_{ q}|^2\ ,
\eqn\Zdef
$$
where we take for the charge vectors $\vec q$
the roots $\alpha\in\L_R(G)$ in Dynkin basis.

The Cartan sub-algebra variables $a_k$ are not gauge invariant and in
particular not invariant under discrete Weyl transformations.
Therefore, one introduces other variables for parametrizing the
classical moduli space, which are given by Weyl invariant Casimirs
$u_k(a)$. These variables parametrize the Cartan sub-algebra modulo
the Weyl group, ie, $\{u_k\}\cong \IC^{n-1}/\weyA$, and transform
under the the anomaly free global $\ZZ_{2n}$ subgroup of $U(1)_R$ as
$u_k\to e^{i\pi k/n} u_k$. More precisely, in order to go to the
Casimir variables $u_k(a)$, we first change basis according to
$$
Z_{\alpha_{(i,j)}}(a)\ \equiv\ e_i-e_j\ ,
\ \ \ i\not=j\ ,i,j=1,n,\
\eqn\Zeij
$$
with $\sum e_i=0$. These variables are then
related to the Casimir variables $u_k$ by a Miura transformation:
$$
\prod_{i=1}^n\big(x-e_i(a)\big)\ =\
x^n-\sum_{l=0}^{n-2}u_{l+2}(a)\,x^{n-2-l}
\ \equiv\ \simpA(x,u)\ .
\eqn\WAnSing
$$
Here, $\simpA(x,u)$ is nothing but the {\it simple singularity} \Arn\
associated with $SU(n)$, where
$$
\eqalign{
u_k(a)\ &=\ (-1)^{k+1}\sum_{j_1\not=...
\not=j_k}e_{j_1}e_{j_2}\dots e_{j_k}(a)\cr
&\equiv\ {1\over k}\Tr \langle\phi^{k}\rangle + {\rm products\
of\ lower\ order\ Casimirs}\cr
}\eqn\sympol
$$
are the symmetric polynomials. These are manifestly invariant under
the Weyl group $\weyA\cong S(n)$, which acts by permutation of the
$e_i$. Specifically, in terms of the original variables $a_k$, one
has for the bottom and top Casimirs:
$$
u_2(a)\ =\ \coeff1{2n}\sum_{{{\rm positive}\atop{\rm roots}\
\alpha}} \!\!(Z_{\alpha})^2
\ =\ \shalf\,  \vec a^t\cdot C\cdot\vec a\,\qquad\
u_n(a)\ =\ (-1)^{n+1}\!\!\prod_{{{\rm fund.\ rep}\atop{\rm weights}\
\l}}\!\! Z_{\l}(a)\ ,
\eqn\undef
$$
where $C$ is the Cartan matrix of $SU(n)$.  In addition,
let us note for later reference that for $G=SU(3)$:
$$
\eqalign{
u(a_1,a_2)\ &\equiv\ u_2\ =\ {a_1}^2+{a_2}^2-a_1a_2\cr
v(a_1,a_2) &\equiv\ u_3\ =\ a_1 a_2 (a_1-a_2)\cr
a_1(u,v)\ &\equiv\ e_1(u,v)\ =\ \xi_++\xi_- \cr
a_2(u,v)\ &\equiv\ -e_2(u,v)\ =\
e^{-2\pi i/6}\xi_++e^{2\pi i/6}\xi_-\ ,\ \ {\rm where}\cr
\xi_\pm(u,v)\ &\equiv\ {2^{-1/3}}
\root{{\scriptstyle {3\,}}}\of{v\pm\sqrt{v^2-\coeff4{27} u^3}}\ ,
}\eqn\aedef
$$ and
$$
\eqalign{
Z_1\ &\equiv Z_{(2,-1)}\ =\ 2 a_1-a_2\ =\ e_1-e_3\cr
Z_2\ &\equiv Z_{(-1,2)}\ =\ 2 a_2- a_1\ =\ e_3-e_2\cr
Z_3\ &\equiv Z_{(1,1)}\ \ \ =\ a_1+a_2\ \ \,=\ e_1-e_2\ .\cr
}\eqn\Zdef
$$

{}From the above we know that whenever $e_i=e_j$ for some $i$ and
$j$, there are, classically, extra massless non-abelian gauge bosons,
since $Z_{ \a}=0$ for some root $\a$. For such backgrounds
the effective action becomes singular. The classical moduli space is
thus given by the space of Weyl invariant deformations modulo such
singular regions: $\CM=\{u_k\}\backslash\bifset_0$. Here,
$\bifset_0\equiv\{u_k: \Delta_0(u_k)=0\}$ is the zero locus of the
``classical'' discriminant
$$
\Delta_0(u)\ =\
\prod_{i<j}^n(e_i(u)-e_j(u))^2\ \equiv\
\prod_{{{\rm positive}\atop{\rm roots}\
\alpha}}\!\!(Z_{\alpha})^2(u)\ ,
\eqn\cdiscdefu
$$
of the simple singularity \WAnSing. Specifically, one has up to
normalization:\foot{We will often denote $u_2,u_3,u_4$ by $u,v,w$.}
$$
\eqalign{
&SU(2):\ \ \ \Delta_0\ = u\cr
&SU(3):\ \ \ \Delta_0\ =\ 4u^3-27v^2\cr
&SU(4):\ \ \ \Delta_0\  =\ 27\,{v^4}-4\,{u^3}\,{v^2} + 16\,{u^4}\,w -
144\,u\,{v^2}\,w +
   128\,{u^2}\,{w^2} + 256\,{w^3} \cr
}\eqn\discex
$$
We schematically depicted these singular loci $\Delta_0(u)=0$
in \lfig\figdisc.

\figinsert\figdisc{Singular loci $\bifset_0$ in the classical moduli
spaces $\CM$ of pure $SU(n)$ \nex2 Yang-Mills theory. They are
nothing but the bifurcation sets of the type $A_{n-1}$ simple
singularities, and reflect all possible symmetry breaking patterns in
a gauge invariant way (for $SU(3)$ and $SU(4)$ we show only the real
parts). The picture for $SU(4)$ is known in singularity theory as the
``swallowtail''. }{1.2in}{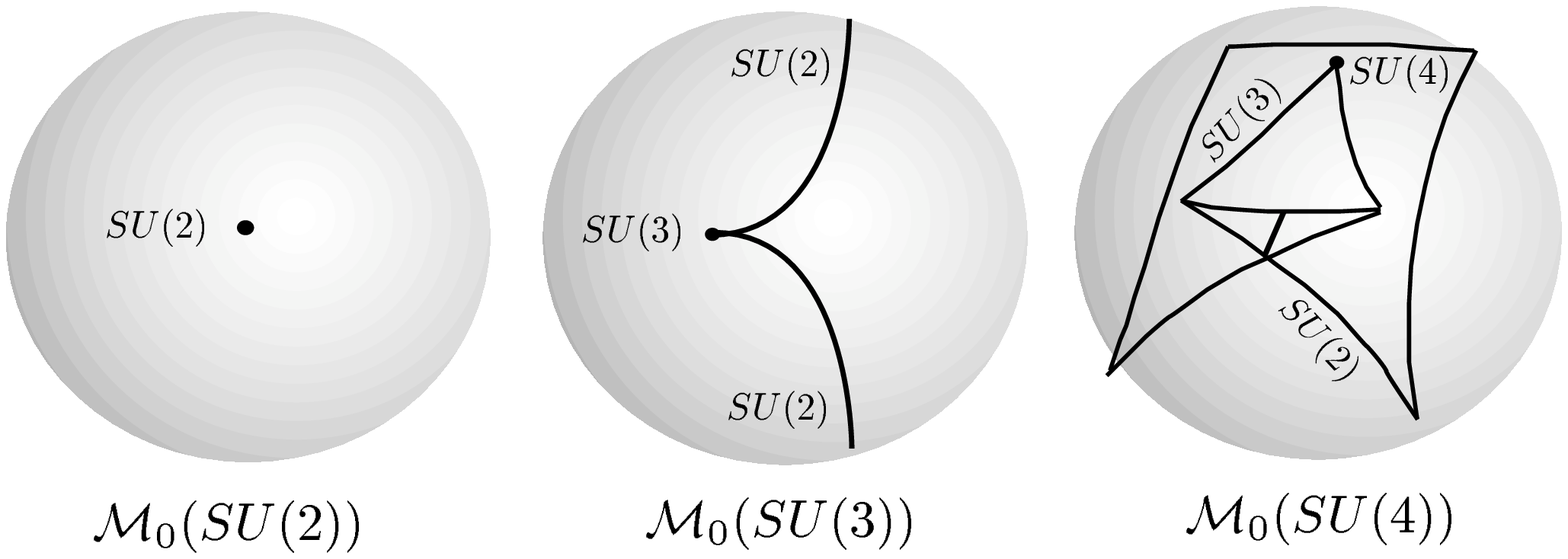}

The discriminant loci $\bifset_0$ are generally given by intersecting
hypersurfaces of complex codimension one. On each such surface one
has $Z_{\alpha}=0$ for some pair of roots $\pm\a$, so that
there is an unbroken $SU(2)$, and the Weyl group action $r_{\a}:\,
Z_{\alpha}\to-Z_{\alpha}$ is singular. On intersections of
these hypersurfaces one has, correspondingly, larger unbroken gauge
groups. All planes together intersect in just one point, namely in
the origin, where the gauge group $SU(n)$ is fully restored. Thus,
all possible classical symmetry breaking patterns are encoded in the
discriminants of $\simpA(x,u)$.

Finally, note that for a general singularity with $n$ variables, the
discriminant locus $\bifset_0$ coincides with what is called the
level bifurcation set of a singularity $\cW(x_i,u)$, because on it
the level surface
$$
V_u\ =\ \big\{\,x_i\,:\,\cW(x_i,u)=0,\ ||x||\leq\epsilon\big\}
\eqn\levsurf
$$
becomes singular \Arn. More specifically, $V_u$ becomes singular in
that certain homology cycles $\nu\in H_{n-1}(V_u,\ZZ)$ shrink to
zero. Such cycles $\nu$ are called vanishing cycles. For the case at
hand, the level surface determined by $\simpA(x,u)=0$ is zero
dimensional and given by the set of points
$V_u=\{e_i,\,i\!=\!1,\dots,n\}$. This space is singular if any two of
the $e_i(u)$ coincide, and indeed, the vanishing cycles are given by
the differences, $\nu_{i,j}=e_i-e_j$, ie., by the central charges
$Z_{\a}$. They generate the root lattice: $H_0(V_u,\ZZ)\cong\L_R$.

What we learn is that the classically massless
non-abelian gauge bosons are directly related to vanishing cycles of
level surfaces. This reflects an apparent general property of BPS
states and, as will be explained below, generalizes in particular to
massless magnetic monopoles and dyons in the exact quantum theory,
where the relevant ``level'' surfaces are given by special Riemann
surfaces.

\section{Classical and semi-classical monodromy$^*$}
\footnote {} {* While
preparing the manuscript, we received the preprint \DS\ with
results that overlap with some results of this section.}

Since the map $u_k\to a_k$ is multi-valued and since $u_k$ are Weyl
group invariant, closed paths in $\cM_0$ space will in general be
closed in $\{a_k\}$-space only up to Weyl transformations. This means
that the classical part of the monodromy group is given by the
corresponding Weyl group, which is a well-known fact in the
theory of simple singularities \Arn. More precisely, the singular
locus $\bifset_0$ has various branches that are the images of the
lines $Z_{\a}=0$ in $a$-space, and encircling such a branch will
induce a classical monodromy given by a Weyl reflection corresponding
to the root $\a$. In addition, we have monodromy acting on the
dual magnetic variables,
$$
a_{Di}\ \equiv\ {\del\over\del a_i}\cF(a)\ ,
\eqn\aDdef
$$
which is dual to the monodromy acting on the $a_i$. Hence the total
classical monodromy is
$$
\left({\vec a_{D}\atop\vec a}\right)\ \longrightarrow\
P^{(r)}\shdot
\left({\vec a_{D}\atop\vec a}\right)\ ,
\qquad\ {\rm where}\ \ \
P^{(r)}\ =\ \pmatrix{(r^{-1})^t & 0\cr0 & r\cr}\ ,
\eqn\classmono
$$
with $r\in W_G$. Which specific Weyl transformation actually occurs
depends of course on the specific closed path in $\cM_0$. It is clear
that all possible classical monodromies can be generated by loops
associated with the fundamental Weyl reflections. These have
the following matrix representations:
$$
r_i\ =\ \bfone - \a_i\otimes\l_i\ ,\ \ \ i=1,...,n-1\ ,
\eqn\ridef
$$
where $\a_i,\l_i$ are the simple roots and fundamental weights
in Dynkin basis.

In addition to the classical monodromy \classmono, there is a
semi-classical contribution (``$\theta$-shift'') from the logarithmic
piece of the one-loop effective action. Specifically, the general
formula for the one-loop correction to the gauge coupling constant is
$
\delta({1\over g^2})_{\a\b}\sim\
\Tr(T_\a T_\b\ln[m^2(a)])
$,
where $m^2(a)$ is the vev-dependent mass matrix of the non-abelian
gauge bosons, and $T_\a$ denotes a generator in the adjoint
representation. From this one obtains the one-loop correction to the
prepotential as follows:
$$
\cF_\onel(a(u))\ =\ {i\over 4\pi}
\sum_{{{\rm positive}\atop{\rm roots}\
\alpha}}(Z_{\alpha})^2\ln[(Z_{\alpha})^2/\L^2] \ .
\eqn\clasF
$$
The scale parameter $\Lambda$ is arbitrary and reflects the breakdown
of conformal invariance at the quantum level. Note that the gauge
coupling constant, $\tau_{ij}=\del_{a_i}\del_{a_j}\cF$, blows up
logarithmically precisely when $\Delta_0=0$, ie., whenever there are
massless charged fields in the theory that lead to an IR divergence.
Note also that even though $\cF_\onel$ is manifestly Weyl group
invariant, it does not have a simple form in terms of the Casimir
variables $u_k$.

The effect of $\cF_\onel$ is to modify the classical monodromy by a
perturbative quantum piece. That is, the monodromy around a singular
line in $\cM_0$ that is the image of some singular line
$Z_{\a}=0$ in weight space, will induce a $2\pi i$ phase
contribution from $\ln(Z_{\alpha})^2$ in $\del_a\cF_\onel(a(u))$.
Though there exist in general more complicated paths, it suffices
\DS\ to consider the following ``fundamental'' monodromies that
generate the semi-classical monodromy group:
$$
M_\sclass^{(r_i)}\ =\ P^{(r_i)}\cdot {T_{\a_i}}^{-1}
\eqn\scMono
$$
with theta-shifts given by\foot{Note that the off-diagonal terms can
be modified at wish by a change of homology basis, ie., $a_{Di}\to
a_{Di}+ h_{ij}a_j$. A different basis was used in the discussion
of the semi-classical monodromy in \doubref\KLTYa\KLTYb.}
$$
T_{\a_i}\ =\
\pmatrix{\bfone &\a_i\otimes\a_i\cr0&\bfone\cr}\ .
\eqn\thetashit
$$

For our discussion of the Picard-Fuchs equations, we will be
particularly interested in the monodromy associated with a large loop
(around ``infinity'') in a given $u_l$-plane, the other $u_k$ being
held fixed. Such a monodromy is given by a certain product of the
generators \scMono, depending on which sub-set of branches of
$\bifset_0$ is actually encircled by the large loop. Though the
precise monodromy transformation depends on the chosen basepoint
$u_0$, the conjugacy class of the Weyl transformation just depends on
the given $u_l$-plane. That is, for large loops $u_l\to e^{2\pi i
t}u_l$, where $t\in[0,1]$, the classical Weyl monodromy is of order
$l$. In particular, for large loops in the $u_2$-plane, the monodromy
at infinity is given by a single Weyl reflection, while for large
loops in the top Casimir plane, $u_n$, the monodromy is given by a
Coxeter element.

One can also consider monodromy induced by a rotation of the quantum
scale $\L$. Indeed, as will be discussed later, one may view $\L$ as
an additional coupling constant that may be used to compactify the
moduli space. A rotation $\L^{2n}\to e^{2\pi i t}\L^{2n}$,
$t\in[0,1]$, induces a mere $\theta$-shift, and from \clasF\ and
\undef\ one obtains the following matrix representation:
$$
T\ =\
\pmatrix{\bfone &C\cr0&\bfone\cr}\ .
\eqn\quantmono
$$
This may be viewed as ``pure'' quantum monodromy. Note that from the
point of view of $R$-symmetry, only $\theta$-shifts associated with
$T^2$ are allowed.

To give an example, we have for $G=SU(3)$ the following
semi-classical
monodromies, corresponding the three singular lines in $\cM_0$:
$$
\eqalign{
M_\sclass^{(r_1)}\ &=\ \pmatrix{ -1 & 0 & 4 & -2 \cr 1 & 1 & -2 & 1
\cr 0 & 0 & -1 & 1 \cr 0 & 0 & 0 & 1 \cr }\cr M_\sclass^{(r_2)}\ &=\
\pmatrix{ 1 & 1 & 1 & -2 \cr 0 & -1 & -2 & 4 \cr 0 & 0 & 1 & 0 \cr 0
& 0 & 1 & -1 \cr }\cr M_\sclass^{(r_3)}\ \equiv\
M_\sclass^{(r_2)}M_\sclass^{(r_1)}(M_\sclass^{(r_2)})^{-1}\ &=\
\pmatrix{ 0 & -1 & 1 & 4 \cr -1 & 0 & -2 & 1 \cr 0 & 0 & 0 & -1 \cr
0 & 0 & -1 & 0 \cr }
}\eqn\SuThreeScMo
$$
These three singular lines correspond of course to the three ways of
embedding $SU(2)\hookrightarrow SU(3)$, and indeed, the matrices
appropriately contain the semi-classical monodromy in
\suIImon\ of $SU(2)$ Yang-Mills theory as sub-matrices. They are
related via conjugation by the Coxeter element
$$
U\ =\
\pmatrix{ -1 & -1 & 1 & -2 \cr 1 & 0 & -2 & 1 \cr 0 & 0 & 0 & -1 \cr
  0 & 0 & 1 & -1 \cr  }\ ,\qquad\ \ \ U^3=\bfone\ ,
\eqn\Ucox
$$
which represents the global symmetry of $\ZZ_3$ rotations acting on
$u$.
Since all three lines cut the $v=const.$-plane, but only two lines
cut the $u=const.$-plane, we have two different kinds of monodromies
at ``infinity'':
$$
\eqalign{
&u-{\rm plane}:\ \ M_{\infty,u}^{(r_2)}\ \equiv\
M_\sclass^{(r_3)}M_\sclass^{(r_2)}M_\sclass^{(r_1)}\ =\
\pmatrix{ 1 & 1 & -3 & 0 \cr 0 & -1 & -6 & 6 \cr 0 & 0 & 1 & 0 \cr 0
   & 0 & 1 & -1 \cr  }
\cr
&v-{\rm plane}:\ \ M_{\infty,v}^{(r_{cox})}\ \equiv\
M_\sclass^{(r_1)}M_\sclass^{(r_2)}\ \ =\
\pmatrix{ -1 & -1 & 1 & 4 \cr 1 & 0 & -2 & 1 \cr 0 & 0 & 0 & -1
   \cr 0 & 0 & 1 & -1 \cr  }
}\eqn\infmonodr
$$
(up to conjugation, depending on base point and offset of the chosen
plane).

%%%%%%%%%%%%%%%%%%%%%%%%%%%%% %%%%%%%%%%%%%%%%%%%%%%%%%%%%%%%

%%%%%%%%%%%%%%%%%%%%%%%%%%%%% %%%%%%%%%%%%%%%%%%%%%%%%%%%%%%%
\chapter{Quantum Yang-Mills Theory}
\section{Hyperelliptic curves and quantum moduli spaces}

The issue is to construct auxiliary curves $\cC$ whose moduli spaces
give the supposed quantum moduli spaces of $SU(n)$ Yang-Mills
theories.
Such curves  were found in \doubref\KLTYa\AF\ and are given by:
$$
\cC:\ \ y^2=p(x)\ =\ \left(W_{A_{n-1}}(x,u_i)\right)^2-\Lambda^{2n}\
\equiv\
(x^n-\sum_{i=2}^n u_i x^{n-i})^2-\Lambda^{2 n}\ .\eqn\aaa
$$
Since $p(x)$ factors into $W_{A_{n-1}}\pm\L^n$, the situation is in
some respect like two copies of the classical theory, with the top
Casimir $u_n$ shifted by $\pm\L^n$. Specifically, the points $e_i$ of
the classical level surface split,
$$
e_i(u_k)\ \to\ e_i^\pm(u_k,\L) \equiv\
e_i(u_2,,...,u_{n-1},u_n\pm\L^n)\ ,
\eqn\eipm
$$
and become the $2n$ branch points of the hyperelliptic curve \aaa.
The curve can thus be represented by the two-sheeted $x$-plane with
cuts running between pairs $e_i^+$ and $e_i^-$. In addition, the
``quantum'' discriminant, whose zero locus $\bifset_\L$ gives the
singularities in the quantum moduli space $\cM_\L$, is easily seen to
factorize as follows:
$$
\eqalign{
\Delta_\Lambda(u_k,\L)\ &\equiv\
\prod_{i<j}(e_i^+-e_j^+)^2(e_i^--e_j^-)^2 \ =\ {\rm
const.}\,\L^{2n^2}
\delta_+\,\delta_-\ ,\ \ {\rm where}\cr \delta_\pm(u_k,\Lambda)\ &=\
\Delta_0(u_2,...,u_{n-1},u_n\pm\L^n)\ ,}
\eqn\DLdef
$$
is the shifted classical discriminant \cdiscdefu. Thus,
$\bifset_\L$ consists of two copies of $\bifset_0$, shifted by
$\pm\L^n$ in the $u_n$ direction. Obviously, for $\L\to0$, the
classical moduli space is recovered: $\bifset_\L\to \bifset_0$. That
is, when the quantum corrections are switched on, a single isolated
branch of $\bifset_0$ (associated with massless gauge bosons of a
particular $SU(2)$ subgroup) splits into two branches of $\bifset_\L$
(describing massless Seiberg-Witten monopoles related to this
$SU(2)$).

\ni Specifically, for $G=SU(3)$ the curve is
$$
y^2=p(x)=(x^3-u x-v)^2-\Lambda^6\ ,\eqn\fff
$$
and this leads to the following quantum discriminant:
$$
\Delta_\L=\Lambda^{18}(4 u^3-27(v+\Lambda^3)^2)
(4 u^3-27(v-\Lambda^3)^2)\ .\eqn\ggg
$$
The corresponding singular locus $\bifset_\L$ of the quantum moduli
space is
depicted in \lfig\figQantMs. Explicit expressions for the branch
points $e_i^\pm(u,v,\L)$ can easily be inferred from \aedef\ and
\eipm.

\figinsert\figQantMs{Quantum moduli space for $G=SU(3)$ at real $v$.
The six lines are the singular loci where $\Delta_\Lambda=0$ and
where certain dyons become massless. Asymptotically, for $\L\to0$ the
classical moduli space is recovered. Each of the six pairs of
outgoing lines represents a copy of the $SU(2)$ strong coupling
singularities. The various markings of the lines indicate how the
association with particular monodromy matrices changes when moving
through the cusps, and $u_0$ is the basepoint that defines our
monodromies. }{4.2in}{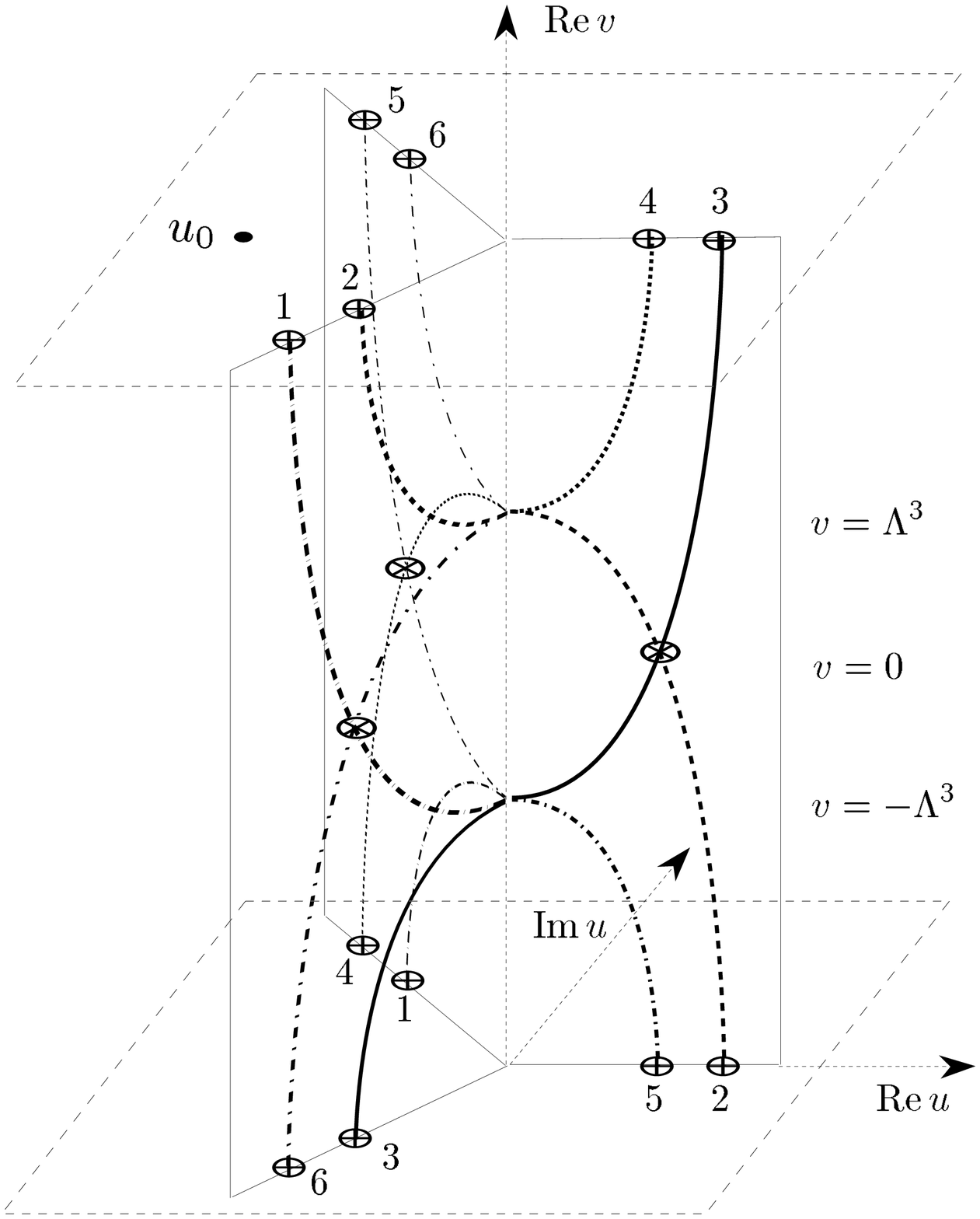}

\goodbreak
The genus of the hyperelliptic curve $\cC$ \aaa\ is equal to $g=n-1$,
so that its $2n-2$ periods can naturally be associated with
$$
\vec \pi\ \equiv\ \left({\vec a_{D}\atop\vec a}\right)\ .
\eqn\pidef
$$
More precisely, on such a curve there are $n-1$ holomorphic
differentials (abelian differentials of the first kind) \FK\
$\omega_{n-i}={x^{i-1}\,dx\over y},\, i=1,\dots,g$, out of which one
can construct $n-1$ sets of periods $\int_{\gamma_j}\omega_i$. (Here
$\gamma_j,\, j=1,\dots,2g$, is any basis of $H_1(\Sigma_{g},{\bf
Z})$.) All periods together can be combined in the
$(g,2g)$-dimensional period matrix
$$
\Pi_{ij}=\int_{\gamma_j}\omega_i\ .
\eqn\ccc
$$
If we chose a symplectic homology basis, i.e. $\a_i=\gamma_i,\,
\b_i=\gamma_{g+i},\,i=1,\dots,g$, with intersection pairing\foot{We
use the convention that a crossing between the cycles $\a$, $\b$
counts positively to the intersection $(\a\cap\b)$, if looking in the
direction of the arrow of $\a$ the arrow of $\b$ points to the
right.}
$(\a_i\cap\b_i)=\delta_{ij},\,(\a_i\cap\a_j)=(\b_i\cap\b_j)=0$, and
if we write $\Pi=(A,B)$, then $\tau\equiv A^{-1}B$ is the metric on
the quantum moduli space. By Riemann's second relation Im$\tau\equiv
8\pi^2/g_{eff}^2$ is positive, which is important for unitarity of
the
effective $N=2$ supersymmetric gauge theory.

The precise relation between the periods and the components of the
section $\vec\pi$ is given by:
$$
\eqalign{
A_{ij}\ &=\ \int_{\a_j}\!\omega_i\ =\
{\partial\over\partial u_{i+1}}\, a_j\cr
B_{ij}\ &=\ \int_{\b_j}\!\omega_i\ =\
{\partial\over\partial u_{i+1}}\, a_{D_j}
}\eqn\ddd
$$
(where $i,j=1,\dots,n-1$). From the explicit expression \aaa\ for the
family of hyperelliptic curves, one immediately verifies that the
integrability conditions $\partial_{i+1} A_{jk}=\partial_{j+1}
A_{ik},\,\partial_{i+1} B_{jk}=\partial_{j+1} B_{ik}$ are satisfied.
It also follows that $\tau_{ij}\equiv \del_{a_i}\del_{a_j}\cF(a)$.
This reflects the special geometry of the quantum moduli space, and
implies that the components of $\vec \pi$ can directly be expressed
as integrals over a suitably chosen abelian differential of the
second kind $\lambda$:
$$
a_{D_i}=\int_{\b_i}\lambda\,,\qquad a_i=\int_{\a_i}
\lambda\eqn\eee
$$
Indeed, one verifies that e.g.,
$$
\lambda={\rm const.}\,{1\over2\pi i}{
\Big({\partial \over\partial x}W_{A_{n-1}}(x,u_i)\Big)}{x\,dx\over y}
\eqn\Lamdef
$$
does the job \AF\ (as well as the choice of $\l$ given in \KLTYa).
The normalization is fixed by matching $a_i$ to the Casimirs $u_k$ in
the semi-classical limit (c.f. below).

In order to compute the exact quantum effective action $\cF(A)$, one
first needs to determine the section $\vec \pi=(\vec a_D,\vec a)^t$
in terms of the coordinates $u_k$. One way to do this would be to
evaluate the integrals \eee\ explicitly. This was done in ref.\
\MDSS\ to lowest order near the point in moduli space where there are
massless dyons. However, it is rather difficult to perform the period
integrals explicitly or, at least, in a series expansion to higher
order. This would be needed for obtaining the instanton corrections
in the effective Lagrangian.

In fact, the integrals $\int_{\gamma_i}\omega$, where $\omega$ is any
abelian diffential of the second kind, satisfy as functions of the
moduli $u_i$ a system of partial differental equations, namely the
Picard-Fuchs equations. It turns out that it is relatively
straightforward to derive and solve these equations, and this will be
discussed below in \hsect{5}. Note, though, that this will
not save us completely from having to evaluate some integrals,
because that will be necessary in order to identify the correct
linear
combinations of the solutions. However, for this
one needs to compute the integrals only to low order.
This is done in Appendix A, and will be used in \hsect{6}.

%%%%%%%%%%%%%%%%%%%%%%%%%%%%%%%%%
\section{BPS states, vanishing cycles and Picard-Lefshetz monodromy}
%%%%%%%%%%%%%%%%%%%%%%%%%%%%%%%%%

We will now discuss how one can obtain the quantum numbers of the
various massless BPS dyons, as well as the associated strong coupling
monodromies. The basic point is to relate all quantum numbers and
monodromy
properties to properties of the homology cycles $\vec\a,\vec\b\in
H_1(\Sigma_g,\ZZ)$, which are involved in the definition of the
periods $\vec\pi$ \eee. Specifically, we expect certain dyons to be
massless on the various branches of the singular locus $\bifset_\L$
in the quantum moduli space. On any such isolated branch, the surface
$\Sigma_g$ becomes singular in that a particular homology cycle $\nu$
vanishes; for a schematic sketch, see \lfig\figdegen. Now, any such
cycle can be expanded in terms of the basis cycles as
follows:
$$
\nu\ =\ \vec q\cdot\vec\a + \vec g\cdot\vec\b\ ,\ \qquad
q_i,g_i\in\ZZ\ .
\eqn\defVC
$$
Since this cycle vanishes, it immediately follows that
$$
\eqalign{
0\ =\ \int_\nu\l\
&=\ \big(\vec q\int_{\vec\a}\,+\,\vec g\int_{\vec\b}\big)\,\l\cr
&=\ \vec q\cdot\vec a + \vec g\cdot\vec a_D\cr
&\equiv Z_{(\vec q,\vec g)}\ .
}\eqn\Zqg
$$
Here, $Z$ is the central charge that enters in the BPS mass formula:
$m^2=2|Z|^2$. This means that on the branch of the singular locus
where some cycle $\nu$ as defined in \defVC\ vanishes, a dyon with
(magnetic,electric) charges equal to $\vec\nu=(\vec g,\vec q)$
becomes massless. Clearly, under a change of homology basis, the
charges change as well, but this is nothing but a duality rotation.
What remains invariant is the intersection number
$$
\nu_i\cap\nu_j\ =\ \vec\nu^t\cdot\Omega\cdot\vec\nu\
= \vec g_i\cdot \vec q_j-\vec g_j\cdot \vec q_i\ \in\
\ZZ\ , \eqn\dzw
$$
where
$$
\Omega\ =\ \pmatrix{0&\bfone\cr-\bfone&0\cr}\eqn\intmet
$$
is the standard symplectic metric. Note that \dzw\ represents the
well-known Dirac-Zwanziger quantization condition for the possible
electric and magnetic charges. The vanishing of the r.h.s.\ is
required for two dyons to be local with respect to each other and, in
particular, to be able to condense simultaneously \thooft. Thus, only
states that are related to non-intersecting vanishing cycles are
mutually local, and can be simultaneously represented by a local
effective lagrangian. For example, since monopoles are associated
with $\b$-cycles and the gauge bosons with (combinations of)
$\a$-cycles, these fields can in general not be represented together
in a local lagrangian.

\figinsert\figdegen{On the singular locus $\bifset_\L$, the level
surface degenerates by pinching of vanishing cycles $\nu$. The
coordinates of any such cycle with respect to some symplectic basis
of $H_1(\cC,\ZZ)$ gives the electric and magnetic quantum numbers of
the corresponding massless dyon.}{2.0in}{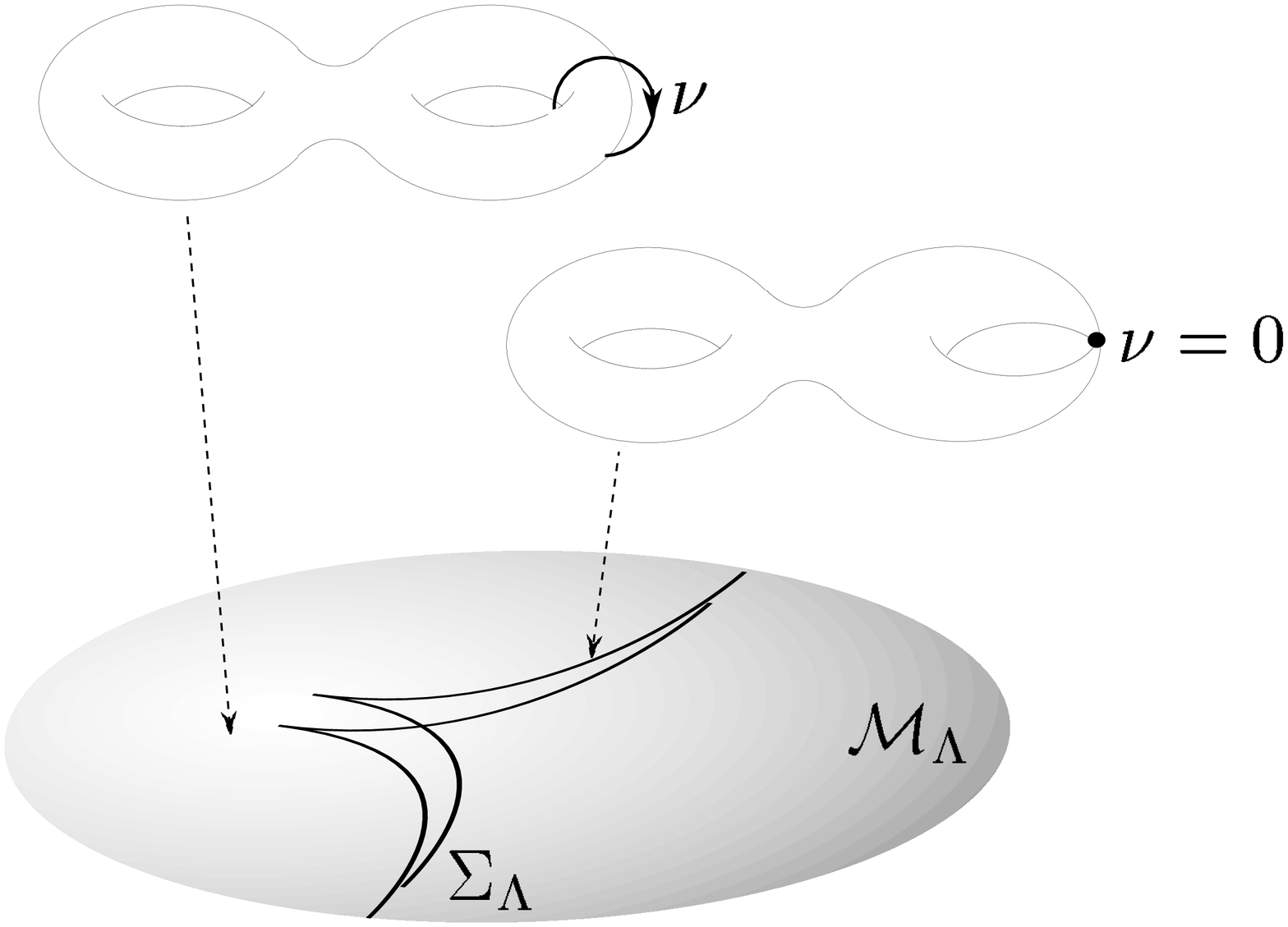}

Note that according to \Zdef, the electric charges are root lattice
vectors in Dynkin basis, $\vec q\in\L_R$. In order for products $\vec
q_i \vec g_j$ to make sense, it follows that $\vec g$ are vectors in
the dual, ie., simple root basis, such that $\vec q\,\vec g\equiv
q_i\langle\l_i,\a_j\rangle g_j$. The $\a$-cycles can thus be thought
of
as to generate the weight lattice and the $\b$-cycles to generate the
root lattice, so that the period lattice can be interpreted as
$H_1(\cC,\ZZ)\cong\L_W\oplus\tau\L_R$. For our particular curves
\aaa, the $\a$-cycles do not vanish anywhere in the moduli space, so
that the gauge bosons (or any other purely electrically charged
states) are never massless. This means that the Milnor lattice
generated by the vanishing cycles is a sub-lattice of the above
period lattice.

If we like to represent both electric and magnetic charges in the
same basis, then, of course, the metric \intmet\ changes. For
example, we can take the electric charges in the simple root basis as
well, by conjugation with
$$
W\ =\ \pmatrix{\bfone&0\cr0&C}\ ,\eqn\Wdef
$$
where $C$ is the Cartan matrix. In this basis, the semi-classical
monodromies \scMono\ and intersection metric are
$$
M^{(r_i)}_W\ =\ \pmatrix{(r^{-1})^t & \a_i\otimes\l_i\cr0 & r\cr}
\ ,\qquad\ \ \Omega_W\ =\ \pmatrix{0&C\cr-C&0\cr}\ .
\eqn\WscMono
$$
This basis corresponds to the one which was used in the first paper
of
Seiberg and Witten and is appropriate to the $\Gamma(2)$
curve \SWonecurve.

We now turn to the monodromies that arise when we loop around the
various branches of $\bifset_\L$. We noted that there is a
particular vanishing cycle $\nu$ associated with any such branch. The
monodromy action on any given cycle $\d\in H_1(\Sigma_g,\ZZ)$ is
very simply determined in terms of this vanishing cycle, by
means of the Picard-Lefshetz formula \Arn:
$$
M_\nu:\ \ \ \d\ \longrightarrow\ \d - (\d\cap\nu)\,\nu\ .
\eqn\PicLef
$$
{}From this one can find for a vanishing cycle of the form \defVC\
the following monodromy matrix \KLTYb,
$$
{M}_{(g, q)}= \pmatrix{\bfone + \vec q\otimes \vec g& \vec
q\otimes \vec q\cr -\vec g\otimes \vec g&\bfone -\vec g\otimes \vec
q}\ \in\ Sp(2n-2,\ZZ)\ , \eqn\monmatrix
$$
which obeys $(\vec g, \vec q){M}_{( g, q)}=(\vec g, \vec q)$. Under a
change of basis, one has $S^{-1}{M}_{\nu}S=M_{\vec\nu\cdot S}$. Also
observe that for $\nu_i\cap\nu_j=0$ the corresponding monodromies
commute: $[M_{\nu_i},M_{\nu_j}]=0$, as it should be for two mutually
local states. The actual charges $(\vec g,\vec q)$ of the $SU(n)$
curves \aaa\ will generically be given by root vectors in simple root
and Dynkin bases, respectively.

As far as the semi-classical monodromies are concerned, we already
noted that the $\a$-cycles do not vanish anywhere in the quantum
moduli space. However, we can formally compactify the moduli space by
considering $\L$ as an extra modulus, and study monodromy around
$\L=0$. For $\L\to0$, the $\a$-cycles vanish since: $e^+_i\to
e^-_i=e_i$, and the classical level surface $V_u=\{e_i\}$ is
recovered. Under a $2\pi$ rotation of $\L^{2n}$ (which leaves the
curve \aaa\ invariant), $e^+_i$ simply exchanges with $e^-_i$, if we
take $\L$ sufficiently small as compared to $u_n$.
Accordingly, since the $\a$-cycles correspond to the weights of the
fundamental representation of $SU(n)$, the monodromy corresponding to
this simultaneous braid is
$$
\prod_{{{\rm fund.\ rep}\atop{\rm weights}\
\l_i}}M_{(0, \l_i)}\ =\ T
$$
and thus we reproduce the quantum monodromy \quantmono\ directly from
the curves \aaa. The $\theta$-shifts \thetashit\ associated with the
classical monodromies are similarly given by $T_{\a_i}=
M_{(0, \a_i)}$.

Note that the Picard-Lefshetz formula \PicLef\ directly expresses the
correct logarithmic monodromy property of the corresponding
$\b$-function, and thus automatically guarantees a consistent
physical picture. That is, near the vanishing of some
$\nu=\vec\nu\cdot\vec\g$ (where $\vec\g\equiv(\vec\b,\vec\a)$), the
monodromy shift of the gauge coupling, when expressed in suitable
local variables, is
$$
\Delta\tau_{ij}\ =
-(\g_i^*\cap\nu)\,{\del\over\del\pi_j}\int_\nu\l \ \equiv\
-\sum\nu_k\,(\g_i^*\cap\g_k)\,{\del\over\del\pi_j}Z_\nu\ =\
-\nu_i\,\nu_j\
\eqn\tauhshft
$$
where $Z_\nu\equiv\vec\nu\cdot\vec\pi$ and $\g^*$ is the cycle dual
to $\g$. This is indeed the monodromy associated with the
corresponding one-loop effective action near the singular line
$\bifset_\L^{(\nu)}$:
$$
\cF_\nu\ =\ \Coeff1{4\pi i}{Z_\nu}^2\ln\big[{Z_\nu\over\L}\big]\ .
$$

%%%%%%%%%%%%%%%%%%%%%%%%%%%%%%%%%
\section{Strong coupling monodromies and dyon charge
spectrum for $G=SU(3)$}
%%%%%%%%%%%%%%%%%%%%%%%%%%%%%%%%%

We now consider $G=SU(3)$ in some more detail, extending our
previous work discussed in \doubref\KLTYa\KLTYb. We first need to fix
some symplectic basis for the homology cycles. It turns out that a
convenient basis is the one depicted in \lfig\figxplane, since
it will directly reproduce the semi-classical monodromies and
is adapted to the vanishing cycles.

\figinsert\figxplane{The genus two curve for $G=SU(3)$ is represented
as branched $x$-plane with cuts linking pairs of roots \eipm\ of
$p(x)=0$. The locations of the cuts refer to $u=0$, $\Im v=0,\, \Re
v>1$.
We depicted our choice of homology basis that is adapted to
the vanishing $\b$-cycles. The cycles $\a_i$ can be associated with
the fundamental weights, and $\b_i$ with the simple roots.
}{2.0in}{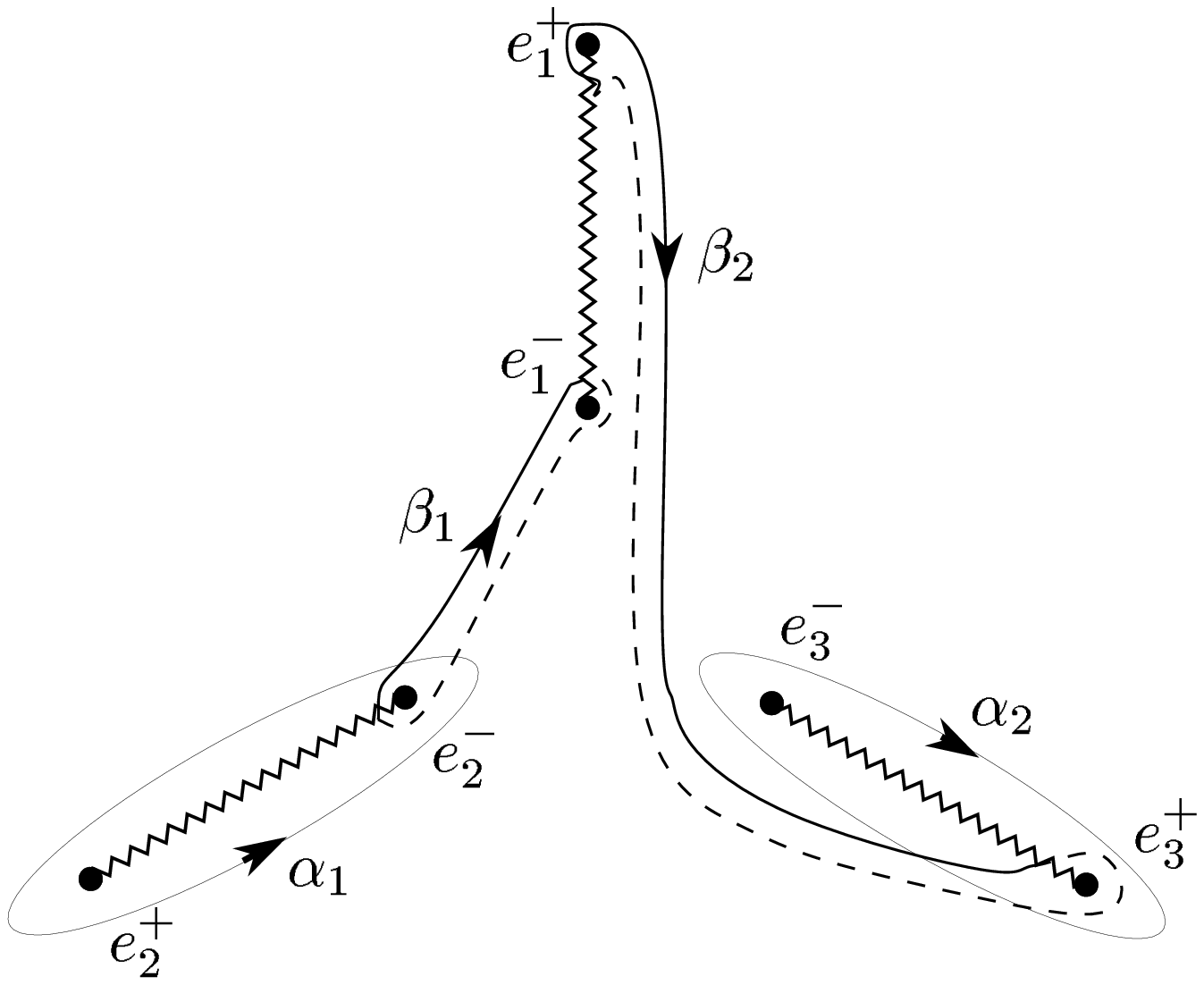}

We now consider the monodromies generated by loops around the six
singular lines in the quantum moduli space \lfig\figQantMs. With
reference to results by Zariski and van Kampen (c.f., \CDFLLR\ and
references therein) it suffices to study loops in a generic complex
line through the base point. Across cusps and nodes the monodromies
are related through the ``van Kampen relations''. Specifically, we
fix a plane in $\cM_\L$ at ${\rm Re}(v)=const>\Lambda^3,\ {\rm
Im(}v)=0$, as well as a base point $u_0$, and consider a family of
loops as given in figure \lfig\figuplane. By carefully tracing the
effects of loops in moduli space on the motions of the branch points
in the $x$-plane, we find the corresponding vanishing cycles to be as
depicted in \lfig\figvancyc.

\figinsert\figuplane{Loops $\g_i$ in the $u$-plane at ${\rm
Re}(v)=const>\Lambda^3,\ {\rm Im(}v)=0$, starting from
the base point $u_0$ (cf., \lfig\figQantMs).
The composite loops $r_i$ give the semi-classical monodromies
\SuThreeScMo.}{2.3in}{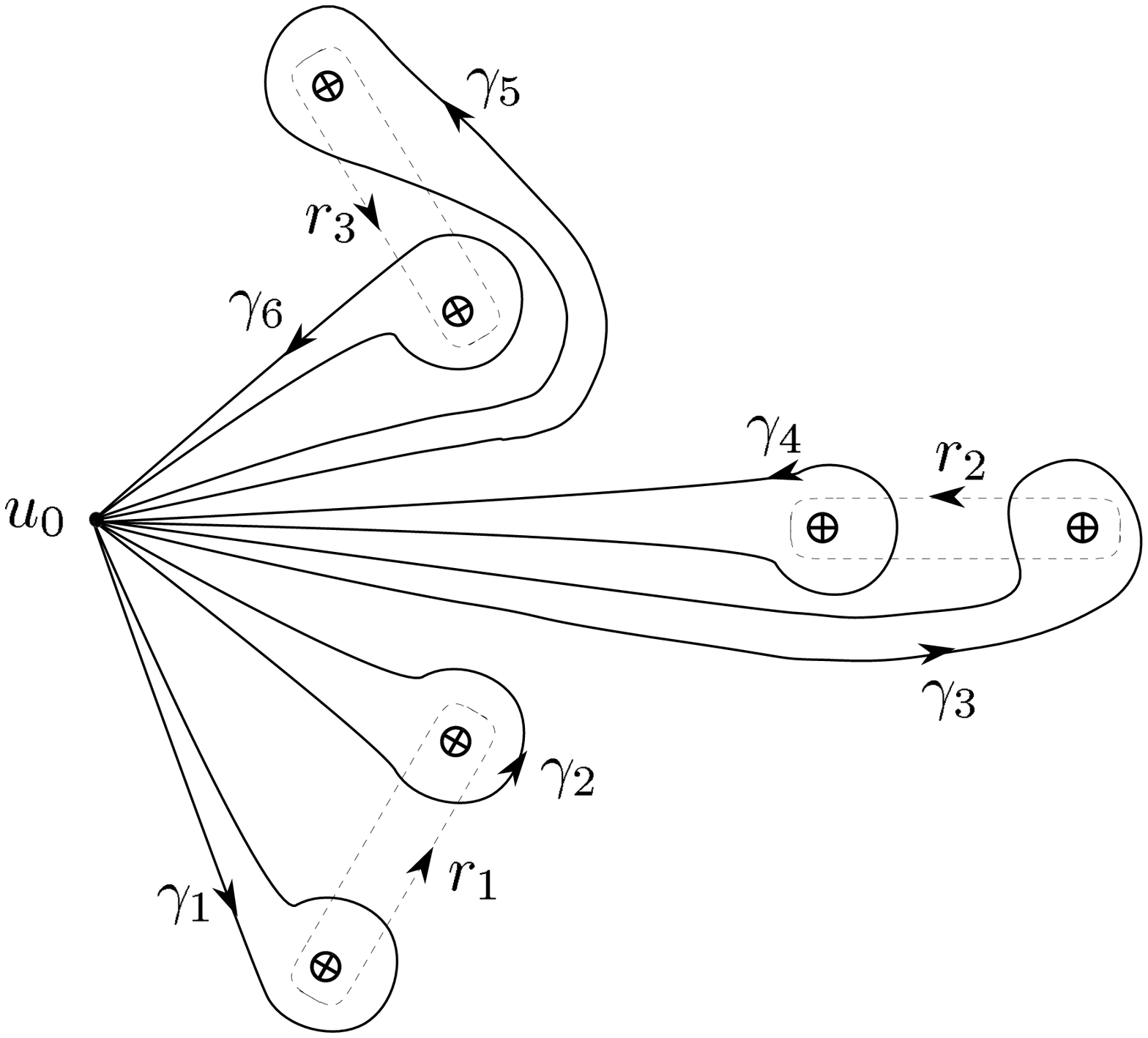}
\vskip-.5cm
\figinsert\figvancyc{Vanishing cycles $\nu_i$ in the $x$-plane,
associated with the loops $\g_i$ in \lfig\figuplane. We have depicted
here only the paths on the upper sheet, and not the return paths on
the lower sheet.}{2.3in}{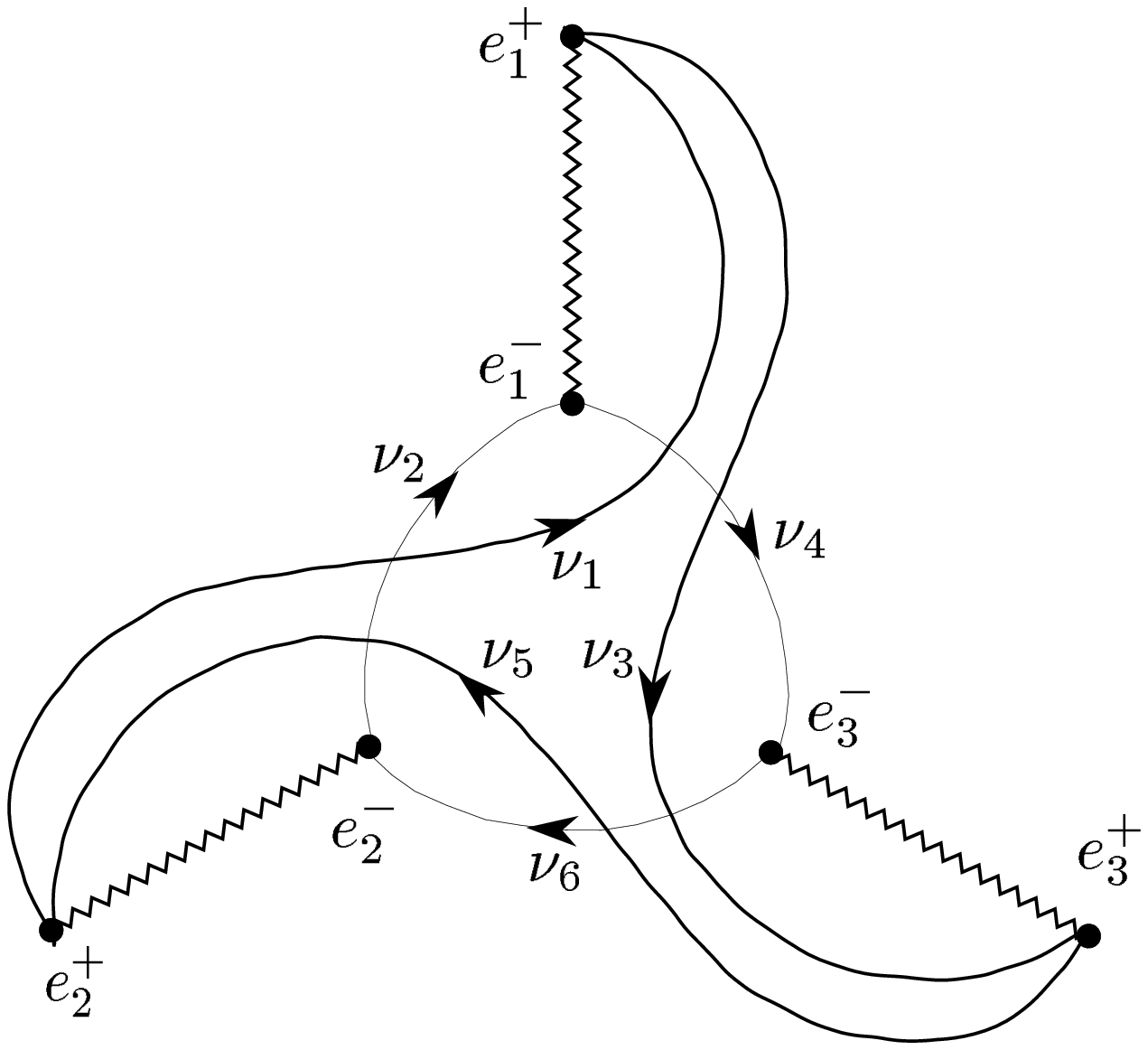}

According to what we said above, the quantum numbers of the dyons
that become massless at the various singular lines in moduli space
can be directly obtained from \lfig\figvancyc, by comparing the
vanishing cycles with the basis cycles in \lfig\figxplane.
Our basis is such that the massless excitations at
$u=(\coeff{27}4)^{1/3}\L^2$
are pure monopoles with charges $(\vec g,\vec q)$ equal to
$(1,0,0,0)$, and $(0,1,0,0)$, respectively. (For $G=SU(n)$ one can
always choose a basis for the $\b$-cycles so as to have $n-1$
massless monopoles with unit charges at a given node of $n-1$
intersecting singular lines.)

The quantum numbers of the other massless dyons then simply follow
from the global $\ZZ_3$ symmetry $U$ \Ucox. More precisely, the
charges
$\nu_i\equiv(\vec g_i,\vec q_i)$ and monodromies associated with all
the six singular lines in $\cM_\L$ are given by matrices \monmatrix\
with
$$
\eqalign{
&M_{\nu_1}\ =\ M_{(1, 0, -2, 1)}\ \equiv\
\pmatrix{ -1 & 0 & 4 & -2 \cr 1 & 1 & -2 & 1 \cr -1 & 0 & 3 & -1 \cr
  0 & 0 & 0 & 1 \cr  }\cr
&M_{\nu_2}\ =\ M_{(1, 0, 0, 0)}\ \equiv\ \ \
\pmatrix{ 1 & 0 & 0 & 0 \cr 0 & 1 & 0 & 0 \cr -1 & 0 & 1 & 0 \cr 0 &
  0 & 0 & 1 \cr  }\cr
&M_{\nu_3}\ =\ M_{(0, 1, 0, 0)}\ \equiv\ \ \
\pmatrix{ 1 & 0 & 0 & 0 \cr 0 & 1 & 0 & 0 \cr 0 & 0 & 1 & 0 \cr 0 &
  -1 & 0 & 1 \cr  }\cr
}$$ $$\eqalign{
&M_{\nu_4}\ =\ M_{(0, 1, -1, 2)}\ \ \equiv\ \ \
\pmatrix{ 1 & -1 & 1 & -2 \cr 0 & 3 & -2 & 4 \cr 0 & 0 & 1 & 0 \cr 0
   & -1 & 1 & -1 \cr } \cr
&M_{\nu_5}\ =\ M_{(-1, -1, 2, -1)}\ \equiv\
\pmatrix{ -1 & -2 & 4 & -2 \cr 1 & 2 & -2 & 1 \cr -1 & -1 & 3 & -1
   \cr -1 & -1 & 2 & 0 \cr  }\cr
&M_{\nu_6}\ =\ M_{(-1, -1, 1, -2)}\ \equiv\
\pmatrix{ 0 & -1 & 1 & -2 \cr 2 & 3 & -2 & 4 \cr -1 & -1 & 2 & -2
   \cr -1 & -1 & 1 & -1 \cr}
\ .}\eqn\sixmonodroms
$$
They form indeed two orbits under conjugation by $U$,
and even form a single orbit under
$$
A\ =\
\pmatrix{ -1 & -1 & 2 & -1 \cr 0 & 1 & -1 & 2 \cr -1 & -1 & 1 & -1
   \cr 0 & -1 & 0 & -1 \cr  }\ \equiv\ U^{-1}V^{-1}\ ,
\eqn\Adef
$$
with $A^2=U$ and $A^6=U^3=V^2=\bfone$. These global $R$-symmetries
act
simultaneously on the moduli space in \lfig\figQantMs\ and on the
vanishing cycles in \lfig\figvancyc. That is, $U$ rotates the moduli
space by $e^{2\pi i/3}$ and $V$ represents $v\to-v$. In fact, $A$ is
the
generalization of the global $\ZZ_4$ symmetry of $G=SU(2)$ \SWa. Note
that these matrices do not belong to the monodromy group.

{}From \sixmonodroms\ we see that the magnetic and electric quantum
numbers of the massless
dyons are indeed given by root vectors (in simple root and Dynkin
bases,
respectively). This is in accordance with semi-classical stability
\BN. Though the above quantum numbers can formally be changed via
monodromy to other points in the root lattices, one expects
\doubref\SWa\MDSS\ dyons with higher charges to become unstable when
crossing lines of marginal stability. Nevertheless, from a group
theoretical point of view, the possible charges are given by root
lattices, and this mirrors the special properties of the
lattice generated by the vanishing cycles of the curve \fff.
One may also perform a basis change via conjugation with $W$ \Wdef,
under which $M_{\nu_i}$ remain integral. The matrices then contain as
sub-matrices the $SU(2)$ strong coupling monodromies in the basis of
\SWa, which are the monodromies of the isogenous $SU(2)$ curve
\SWonecurve. This suggests that there might be a similar isogenous
curve for $G=SU(3)$, or even all $SU(n)$.

It has been observed \AF that the charge vectors of each pair of
lines that intersect in the nodes at $v=0$ satisfy
$\vec\nu_i^t\,\Omega\, \vec\nu_j=0$, since the cycles do not
intersect (these pairs are given by
$(\vec\nu_1,\vec\nu_6),\,(\vec\nu_2,\vec\nu_3)$ and
$(\vec\nu_4,\vec\nu_5)$). Thus, at any of the three nodes, two
mutually local dyons become massless and both dual $U(1)$'s are
weakly coupled. We will find below in \hsubsect{6.2}\ that the exact
beta function of the effective dual theory indeed reflects two
massless monopole hypermultiplets. On the other hand, near the cusps
at $u=0, v=\pm\L^3$, three dyons become massless simultaneously,
though they are not local with respect to each other. The effective
theory near these regions has recently been discussed in \PAMD.

The strong coupling monodromies \sixmonodroms\ contain, as expected,
the strong coupling monodromies \suIImon\ of $G=SU(2)$ embedded as
sub-matrices. Also, the monodromies of the three pairs of lines that
go out to $v\to+\infty$ reproduce precisely the $SU(3)$
semi-classical monodromies \SuThreeScMo:
$$
\eqalign{
&M_{\nu_2}\cdot M_{\nu_1}\ =\ M_\sclass^{(r_1)}\cr
&M_{\nu_4}\cdot M_{\nu_3}\ =\ M_\sclass^{(r_2)}\cr
&M_{\nu_6}\cdot M_{\nu_5}\ =\ M_\sclass^{(r_3)}\ ,
}\eqn\MonoFit
$$
(c.f., \lfig\figuplane) and thus: $M_{\nu_2}
M_{\nu_1}M_{\nu_6}M_{\nu_5}M_{\nu_4} M_{\nu_3}
=M_{\infty,u}^{(r_3)}$. (For the three pairs that go out
$v\to-\infty$ the same is true up to a change of basis). Note that
this semi-classical monodromy is defined by a loop around infinity in
the $u$-plane located at $\Im v=0,\,\Re v>1$, as shown in
\lfig\figQantMs. For the same loop in a $u$-plane located at $-1<\Re
v<1$, the monodromy is different because some of the singular lines
interchange at the cusp at $\Re v=1$. This monodromy is given by
$$
\tilde M_{\infty,u}^{(r_3)}\ =\ M_{\nu_6}
M_{\nu_1}M_{\nu_4}M_{\nu_5}M_{\nu_2} M_{\nu_3}\ =\
\pmatrix{ 0 & 5 & -3 & 6 \cr -1 & -3 & 6 & -3 \cr 0 & -1 & 0 & -1
   \cr 1 & 3 & -7 & 3 \cr }\ .
\eqn\tildeM
$$
%%%%%%%%%%%%%%%%%%%%%%%%%%%%% %%%%%%%%%%%%%%%%%%%%%%%%%%%%%%%

%%%%%%%%%%%%%%%%%%%%%%%%%%%%% %%%%%%%%%%%%%%%%%%%%%%%%%%%%%%%
%%%%%%%%%%%%%%%%%%%%%%%% %%%%%%%%%%%%%%%%%%%%%%%%%%
\chapter{Picard-Fuchs Equations for $G=SU(3)$}

\section{Derivation}
%%%%%%%%%%%%%%%%%%%%%%%% %%%%%%%%%%%%%%%%%%%%%%%%%%

Starting from the $SU(3)$ curve \fff, we will first
derive the system of equations for the periods
$\int_{\gamma_i}\!\!w=\partial_{v}\pi$. Subsequently, we will
obtain the system for $\partial_{u}\vec \pi$ as well as directly for
$\vec \pi=(\vec a_D,\vec a)^t$ itself.

There is a systematic, though tedious way to set up the Picard-Fuchs
equations, by considering derivatives of $\omega_2={dx\over
y}\equiv{dx\over\sqrt{p(x)}}$
with respect to $u$ and $v$. These
produce terms of the form $\phi(x)\over y^n$ for some polynomials
$\phi(x)$. The idea is to reduce the order of $\phi(x)$, to integrate
by
parts and re-express the result in terms of the abelian
differentials. For this, we will use the fact that the discriminant
\ggg\ can always be written in the form
\VW
$$
\Delta=a(x)p(x)+b(x)p'(x)\eqn\hhh
$$
where, for the $SU(3)$ curve, $a$ and $b$ are polynomials of order
four and five, respectively. These polynomials are straightforwardly
determined to be
$$
a(x)=\sum_{i=0}^4 a_i x^i\,,\qquad b(x)=\sum_{i=0}^5 b_i x^i
$$
with
$$\eqalign{
a_0&=-729\L^6+216 u^3-16u^6+729 v^2+108 u^3 v^2\cr
a_1&=9uv(-135\L^6-4 u^3+27 v^2)\cr
a_2&=18 u^2(-27\L^6+4u^3-27 v^2)\cr
a_3&=27 v(27\L^6+4 u^3-27 v^2)\cr
a_4&=18 u(27\L^6-4 u^3+27 v^2)}\eqn\ab
$$
and
$$\eqalign{
b_0&= 2u^2 v(-81\L^6+4 u^3-27 v^2)\cr
b_1&= (-27\L^6+4 u^3-27v^2)(-9\L^6+4 u^3+9 v^2)/2\cr
b_2&= 3uv(243\L^6+4 u^3 -27 v^2)/2\cr
b_3&= 5 u^2(27\L^6-4 u^3+27 v^2)\cr
b_4&= 9 v(-27\L^6-4 u^3+27 v^2)/2\cr
b_5&= -3u(27\L^6-4 u^3+27 v^2)}
$$
We can thus write under the integral sign:
$$
{\phi(x)\over y^n}=
{1\over\Delta}{a\phi+{2\over n-2}(b\phi)'\over y^{n-2}}\ .
\eqn\iii
$$
If the order of the polynomial in the numerator is equal to or
exceeding the power of $p'(x)$, we can reduce it by expressing the
highest power in terms of $p$ or $p'$ and lower powers and
integrating by parts. This procedure allows one, after some work, to
find a set of two second order differential equations that are
satisfied by the periods
$\partial_v \vec \pi$: $\tilde{\cal L}_i(\partial_v\vec\pi)=0$ with
$$\eqalign{
\tilde{\cal L}_1
&=(27\Lambda^6-4 u^3-27 v^2)\p_u^2-12
u^2 v\p_u\p_v-12 u^2\p_u-21 uv\p_v-4u\cr
\tilde{\cal L}_2
&=(27\Lambda^6-4 u^3-27 v^2)\partial_v^2-
36 uv\p_u\p_v-36 u\p_u-63 v\p_v-12
}\eqn\ii
$$
Note that these equations imply $(u\p_v^2-3\p_u^2)\p_v\vec \pi=0$,
which can also be verified directly. In fact,
$(u\partial_v^2-3\partial_u^2) {1\over y}=\partial_x{W_{A_2}\over
y^3}$. If we introduce the gauge and ${\bf Z}_{12}$ invariant
dimensionless moduli $\alpha={4u^3\over 27\Lambda^6},\,
\b={v^2\over\Lambda^6}$, the Picard-Fuchs equations take the form
$$
\eqalign{
\tilde{\cal L}_1&=\a(1-\a)\p_\a^2-\b^2\p_\b^2-2\a\b\p_\a\p_\b
+{1\over3}(2-5\a)\p_\a-{5\over3}\b\p_\b-{1\over9}\cr
\tilde{\cal L}_2&=\b(1-\b)\p_\b^2-\a^2\p_\a^2-2\a\b\p_\a\p_\b
+({1\over2}-{5\over3}\b)\p_\b-{5\over3}\a\p_\a-{1\over9}\ .
}\eqn\jjj
$$
Expressed in terms of the logarithmic derivatives
$\theta_\a=\a\partial_\a,\,\theta_\b=\b\partial_\b$,
the two differential operators become
%$$\eqalign{
%\tilde{\cal L}_1&=\ta(\ta-{1\over3})-\a(\ta+\tb+{1\over3})^2\cr
%\tilde{\cal
% L}_2&=\tb(\tb-{1\over2})-\b(\ta+\tb+{1\over3})^2}\eqn\kkk
%$$
$$
\eqalign{
\tilde{\cal L}_1&=\ta(\ta+c-1)-\a(\ta+\tb+a)(\ta+\tb+b)\cr
\tilde{\cal L}_2&=\tb(\tb+c'-1)-\b(\ta+\tb+a)(\ta+\tb+b)\ ,}
\eqn\mmm
$$
with $a=b=\coeff13,c=\coeff23,c'=\coeff12$. The system \mmm\ is in
fact the generalized hypergeometric system
$F_4(a,b;c,c';\alpha,\beta)$ of Appell \doubref\Appell\Yoshida.
Unfortunately, not much appears to be known in the literature about
analytic continuation and non-trivial transformations of its
solutions, especially for the present special values of $(a,b;c,c')$,
so that we have to work out the solutions ourselves.

If we write the differential equations \jjj\ in the form
$\tilde {\cal L}_i\p_v\vec \pi=0$, we can pull the partial derivative
operator through to get
$\p_v{\cal L}_i\vec \pi=0$ with
$$\eqalign{
{\cal L}_1&=(27\Lambda^6-4 u^3-27 v^2)\p_u^2
                   -12 u^2 v\p_u\p_v-3 u v \p_v-u\cr
{\cal L}_2&=(27\Lambda^6-4 u^3-27 v^2)\p_v^2-36 u v \p_u\p_v-9
v\p_v-3\
}\eqn\lll
$$
and also $(u\partial_v^2-3\partial_u^2)\vec\pi=0$.
When we express this in terms of the variables $\a,\b$,
we find that this system (that is satisfied
directly by the sections $a_{D_i},a_i$) is equivalent to an Appell
system of type $F_4({1\over6},{1\over6};{1\over3},{1\over2};\a,\b)$.
%\foot{This is analogous to $G=SU(2)$, where the
%relevant hypergeometric indices are $(-\coeff14,-\coeff14,\coeff12)$
%(c.f., \hsect{2}). This pattern seems to continue for general
% $SU(n)$
%with indices of the form
%$(-\coeff1{2n},...,-\coeff1{2n},1-\coeff1n,...,1-\coeff12)$.}
%$F_4(-{1\over6}, -{1\over6}; {2\over3}, {1\over2}; \a,\b)$.
Similarly, we can find the Picard-Fuchs operators also for the
periods $\partial_u\vec \pi$; they constitute the system

In order to solve the Picard-Fuchs equations, note that they are
always Fuchsian \doubref\Deligne\Yoshida\ and thus have only regular
singularities. These can be described as follows \Yoshida. Denote the
linear partial differential operators of degree $m$, defined in a
neighborhood $U$ of a point $z$ of the $g$ dimensional moduli space
${\cal M}$ by ${\cal L}_i=\sum_{|p|\leq m} a_1^p(z)\left({d\over
dz}\right)^p$ where we have used the notation $\left({d\over
dz}\right)^p=\prod_{j=1}^{g} {\p\over\p z_j}^{p_j}.$ They define a
left ideal $I$ in the ring of partial differential operators on $U$.
We now introduce the symbol of ${\cal L}_i$:
$\sigma({\cal
L}_i)=\sum_{|p|=m}a_i^p(z)\xi_1^{p_1}\cdots\xi_g^{p_g}$, where
$\xi_1,\dots,\xi_g$ is a coordinate system in the fiber of the
cotangent bundle $T^*U$ at z. The ideal of symbols is defined by
$\sigma(I)=\lbrace\sigma({\cal L})|{\cal L}\in I\rbrace$. The
singular locus is then $\Delta(I)=\pi(Ch(I)-U\times\lbrace
0\rbrace)$, where the characteristic variety $Ch(I)$ is the
subvariety in $T^*U$ specified by the ideal of symbols, and $\pi$ is
the projection along the fiber of $T^*U$. The fact that $\sigma(I)$
is generated by $\sigma({\cal L}_i)$ is a special property of
Picard-Fuchs
systems, and this is not the case, for example, for the Appell system
$F_1$.

Let us now find the singular locus of the general system
$F_4(a,b;c,c';\a,\b)$ \mmm, for which the symbols turn out to be
independent of the parameters $(a,b,c,c')$:
$$\eqalign{
\sigma_1=\a^2\xi_\a^2-\a(\a\xi_\a+\b\xi_\b)^2\cr
\sigma_2=\b^2\xi_\b^2-\b(\a\xi_\a+\b\xi_\b)^2}\ .\eqn\nnn
$$
It is straightforward to find the discriminant to be
$\Delta(I):\,\a\b(1+\a^2+\b^2-2(\a\b+\a+\b))=0$. This coincides, up
to the factor $\a\b$, with the quantum discriminant $\Delta_\L$ \ggg\
of the $SU(3)$ curve (note that if we had computed the singular locus
of the Picard-Fuchs equations in the form \ii, we would have gotten
the discriminant given in eq.\ggg\ only when taking also into account
the vanishing of $(u\partial_v^2-3\partial_u^2)\del_v\vec\pi$). The
additional lines $\a=0$ and $\b=0$ are due to the change of variables
$(u,v)\to(\a,\b)$.

The formalism described in this section applies to every
hyperelliptic curve and can be used to obtain the Picard-Fuchs
equations for all $SU(n)$. Let us point out here some features of the
Picard-Fuchs system, which are due to the special symmetries of our
curves and hold for all $SU(n)$. We start by characterizing
differential operators, which are pure second order in the
derivatives. These are given by
$$
\eqalign{
{\cal L}^{SU(n)}_{p,q,r,s}&=\partial_{u_{p}} \partial_{u_{q}}-
\partial_{u_{r}} \partial_{u_{s}},\qquad {\rm with\ \ } p+q=r+s \cr
{\cal L}_0^{SU(n)}&=
\sum_{k=1}^{n-2}
k u_{n-k} \partial_{u_{n}}\partial_{u_{n-k+1}} -
n \partial_{u_2} \partial_{u_{n-1}}, {\rm for \ \ } n > 2 }
$$
and follow directly from ${\cal L}^{SU(n)}_{p,q,r,s}\left({1\over
y}\right)=0$ and ${\cal L}_0^{SU(n)}\left({1\over y}\right) =
\partial_x \del_{u_n}\!\left(1\over y\right)$. In general, exact
forms
of the type $\partial_x \left(x^r \cW_{A_{n-1}}\over y^3\right)$ lead
to
differential operators involving first and second order
derivatives, namely
$$
{\cal L}^{SU(n)}_r=\sum_{k=1}^{n-2} k u_{n-k}
\partial_{u_{n-r+1}}
\partial_{u_{n-k}}-n \partial_{u_{n-r-1}}\partial_{u_{2}}+
r \partial_{u_{n+1-r}}\qquad {\rm for \ \ } r < n-2\ .
$$
For instance, for $SU(3)$ the complete system of PF-operators is
given by ${\cal L}_0^{SU(3)}= u\partial^2_v-3 \partial_u^2$ and any
one of the operators from \ii. Similarly, for $SU(4)$ the complete
system consists of
${\cal L}^{SU(4)}_{2,4,3,3}$,
${\cal L}^{SU(4)}_0$,
${\cal L}^{SU(4)}_1$ and
$$
\eqalign{
&{\cal L}=
(64 u^2 w^2 - 32 u v^2 w + 9 v^4 - 64 u^2) \partial^2_w+
4 v (12 u^3- 9 v^2+ 32 u w)\partial_u \partial_v +\cr &
8 u (2 u^3 + 9 v^2 + 16 u w) \partial_u^2+
2 u (64 u w- 9 v^2)\partial_w +
108 u^2 v \partial_v+
64 u^3 \partial_u
36 u^2,}$$ where we set $\Lambda=1$. Completeness can be
checked by calculating the rank of the linear system of
symbols. For generic values of the moduli the rank has
to be maximal, while the rank drops precisely at the
principal locus (all $\xi_i\ne 0$) of the discriminant.

%%%%%%%%%%%%%%%%%%%%%%%% %%%%%%%%%%%%%%%%%%%%%%%%%%
\section{Solutions in the semi-classical regime}
%%%%%%%%%%%%%%%%%%%%%%%% %%%%%%%%%%%%%%%%%%%%%%%%%%

We will be particularly interested in solving the Picard-Fuchs
equations in the semi-classical regime, to which we so far, somewhat
vaguely, referred to as ``infinity'' in the moduli space $\cM_\L$. We
now like to make precise what we mean by this, by compactifying the
moduli space $(\a,\b)\in{\IC}^2$ to ${\IP}^2$, by adding a line
$\g=0$ at infinity (where $\g\equiv 27\L^6$). This makes contact with
our discussion of the semi-classical monodromy around $\L=0$ in
\hsubsect{3.2}. In terms of homogeneous coordinates
$(\a:\b:\c)\in{\IP}^2$, we get for the discriminant
$$
\Delta(I)=\a\b\c(\a^2+\b^2+\c^2-2(\a\b+\b\c+\a\c))\ .\eqn\ppp
$$
We have thus three singular lines which intersect with each other in
three points $P_i$, as well with the discriminant locus $\bifset_\L$
in three points $Q_i$. We have sketched the singular locus of the
system $F_4$ in \lfig\figF.

\figinsert\figF{The singular locus $\bifset(I)$ of the system $F_4$
for the compactification of the $(\a,\b)$-plane to $\IP^2$, with
$\a=4u^3$, $\b=27v^2$ and $\g=27\L^6$. The semi-classical regions
correspond to the neighborhoods of $P_2$ and $P_3$, and the magnetic
dual semi-classical region to $Q_1$. A full set of two power series
and two logarithmic solutions can be found only in these regions. The
cusp point is mapped to $Q_2$, where the theory is badly behaved, and
$Q_3$ represents the intersection of the six singular lines with
infinity; at the origin, $P_1$, nothing special happens.}
{2.5in}{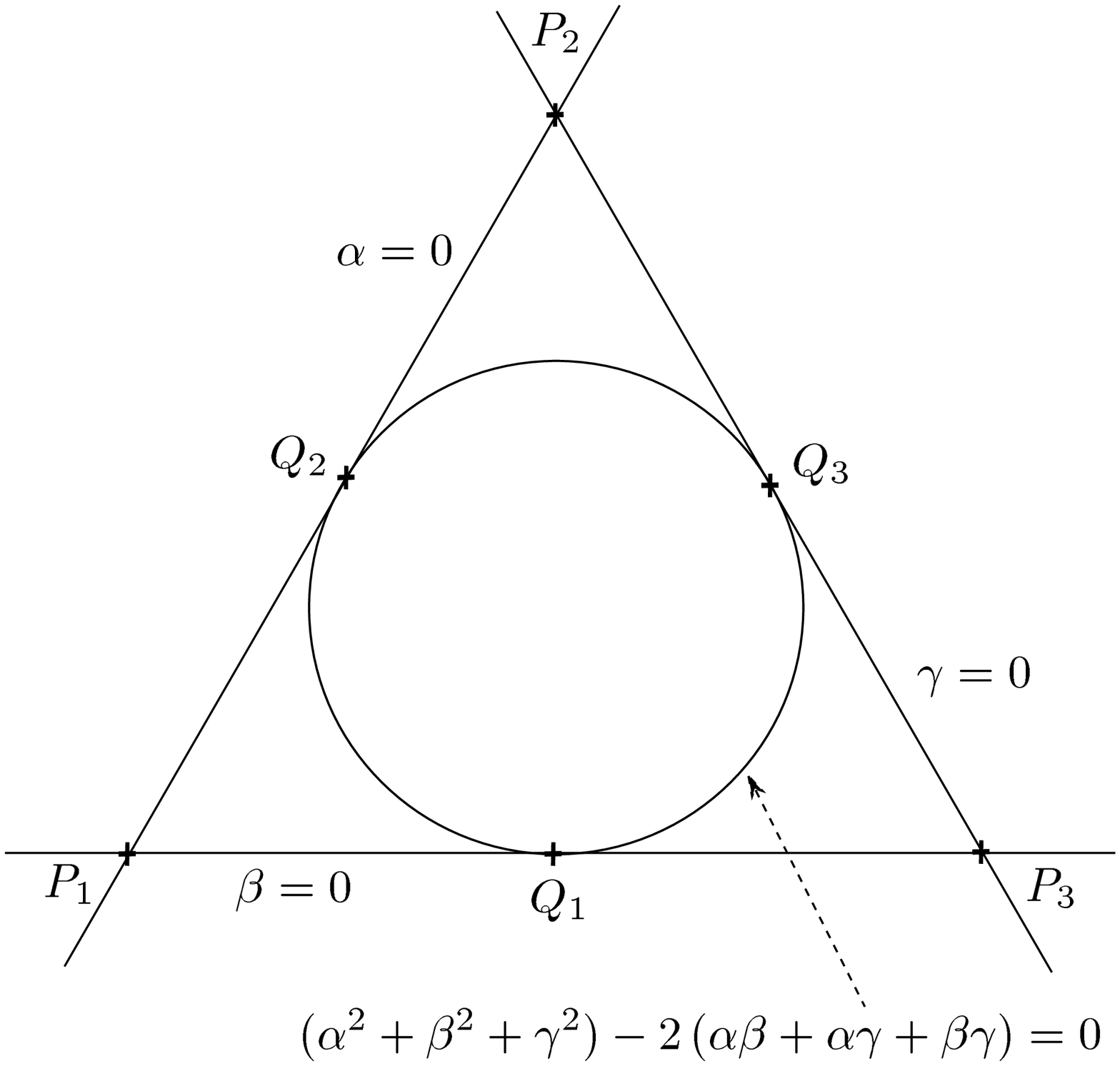}

We can cover ${\IP}^2$ with three coordinate patches (centered
on the points $P_i$), and in each such patch there is a set of
preferred inhomogeneous coordinates, ie.,
$$
\eqalign{
&P_1:\ \ ({\a\over\c}:{\b\over\c}:1)\equiv(x_1:y_1:1)\cr
&P_2:\ \ ({\a\over\b}:1:{\c\over\b})\equiv(x_2:1:y_2)\cr
&P_3:\ \ (1:{\b\over\a}:{\c\over\a})\equiv(1:x_3:y_3)\ .\cr
}\eqn\patches
$$
We thus have two natural coordinate patches corresponding to
semi-classical ``infinity'' in the moduli space, one roughly given by
large $u$, the other one by large $v$, just as mentioned in
\hsubsect{3.2}.

So far we have written the Picard-Fuchs equations in terms of the
coordinates appropriate for $P_1$. In the more interesting two
patches at infinity, the Picard-Fuchs equations are
($\theta_i=x_i\partial_{x_i},\,\theta'_i=y_i\partial_{y_i}$):
$$\eqalign{
P_2:\qquad {\cal L}_1&=y_2\t_2(\t_2+c-1)-x_2(\t'_2-a)(\t'_2-b)\cr
{\cal L}_2&=y_2(\t_2+\t'_2)(\t_2+\t'_2-c'+1)-(\t'_2-a)(\t'_2-b)\cr
\noalign{\vskip.3cm} P_3:\qquad {\cal
L}_1&=y_3(\t_3+\t'_3)(\t_3+\t'_3-c+1)-(\t'_3-a)(\t'_3-b)\cr {\cal
L}_2&=y_3\t_3(\t_3+c'-1)-x_3(\t'_3-a)(\t'_3-b)}
\eqn\qqq
$$
with $(a,b,c,c')=(-\coeff16,-\coeff16,\coeff23,\coeff12)$. Note that
for these equations we have $a=b$ and $1-c,\,1-c',\,a$ all distinct
and no pair differing by integers. Thus, we can expect them to be
solved by power and logarithmic series ans\"atze around the origin in
each patch:
$$
\o i\ul i(x_i,y_i)=\sum_{n,m\geq 0}c_i(n,m)x_i^{n+s}
y_i^{m+t}\eqn\rrrr
$$
We find the following solutions to the indicial equations:
$$\eqalign{
P_1:\qquad (s,t)&=(0,0),\,(0,1-c'),\,(1-c,0),\,(1-c,1-c')\cr
\noalign{\vskip.2cm}
P_2:\qquad (s,t)&=(0,a),\,(0,b),\,(1-c,a),\,(1-c,b)\cr
\noalign{\vskip.2cm}
P_3:\qquad (s,t)&=(0,a),\,(0,b),\,(1-c',a),\,(1-c',b)}\eqn\sss
$$
With this it is possible to solve the recursion
relations for the coefficients $c_i(n,m)$ in \rrrr. The
solutions in the patch $P_1$ are simply given by four power series,
which is not very interesting. On the other hand, for each of the two
semi-classical patches at infinity, we find two series and two
logarithmic solutions, and this is precisely what one expects on
physical grounds.

More precisely, for the power series solutions in the patches at
infinity one finds (we have set $a=b$ but have left
$a,c,c'$ arbitrary and normalized $c_i^{(s,t)}(0,0)=1$)
$$\eqalign{
P_2:\qquad c_2^{(0,a)}(n,m)&={(a)_{n+m}\,(a+1-c')_{n+m}
                          \over(1)_n (1)_m^2(c)_n}\cr
           c_2^{(1-c,a)}(n,m)&={(a+1-c)_{n+m}
                          \,(a+2-c-c')_{n+m}\over (1)_n
(1)_m^2(2-c)_n}\cr
P_3:\qquad c_2^{(0,a)}(n,m)&={(a)_{n+m}\,(a+1-c)_{n+m}
                          \over(1)_n (1)_m^2(c')_n}\cr
           c_2^{(1-c',a)}(n,m)&={(a+1-c')_{n+m}
       \,(a+2-c-c')_{n+m}\over (1)_n (1)_m^2(2-c')_n}\ .
}\eqn\ttt
$$
%In terms of the generalized Gaussian sum
%$$
%F_4(a,b;c,c';x,y)=\sum_{n,m\geq0}{(a)_{n+m} (b)_{n+m}\over
%(1)_n (1)_m (c)_n (c')_m}x^n y^m
%$$
%the two power series solutions in $P_2$ can be written as
%$$\eqalign{
%\o1\ul2(x_2,y_2)&=y_2^a\,F_4(a,1+a-c';c,1;x_2,y_2)\cr
%\o2\ul2(x_2,y_2)&=x_2^{1-c}
% y_2^a\,F_4(1+a-c,2+a-c-c';2-c,1;x_2,y_2)\ ,}
%$$
%and likewise in $P_3$:
%$$\eqalign{
%\o1\ul3(x_3,y_3)&=y_3^a\,F_4(a,1+a-c;c',1;x_3,y_3)\cr
%\o2\ul3(x_3,y_3)&=x_3^{1-c'}
% y_3^a\,F_4(1+a-c',2+a-c-c';2-c',1;x_3,y_3)}
%$$

In addition to the above power series solutions, we find logarithmic
solutions of the form
$$
(2\pi i)\Omega\ul
i(x_i,y_i)=\sum_{n,m\geq0}d_i(n,m)x_i^{n+s}y_i^{m+t}
+\log y_i\,\Big\{\sum_{n,m=\geq0}c_i(n,m)x_i^{n+s}
y_i^{m+t}\Big\}\eqn\uuu
$$
One verifies that
$$
d_i(n,m)={\partial\over\partial\rho}c_i(n,m,\rho)|_{\rho=0}
$$
where
$$\eqalign{
c_2(n,m,\rho)&={(s+t+\rho)_{n+m}(s+t+\rho+1-c')_{n+m}\over
(1)_n (1+\rho)_m^2 (c+2 s)_n}\cr
c_3(n,m,\rho)&={(s+t+\rho)_{n+m}(s+t+\rho+1-c)_{n+m}\over
(1)_n (1+\rho)_m^2 (c'+2 s)_n}\ .}
$$
With the definition $\psi(x)=\Gamma'(x)/\Gamma(x)$ we can write
$$\eqalign{
d_2(n,m)&=c_2(n,m)\Bigl\lbrace
2\psi(1)-2\psi(m+1)+\psi(n+m+s+t)-\psi(s+t)\cr
&\qquad\qquad\qquad+\psi(n+m+s+t+1-c')-\psi(s+t+1-c')\Bigr\rbrace\cr
d_3(n,m)&=c_3(n,m)\Bigl\lbrace
2\psi(1)-2\psi(m+1)+\psi(n+m+s+t)-\psi(s+t)\cr
&\qquad\qquad\qquad+\psi(n+m+s+t+1-c)-\psi(s+t+1-c)\Bigr\rbrace\ ,}
$$
where we have chosen the normalization $d(0,0)=0$.
$$
\eqn\tttlog
$$

%%%%%%%%%%%%%%%%%%%%%%%% %%%%%%%%%%%%%%%%%%%%%%%%%%
\section{Solutions in the magnetic dual semi-classical regime $Q_1$}
%%%%%%%%%%%%%%%%%%%%%%%% %%%%%%%%%%%%%%%%%%%%%%%%%%

On the discriminant locus, only the patch $Q_1$ has an easy physics
interpretation in that we can find two series and two logarithmic
solutions. This reflects the fact that on $\bifset_\L$, only near
$Q_1$ the theory is weakly coupled, in suitable dual local
variables. The tricky point is to find good variables for which we
really do have two series and two logarithmic solutions. We find that
$$
\d_\pm\ =\ (1-\a+\b\pm2\sqrt \b)\ =\
\prod_{i<j}^3(e_i^\pm-e_j^\pm)^2\
\eqn\dpmdef
$$
($\a={4u^3\over27\Lambda^6},\,\b={v^2\over\L^6}$)
are suitable variables, since they vanish precisely on the
discriminant (cf., \DLdef), and also incorporate all three
intersection points simultaneously. In terms of these variables,
the Picard-Fuchs operators for the system
$F_4(-\coeff16,-\coeff16;\coeff23,\coeff12;\a,\b)$ become:
$$
\eqalign{
&\cL_1 = \Big\{\Coeff{1}{2}\d_-\!\left( 2\!-\!\d_-\!\!-\!\d_+\!
\right)
    {\del^2\over\del_{\d_-\!}^2}\!+\!
  \Coeff{1}{2}\left( 2\!-\!\d_-\!\!-\!\d_+\! \right) \d_+\!
   {\del^2\over\del_{\d_+\!}^2}\!-\!
  \Coeff{1}{24}\left( 9\d_-\!\!+\!7\d_+\! \right)
   {\del\over\del_{\d_-\!}}
\cr&-
  \Coeff{1}{24}\left( 7\d_-\!\!+\!9\d_+\! \right)
   {\del\over\del_{\d_+\!}} +
  \Coeff{1}{4}\left(
4\d_-\!\!-\!{{\d_-\!}^2}\!+\!4\d_+\!\!-\!6\d_-\!\d_+\!\!-\!
      {{\d_+\!}^2} \right)
   {\del^2\over\del_{\d_+\!}\del_{\d_-\!}}-{1\over {36}} \Big\}
\cr
&\cL_2 = \Big\{ \Coeff{1}{2}\d_-\!\left( 4\!-\!\d_-\!\!-\!\d_+\!
\right)
    {\del^2\over\del_{\d_-\!}^2}\!+\!
  \Coeff{1}{2}\left( 4\!-\!\d_-\!\!-\!\d_+\! \right) \d_+\!
   {\del^2\over\del_{\d_+\!}^2}\!+\!
  \Coeff{1}{24}\left( 28\!-\!9\d_-\!\!-\!7\d_+\! \right)
   {\del\over\del_{\d_-\!}}
\cr&+
  \Coeff{1}{24}\left( 28\!-\!7\d_-\!\!-\!9\d_+\! \right)
   {\del\over\del_{\d_+\!}}\!+\!
  \Coeff{1}{4}\left(  8\d_-\! -16\!-\!{{\d_-\!}^2}\!+\!8\d_+\!\!-\!
      6\d_-\!\d_+\!\!-\!{{\d_+\!}^2} \right)
   {\del^2\over\del_{\d_+\!}\del_{\d_-\!}}-{1\over {36}} \Big\}
}\eqn\nodeL
$$
The solutions have the general form \rrrr,\uuu; though we did not
succeed to obtain them in a closed form, say in terms of $F_4$
functions, we can easily compute them up to arbitrary order.
Specifically, the first terms are:
$$
\eqalign{
\o1\ql &= \d_+\!\Big\{1 + \Coeff{1}{18}\,\d_+\! +
\Coeff{25}{3888}\,{{\d_+\!}^2} +
  \Coeff{7}{24}\,\d_-\! + \Coeff{377}{3456}\,\d_+\!\,\d_-\! +
  \Coeff{25289}{746496}\,{{\d_+\!}^2}\,\d_-\! +...\Big\}
\cr
\O1\ql &= \o1\ql\ln \d_+ \ + \
\Big\{36 + \d_+\! \cr &\ \qquad\qquad
+ \Coeff{5}{36}\,{{\d_+\!}^2} - \Coeff{1}{48}\,{{\d_-\!}^2}+
 \Coeff{13}{24}\,\d_+\!\,\d_-\! +
  \Coeff{1609}{6912}\,{{\d_+\!}^2}\,\d_-\! +...\Big\}\ .
}\eqn\Qonesol
$$
The remaining solutions, $\o2\ql$ and $\O2\ql$, are given by
exchanging $\d_+$ with $\d_-$ in \Qonesol.

Finally, as far as the solutions near the cusp points are concerned,
we face the problem of finding appropriate variables near $Q_2$. We
tried various compactifications of the moduli space, but could only
find two series solutions. Actually, the non-local physics in these
regions suggests that one cannot find there two series plus two
logarithmic
solutions, and thus a sensible prepotential $\cF$, at all.

%%%%%%%%%%%%%%%%%%%%%%%%%%%%% %%%%%%%%%%%%%%%%%%%%%%%%%%%%%%%

%%%%%%%%%%%%%%%%%%%%%%%%%%%%% %%%%%%%%%%%%%%%%%%%%%%%%%%%%%%%
%%%%%%%%%%%%%%%%%%%%%%%% %%%%%%%%%%%%%%%%%%%%%%%%%%
\chapter{The exact quantum low energy effective action for $G=SU(3)$}
\section{Semi-classical regime}
%%%%%%%%%%%%%%%%%%%%%%%% %%%%%%%%%%%%%%%%%%%%%%%%%%

Let us write and normalize the solutions \ttt,\ \tttlog\ of the
Appell system
$F_4(-\coeff16,-\coeff16,\coeff23, \coeff12)$
in the patch $P_3$ as follows:
$$
\eqalign{
\big(\Coeff1{\sqrt3\L}\big)\,\o1\ul3&=
2^{{2/ 3}} y_3^{-1/6}
F_4(-\Coeff16,\Coeff16,\Coeff12, 1;x_3,y_3)\ \cr&\ \ \sim\
2^{{2/3}}{{{y_3}^{-{1/ 6}}}}\big(1 - {\Coeff1
{36}{y_3}} + ...\big) \cr
\big(\Coeff1{\sqrt3\L}\big)\,\o2\ul3 &=
\Coeff1{3\sqrt3}2^{{2/
3}}{\sqrt{x_3}}{{{y_3}^{-{1\over 6}}}}
F_4(\Coeff13,\Coeff23,\Coeff32, 1;x_3,y_3)\ \cr&\ \ \sim\
\Coeff1{3\sqrt3}2^{{2/
3}}{{{\sqrt{x_3}}{{{y_3}^{-{1/ 6}}}}\big( 1 + {\Coeff4
{27}x_3} +...\big)}} \cr
} \eqn\solthree
$$ and $$
\eqalign{
\big(\Coeff1{\sqrt3\L}\big)\,\O1\ul3&=
(-1)^{-1/6}2^{{2/3}}12\sqrt3\pi {\Gamma(1/3)\over\Gamma(1/6)^2}
F_4(-\Coeff16,-\Coeff16,\Coeff12,
\Coeff23;\Coeff{x_3}{y_3},\Coeff1{y_3})
\cr &\ \
+\big((i  - \sqrt3) \pi + 4 \ln2 +3\ln 3-5\big)\,\o1\ul3
\cr &\ \ \sim\
2^{{2/ 3}}{{{y_3}^{-{1/
6}}}}\big(1 + {\Coeff1{36}{y_3}} +...\big) + \o1\ul3\ln y_3\cr
\big(\Coeff1{\sqrt3\L}\big)\,\O2\ul3&=
(-1)^{4/3}2^{-1/3}\Coeff1{3\pi} \sqrt{\Coeff{x_3}{y_3}}\Gamma(1/3)^2
F_4(\Coeff13,\Coeff13,\Coeff32,
\Coeff23;\Coeff{x_3}{y_3},\Coeff1{y_3})
\cr &\ \
+\big(1+(i+\Coeff1{\sqrt3})\pi+3 \ln 3\big)\,\o2\ul3
\cr &\ \ \sim\
\Coeff1{3\sqrt3}2^{{2/3}}{{{y_3}^{-{1/
6}}}}{\sqrt{x_3}}\,\big( 1 + {\Coeff{22}{27}x_3} +...\big) +
\o2\ul3\ln y_3\ ,
}
$$
where $x_3\equiv \coeff{27v^2}{4u^3}$ and $y_3 \equiv
\coeff{27\L^6}{4u^3}$. Here, we introduced the Appell function $F_4$,
which is defined in terms of a generalized Gaussian sum \Appell:
$$
F_4(a,b;c,c';x,y)\ =\ \sum_{n,m\geq0}{(a)_{n+m} (b)_{n+m}\over
(1)_n (1)_m (c)_n (c')_m}x^n y^m\ .
\eqn\appeldef
$$
This sum converges only for $|\sqrt x|+|\sqrt y|<1$.
For values outside this region, one can define $F_4$ by suitable
analytic continuation. Unfortunately, formulas for analytic
continuation and
transformations of $F_4$ do not seem to be thoroughly discussed in
the literature. However, for some purposes we can use the formula
$$
F_4(a,b;c,c';x,y)\ =\ \sum_{k=0}^\infty\,{(a)_m(b)_m\over(c)_m(1)_m}
\, {}_2F_1(a+m,b+m,c';y)\,x^m
\eqn\Ffourcont
$$
to analytically continue to arbitrary $y$, when $x$ is sufficiently
small.
(There exists a similar formula given by exchanging $x$ with $y$ and
$c$ with $c'$ in \Ffourcont).

Matching asymptotically $a_1,a_2$ to the Casimirs $u,v$ (which fixes
the normalization of $\lambda$), we then obtain the following
identification (up to Weyl conjugation) between $\vec\pi$ and the
solutions of $F_4$:
$$
\eqalign{
\ad1 &= -\Coeff i{4\pi}(\O1\ul3+3
\O2\ul3)-\Coeff1\pi(\a_1 \o1\ul3 - \a_2 \o2\ul3)
\cr &\ \
\sim -\Coeff1\pi( \Coeff i{2} + 2 \a_1 ) {\sqrt{u}} - \Coeff1\pi
{{\big( \Coeff{3}{4}i - \a_2 \big)}} \Coeff vu - \Coeff
i{2\pi}\big(\sqrt u+\Coeff32\Coeff vu\big)
\ln\big[\Coeff{27\L^6}{4u^3}\big]
+...\cr
\ad2 &= -\Coeff i{4\pi}(\O1\ul3-3
\O2\ul3)-\Coeff1\pi(\a_1 \o1\ul3 + \a_2 \o2\ul3)
\cr &\ \
\sim -\Coeff1\pi( \Coeff i{2} + 2 \a_1 ) {\sqrt{u}} + \Coeff1\pi
{{\big( \Coeff{3}{4}i - \a_2 \big)}} \Coeff vu - \Coeff
i{2\pi}\big(\sqrt u-\Coeff32\Coeff vu\big)
\ln\big[\Coeff{27\L^6}{4u^3}\big]
+...\cr
a_1 &= \Coeff12{{\o1\ul3}} + \Coeff12{{\o2\ul3}}
\sim {\sqrt{u}} + \Coeff12\Coeff vu +...\cr
a_2 &= \Coeff12{{\o1\ul3}} - \Coeff12{{\o2\ul3}}
\sim {\sqrt{u}} - \Coeff12\Coeff vu +...
}\eqn\solthreeident
$$
Here, $\a_1,\a_2$ are parameters that cannot determined by the
Picard-Fuchs equations. Their values can only be found by comparison
with the asymptotic expansion of the period integrals. This is done
in Appendix A with the result: $\a_1=\coeff54i-i\ln2-\coeff34i\ln3$,
$\a_2=\coeff3{4}i+\coeff94i\ln3$. For a loop around $u=\infty$,
\solthreeident\
indeed gives back precisely the semi-classical monodromy
$M_{\infty,u}^{(r_3)}$ in \infmonodr.

%Similarly, we find for the normalized solutions \ttt\ in the patch
% $P_2$,
%$$
%\eqalign{ \o1\ul2&=\coeff1{\sqrt3}{{{y_2}^{-{1\over 6}}}}\big({1 -
%{{x_2}\over {12}} -{{y_2}\over {18}} - {{5\,x_2\,y_2}\over
%{54}}}+...\big) \sim v^{{1\over3}}+... \cr \o2\ul2&=
%\coeff{2^{-{2\over3}}}{\sqrt3}{{{y_2}^{-{1\over 6}}}}{{x_2}^{{1\over
%3}}}\, \big( 1 +{{x_2}\over {12}} + {{y_2}\over 9} +... \big) \sim
%\coeff13u\,v^{-{1\over3}}+... \cr \O1\ul2&=
%\coeff1{\sqrt3}{{{y_2}^{-{1\over 6}}}}\big(1 + {{x_2}\over 6} +
%{{2\,y_2}\over 9} +...\big) +\o1\ul2\ln (y_2) \cr \O2\ul2&=
%\coeff{2^{-{2\over 3}}}{\sqrt3}{{{y_2}^{-{1\over
% 6}}}}{{x_2}^{{1\over
%3}}}\big( 1 +{{17\,x_2}\over {24}} + {{13\,y_2}\over {18}} +
%...\big) + \o2\ul2\ln (y_2)\ , }\eqn\soltwo
%$$
%where $x_2\equiv \coeff{4u^3}{27v^2}$, $y_2 \equiv \coeff1{27v^2}$,
%the following identifications (up to Weyl conjugation):
%$$
%\eqalign{
%a_{D1}\ &=\
%\Coeff{i}{2\pi}[(\rho\!-\!1)\O1\ul2\!+\!(\rho^2\!-\!1)\O2\ul2\!]
%\!+\![(\rho^2\!+\!2)\!+\!\rho^2\a_1]\o1\ul2
%\!+\![(\rho\!+\!2)\!+\!\rho\,\a_2]\o2\ul2
%\cr
%a_{D2}\ &=\ \Coeff{\sqrt3}{2\pi}(\O1\ul2-\O2\ul2)+\a_1
%\o1\ul2+\a_2\o2\ul2\cr
%a_1\ &=\ \o1\ul2+\o2\ul2\cr
%a_2\ &=\ -\rho^{-1}\o1\ul2-\rho\,\o2\ul2\ ,
%}\eqn\soltwoident
%$$
%where $\rho\equiv e^{2\pi i/3}$ and $\a_1,\a_2$ are so far
%undetermined parameters that we do not need to fix here. For a loop
%around $v=\infty$, \soltwoident\ reproduces the semi-classical
%Coxeter monodromy $M_{\infty,v}^{(r_{cox})}$ in \infmonodr.

We can treat the semi-classical coordinate patch $P_2$ in a similar
way, and find that for a loop around $v=\infty$ the semi-classical
Coxeter monodromy $M_{\infty,v}^{(r_{cox})}$ in \infmonodr\ is
reproduced.

To obtain the prepotential $\cF$, we need to invert the series
$a_i(u,v)$, ensuring good convergence in terms of the Cartan
sub-algebra variables $a_1,a_2$. We can start with either patch $P_2$
or $P_3$, but we will choose $P_3$ for convenience. Since in the
patch $P_3$ the classical Casimir $u_0\equiv{a_1}^2+{a_2}^2-a_1a_2$
is large and $v_0\equiv a_1a_2(a_1-a_2)$ is small, we can expand, for
example, around $(a_2/a_1),(\sqrt3\L/a_1)\sim0$. Note that we have,
essentially, a double expansion in one dimensionful and one
dimensionless parameter. Though the inversion of the double infinite
series $a_i(u,v)$ in \solthreeident\ in a closed form appears is
quite hard, we can explicitly compute the quantum corrected Casimirs
$$
\eqalign{
u(a_1,a_2) &= u_0(a_1,a_2)+27\L^6\Big\{
{1\over {72\,{{a_1}^4}}} +
  {{a_2}\over {36\,{{a_1}^5}}} +
  {{31\,{{a_2}^2}}\over {288\,{{a_1}^6}}} +
%  {{73\,{{a_2}^3}}\over {288\,{{a_1}^7}}} +
%  {{773\,{{a_2}^4}}\over {1152\,{{a_1}^8}}} +
%  {{451\,{{a_2}^5}}\over {288\,{{a_1}^9}}}+
\dots\Big\}+\cO(\L^{12},{a_2\over a_1})\cr
v(a_1,a_2) &= v_0(a_1,a_2)-27\L^6\Big\{
{{a_2}\over {24\,{{a_1}^4}}} +
  {{{{a_2}^2}}\over {12\,{{a_1}^5}}} +
  {{9\,{{a_2}^3}}\over {32\,{{a_1}^6}}} +
%  {{61\,{{a_2}^4}}\over {96\,{{a_1}^7}}} +
%  {{617\,{{a_2}^5}}\over {384\,{{a_1}^8}}} +
%  {{117\,{{a_2}^6}}\over {32\,{{a_1}^9}}}+
\dots\Big\}+\cO(\L^{12},{a_2\over a_1})}
\eqn\quantCas
$$
to any given order. Inserting this into $\ad1(u,v),\ad2(u,v)$ and
integrating $\int\!\!\ad i\,da_i$, we obtain the prepotential in the
form
$$
\cF(a_1,a_2)\ =\ {i\over 2\pi}6u_0
\big(\,\ln\big[{{a_1}\over\sqrt3\,\L}\big] -
\sum_{k=0}^\infty \cF_{6k}(a_1,a_2)\,\L^{6k}\big)\ .
\eqn\Fser
$$
We find that the logarithmic term together with $\cF_0$ gives indeed
precisely the small-$(a_2/a_1)$-expansion of the semi-classical
prepotential \clasF, ie.,
$$
{i\over 2\pi}6u_0\big(\,\ln\big[{{a_1}\over\sqrt3\,\L}\big] -
\cF_0(a_1,a_2)\big)\
=\ {1\over6}\big(\sum_{i=1}^3{Z_i}^2\big)\,\tau_0 +
{i\over 4\pi} \sum_{i=1}^3{Z_i}^2\ln\,[{Z_i}^2/\L^2]\ ,
\eqn\oneloopres
$$
where $Z_i=Z_i(a_1,a_2)$ are the classical central charges \Zdef,
and where
$$
\tau_0\ =\ {i\over2\pi}\big(\ln\big[{4\over27}\big]-9\big)\
\equiv\ {\theta_0\over\pi} + {8\pi i\over{g_0}^2}
\eqn\baretau
$$
is the ``bare'' coupling constant. It is a priori defined up to even
integers, which is a reflection of the quantum monodromy $T^2$
\quantmono\ induced by $2\pi$ rotations of $\L^3$ (actually, since
the curve \fff\ is unchanged even under $2\pi$ rotations of $\L^6$,
we have an ambiguity in $\tau_0$ up to adding integers.) Of course,
other semi-classical monodromies may induce additional integral
matrix shifts for the $\theta$-angle.

In fact, $\tau_0$ can be tuned to an arbitrary complex number by
appropriately rescaling $\L$. This is why we took some effort to
obtain this coupling (by evaluating the integrals in order to fix all
undetermined parameters in \solthreeident), since its imaginary part
needs to be fixed if one eventually wants to relate the present
quantum scale $\L$ (defined by the curve \fff) to the scale used in
some other physical computation.

We see from \oneloopres\ that by writing $\cF$ in terms of the
variables $Z$, we effectively sum up the series in the dimensionless
variable $(a_2/a_1)$. The remaining series in $(\sqrt3\L/a_1)$ can
then be interpreted in terms of non-perturbative quantum
corrections. In total, we find for the
asymptotic $\L$-expansion of the exact quantum prepotential:
$$
\cF(a_1,a_2)\ =\ \cF_\Class(a_1,a_2) + \cF_\onel(a_1,a_2) +
\cF_{{\rm{non\hyp pert}}}(a_1,a_2)\ ,
\eqn\ThatsIt
$$
with
$$
\eqalign{
&\cF_\Class\ \ =\ {1\over2}\, \tau_0\,(\vec a^t\cdot
C\cdot\vec a)\cr &\cF_\onel(a_1,a_2) =\ {i\over 4\pi}
\sum_{i=1}^3{Z_i}^2\ln\,[{Z_i}^2/\L^2]\cr &\cF_{{\rm{non\hyp pert}}}
=\ -{i\over 2\pi} \big(\sum_{i=1}^3{Z_i}^2\big)
\sum_{k=1}^\infty \cF_{6k}(Z)\L^{6k}\ .\cr}
\eqn\Fdefs
$$
The leading ``instanton coefficients'' are given in terms of
symmetric Laurent polynomials in the $Z_i$ as follows:\foot
{The higher $\cF_{6k}$ have no unique form when written in terms
of the $Z$'s, since some combinations vanish, but they are unique
when written in terms of the $a_i$.}
$$
\eqalign{
\cF_{6}(Z)\ &=\ {1\over 4}{1\over {Z_1}^2{Z_2}^2{Z_3}^2}\
\equiv\
{1\over 4}{1\over\Delta_0}
\cr
\cF_{12}(Z)\ &=\ -{1\over 2^6}\Big[
57{1\over {Z_1}^4{Z_2}^4{Z_3}^4}-5
\Big({1\over {Z_1}^6{Z_2}^6}+{\rm cycl.}\Big)\Big]
\cr &\equiv\ {3\over 2^5 }{1\over {\Delta_0}^3}
 (17 {u_0}^3+189 {v_0}^2)
\cr
\cF_{18}(Z)\ &=\ -{3\over 47\, 2^6}\Big[
 1265\Big( {1\over {{{Z_1}^8}{{Z_2}^{10}}}} +
 {1\over {{{Z_1}^{10}}{{Z_2}^8}}} + {\rm cycl.}\Big)
 \cr
&\ - 1492\Big( {1\over {{{Z_1}^6}{{Z_2}^{12}}}} +
 {1\over {{{Z_1}^{12}}{{Z_2}^6}}}
 + {\rm cycl.} \Big)
 +746\Big( {1\over {{{Z_1}^4}{{Z_2}^{14}}}} +
     {1\over {{{Z_1}^{14}}{{Z_2}^4}}} + {\rm cycl.} \Big)
\cr
&\ - 1492\Big(
 {1\over {{Z_1}^2{Z_2}^2{{Z_3}^{14}}}} + {\rm cycl.}
  \Big) - {76701}{1\over {{{Z_1}^6}{{Z_2}^6}{{Z_3}^6}}}\Big]
\cr &\equiv\
{9\over 2^5}{1\over{\Delta_0}^5} \big(3080{{u_0}^6} +
119529{{u_0}^3}{{v_0}^2} + 248589{{v_0}^4}\big)\ ,\ \ {\rm etc.}\ ,
\cr}
\eqn\Instantonterms
$$
where $\Delta_0$ is the classical discriminant \cdiscdefu,
$\Delta_0=4{u_0}^3-27{v_0}^2$. As expected, the non-perturbative
corrections get arbitrarily suppressed in the weak coupling limit,
where $\Delta_0\to\infty$.

Note that the prepotential \ThatsIt\ is manifestly Weyl group
invariant, in contrast to the Weyl non-covariant expansions in
$a_{1,2}$. This confirms that we have correctly resummed the infinite
series in the dimensionless variable. One can check that when
starting from a different expansion that is adapted to the patch
$P_2$, one obtains the same effective prepotential in terms of the
variables $Z$. Thus, the Weyl invariant resummation in terms of the
variables $Z$ simultaneously covers both patches at ``infinity'' in
$\cM_\L$.

%%%%%%%%%%%%%%%%%%%%%%%% %%%%%%%%%%%%%%%%%%%%%%%%%%
\section{Dual magnetic semi-classical regime}
%%%%%%%%%%%%%%%%%%%%%%%% %%%%%%%%%%%%%%%%%%%%%%%%%%

We can compute in a similar fashion the effective action $\cF_D$ in
the dual semi-classical, magnetic regime, ie., in the patch $Q_1$
where two monopoles become simultaneously massless. To lowest order,
this has been done for all $SU(n)$ in ref.\ \MDSS. In contrast,
though our techniques are not practical for computations for general
$n$, they allow an easy determination of the corrections to $\cF_D$
for any given $SU(n)$ group (here $G=SU(3)$) to arbitrary order.

Since the Appell functions in \solthreeident\ cannot easily expanded
or resummed in the variables $\d_\pm$ near the nodes (and appropriate
transformation formulas for $F_4$ do not seem to be known), we prefer
to resort to the explicit series solutions \Qonesol\ to make the
following identifications:
$$
\Coeff1{\sqrt3\,\L}\pmatrix{\ad1\cr\ad2\cr a_1\cr a_2\cr}\ =\
\a_0\pmatrix {i\o2\ql\sim i\d_-(1+...)\cr i\o1\ql\sim i\d_+(1+...)\cr
\coeff1{2\pi}(\O2\ql+\textstyle\sum \a_{2j}\o
j\ql)\sim\coeff1{2\pi}\d_-\!\ln [\d_-]+...\cr
\coeff1{2\pi}(\O1\ql+\textstyle\sum \a_{1j}\o j\ql)\sim\coeff1{2\pi
}\d_+\!\ln [\d_+]+...}\ .
\eqn\Qoneident
$$
As before, the undetermined parameters $\a_{ij}$, as well as the
overall normalization $\a_0$, can be fixed by asymptotically
evaluating the period integrals. This is done in Appendix A, with the
result: $\a_0=-\Coeff{2^{-1/3}}3$, $\a_{11}=\a_{22}=-2\ln2-3\ln3$,
$\a_{12}=\a_{21}=-2-2\ln2$.

Note that even though the solutions in
$\d_\pm\sim\prod_{i<j}^3(e_i^\pm-e_j^\pm)^2$ are symmetric in
$e_i^\pm-e_j^\pm$, the identification \Qoneident\ is valid only for
one given intersection, where $e_i^+-e_j^+=0$ and $e_k^--e_l^-=0$ for
some given $i,j,k,l$. To be specific, we have chosen the node at
$u=({27\over 4})^{1/3}\L^2$, $v=0$, where the lines $\#2$ (where
$e_1^--e_2^-=0$) and $\#3$ (where $e_1^+-e_3^+=0$) of
\lfig\figQantMs\ intersect. Encircling the lines $\d_\pm=0$,
\Qoneident\ clearly reproduces the correct strong coupling
monodromies, given by the matrices $M_{(1,0,0,0)}$ and
$M_{(0,1,0,0)}$ in \sixmonodroms. Note that even though \Qoneident\
represents a good solution at the remaining other two nodes, it does
not represent the correct identification with the period integrals
there. One rather has to conjugate the above basis with the cyclic
transformation $U$ \Ucox\ to obtain the proper identifications with
the period integrals at the other nodes. Of course, since all nodes
are equivalent under this $\ZZ_3$ symmetry, it suffices to study the
situation only at one node.

Just like in the semi-classical region, we can easily invert the
series solutions $\o1\ql$ and $\o2\ql$ and integrate $\int_{\ad
i}\!\!a_i$ to obtain for the dual effective prepotential the
following
result:
$$
\cF_D(a_1,a_2)\ =\ \cF_{D,0}(a_1,a_2) +
\cF_{D,\onel}(a_1,a_2) + \cF_{D,{\rm{thresh.}}}(a_1,a_2)\ ,
\eqn\Fdual
$$ where
$$
\eqalign{
&\cF_{D,0}\ \ \ =\ 18{i\over\pi}\b_0\L
({\ad1}+{\ad2}) \cr&\ \ \ \ \ \ \ \ \qquad\ \ +
{i\over\pi}\big({3\over8}+\Coeff12\ln2+\Coeff34\ln3\big)\,
({\ad1}^2+{\ad2}^2)+\ad1\ad2 \Coeff i\pi\ln2\cr
&\cF_{D,\onel}\ \ =\ {1\over 4
\pi i}\sum_{i=1}^2\big({\ad i}^2\ln\big[{\ad
i\over\b_0\L}\big]\big)\cr
&\cF_{D,{\rm{thresh.}}} =\ {1\over 2\pi i} \sum_{k=1}^\infty
\cF_{D,k}(\ad i)(432\,\b_0\L)^{-k}\ .
}\eqn\FDdefs
$$
with $\b_0= \coeff i{\sqrt3}2^{-1/3}$. Information about the massive
spectrum is encoded in the threshold corrections
$$
\eqalign{
\cF_{D,1}\ &=
-\left( \ad 1 + \ad 2 \right) \,
  \left( 4\,{{\ad 1}^2} - 13\,\ad 1\,\ad 2 +
    4\,{{\ad 2}^2} \right)
\cr
\cF_{D,2}\ &=\
\left( 44\,{{\ad 1}^4} -
    207\,{{\ad 1}^3}\,\ad 2 -
    189\,{{\ad 1}^2}\,{{\ad 2}^2} -
    207\,\ad 1\,{{\ad 2}^3} + 44\,{{\ad 2}^4}
     \right)
\cr
\cF_{D,3}\ &=
-\left( \ad 1 + \ad 2 \right) \,
  \big( 896\,{{\ad 1}^4} -
    7475\,{{\ad 1}^3}\,\ad 2 +
    2399\,{{\ad 1}^2}\,{{\ad 2}^2} \cr &\qquad -
    7475\,\ad 1\,{{\ad 2}^3} + 896\,{{\ad 2}^4}
     \big)\ ,\ \ {\rm etc.}\ .
}\eqn\unclearcoeff
$$
$\cF_D(\ad1,\ad2)$ indeed represents an effective action for two
$U(1)$ gauge fields, and is manifestly symmetric under exchange
of $\ad1$ and $\ad2$. Observe also that the corresponding
$\b$-functions are asymptotically non-free, and reflect the coupling
to fundamental matter fields with charges $(1,0)$ and $(0,1)$. These
correspond to pure magnetic monopoles in the original variables.

%%%%%%%%%%%%%%%%%%%%%%%% %%%%%%%%%%%%%%%%%%%%%%%%%%
\section{Properties of the period matrix, and various dualities}
%%%%%%%%%%%%%%%%%%%%%%%% %%%%%%%%%%%%%%%%%%%%%%%%%%

Analogous to $G=SU(2)$, there exist for $SU(3)$ certain types of
dualities that relate the electric semi-classical regime near
infinity in $\cM_\L$ with the dual magnetic semi-classical regime
near the nodes. To see this, let us first study the physical gauge
and dual gauge couplings, $\tau_{ij}\equiv{\del^2\over
\del_{a_i}\del_{a_j}}\cF(a)$ and $\tau_{D,ij}\equiv{\del^2\over
\del_{\ad i}\del_{\ad j}}\cF_D(a_D)\equiv-(\tau_{ij})^{-1}$, when
expressed in terms of $\a\equiv 4 u^3/27\L^6$, $\b\equiv v^2/\L^6$.

{}From our identifications \solthreeident, we can easily compute the
period matrix $\tau$, which is the exact quantum gauge coupling
constant:
$$
\Pi\ =\ \left(
{\del_u a_1\atop\del_v a_1}
{\del_u a_2\atop\del_v a_2}
\,;\,
{\del_u \ad1 \atop\del_v \ad1 }
{\del_u \ad2 \atop\del_v \ad2 }
\right)\ \equiv\ \Big(\,A\,;\,B\,\Big)\ ,
\qquad  \tau(u,v)\ =\ A^{-1}B\ ,
\eqn\periodma
$$
Explicit expressions
for the periods are collected in Appendix B.
Of course, to make sense of $\tau(u,v)$ over the whole moduli space,
we
must suitably analytically continue the Appell functions.
For sufficiently small $v$, we can resort to \Ffourcont\ in
order to continue to all $u$. (One may also use the dual formula
to continue to all $v$ for small $u$. This might be useful to study
the behavior at the cusps.)

In particular, we may continue $\tau$ to the origin of moduli space.
This serves as a useful consistency check, since $\tau(0,0)$ is
completely fixed by the $\ZZ_6$ symmetry of the curve. That is, we
require that $A\equiv\big({a\ b\atop c\ d}\big)$ in \Adef\ leaves the
period matrix invariant: $\tau=(a\,\tau+b)(c\,\tau+d)^{-1}$. In
addition, we know that the $\ZZ_6$ symmetry acts on the abelian
differentials as follows: $A:\
\coeff{dx}y\to-\xi\coeff{dx}y,\,\coeff{xdx}y\to-\xi^2\coeff{xdx}y$,
where $\xi=e^{2\pi i/6}$. This transformation has determinant equal
to $-1$, and this must be the same as the determinant of
$(c\,\tau+d)$. These conditions, as well as the positivity of
$\Im\tau$, completely determine the period matrix at the origin as
follows:
$$
\tau(0,0)\ =\ \pmatrix{1&0\cr0&-1} + \Coeff i{\sqrt3}\, C\ ,
\eqn\taufix
$$
where $C\equiv \big({2\ -1\atop -1\ 2}\big)$ is the Cartan matrix.

Note that the identifications \solthreeident\ were based on
the comparison with the period integrals of Appendix A. For these
integrals, a basis of cycles was chosen that corresponds to $\Im
v=0,\,\Re v>1$, in order to match the basis given in \lfig\figxplane\
and to reproduce the semi-classical monodromy \infmonodr. If we want
to
analytically continue to $(u,v)=0$, we need to take into account that
for $-1<\Re v<1$ the identifications between $a_i,\ad i$ and the
solutions of the PF equations change. This change of basis is the one
that relates the monodromy $M_{\infty,u}^{(r_3)}$ \infmonodr\
(defined by a loop around infinity in the $u$-plane at $\Re v>1$)
with the monodromy $\tilde M_{\infty,u}^{(r_3)}$ \tildeM\ (defined by
a loop around infinity in the $u$-plane at $\Re v=0$). Taking this
change of basis into account, we indeed find \taufix\ to hold, by
evaluating the period matrix \periodma\ at the origin.

Since non-trivial transformation properties of $F_4$ do not appear to
be known, it is quite hard to find all possible duality symmetries
that may act on the moduli space. What we can do is to consider
transformations that act solely on $\a\equiv4u^3/27\L^6$ when
$\b\equiv
v^2/\L^6=0$. For this, it is helpful to rewrite the periods for
$\b=0$ in the following form:
$$
\eqalign{
\del_u a_1\ &=\ \del_u a_2\ =\
(\sqrt3\L)^{-1}(-1)^{1/6}2^{-2/3}(1-\a)^{-1/6}
{}_2F_1\big(\Coeff16,\Coeff16,1;\Coeff1{1-\a}\big)\cr
\del_v a_1\ &= -\del_v a_2\ =\
(3\L^2)^{-1}(-1)^{1/3}2^{-1/3}(1-\a)^{-1/3}
{}_2F_1\big(\Coeff13,\Coeff13,1;\Coeff1{1-\a}\big)\cr
\del_u \ad1\ &=\ \del_u \ad2\ =\
i(\sqrt3\L)^{-1}\,2^{-2/3}
{}_2F_1\big(\Coeff16,\Coeff16,1;1-\a\big)\cr
\del_v \ad1\ &= -\del_v \ad2\ =
-i(3\L^2)^{-1}\,2^{-1/3}\sqrt3\,
{}_2F_1\big(\Coeff13,\Coeff13,1;1-\a\big)\ .\cr
}\eqn\manifestform
$$
{}From these expressions one can then infer that under
$$
I:\ \ \a\ \longrightarrow\ \tilde\a\ =\ {\a\over\a-1}\ ,
\eqn\fractT
$$
(which just exchanges the arguments of the hypergeometric
functions), the period matrix transforms as follows,
$$
\tau(\a,\b=0)\ =\ C\cdot \tau_D(\tilde\a,\tilde\b=0)\ ,
\eqn\ISOG
$$
provided that $(1-\a)\in\IR^-$. The presence of the Cartan matrix
reflects that the bases of the electric and magnetic degrees of
freedom are given by the Dynkin and the simple root bases,
respectively, as explained in \hsubsect{4.2}. This electric-magnetic
duality obviously generalizes the isogeny transformation
\taumap,\Samap\ for $G=SU(2)$. Though we did not succeed to find an
extension of \fractT\ to non-zero $\b$, we believe that such an
extension does exist, and reflects a transformation property of
$F_4$. If true, it would be very likely that analogously to
$G=SU(2)$, there exist dual, isogenous forms of the hyperelliptic
curve \aaa, for which electric and magnetic degrees of freedom are
exchanged. This is also suggested by the form of the monodromies, as
mentioned in \hsect{4.2}\ and \hsect{4.3}.

In addition, analogous to $G=SU(2)$ there exists (at least for
$\b=0$)
an $S$-duality, which acts like
$$
\tau(\a,\b=0)\ =\ \tau_D(\hat \a,\hat\b=0)\ ,
\eqn\SDUAL
$$
with $\hat \a=I(\tilde S(I(\a)))$, where $I$ is the isogeny map
\ISOG\
and
$$
\tilde S:\ \a\ \longrightarrow\ \Coeff1{(\a^{1/3}-1)^3}\Big\{
8+\a+12(-1)^{2/3}\a^{1/3}-6(-1)^{1/3}\a^{2/3}\Big\}\ .
\eqn\nSDUAL
$$
Probably there exist other dualities as well, for example, a
transformation that relates the cusp patches $Q_2$ and $Q_3$.

%%%%%%%%%%%%%%%%%%%%%%%%%%%%% %%%%%%%%%%%%%%%%%%%%%%%%%%%%%%%

%%%%%%%%%%%%%%%%%%%%%%%%%%%%% %%%%%%%%%%%%%%%%%%%%%%%%%%%%%%%
\chapter{Conclusions and Outlook}

The main result of the present paper are the explicit expressions
\Fdefs\ and \FDdefs\ for the quantum effective prepotential, whose
expansion can easily be determined to any given order in $\L$. It
would be very interesting to compare the instanton corrections with
expressions obtained by some other kind of computation.

There are, of course, many aspects that were not touched upon in the
present paper. Some of the aspects that we neglected were recently
discussed, eg., in \doubref\PAMD\Nf, and need not be repeated here.
We just like comment on a few things.

First, it is clear that similar to $SU(2)$ (where $u(\tau)$ is given
by a modular function of $\Gamma_0(4)$ or $\Gamma(2)$), the variables
$u_k(\tau)$ should be given by certain higher genus modular functions
(whose modular properties include the $S$-duality transformation
\nSDUAL). Such functions are intrinsically defined via lattice sums
of type $\sum(m+n\tau)^l$, with period lattices given here by root
lattices. (For $SU(3)$, these functions could be found by inverting
the expressions given in Appendix B.) One question that arises would
be what the physical interpretation of such lattice sums is.
Certainly one would like to think in terms of partition functions
involving massive excitations, but the counting of states, of which
many are unstable, would probably be subtle. Also, in contrast to
\nex4 supersymmetric theories, $|m+n\tau|^2$ does not give the mass
of a state, so that these lattice functions do not give the mass
spectrum but rather count electric and magnetic charges.

Among the open
points is also the generalization to other groups ($G=SO(2n+1)$ was
recently treated in \DS). Actually, our classical considerations in
\hsect{3.1}\ directly generalize to other Lie algebras. General
formulae for the discriminants \cdiscdefu\ are discussed in the
literature on arrangements of hyperplanes \OrTe. In particular, for
the remaining simply laced Lie algebras of type $D$ and $E$, the
following simple singularities \Arn\ are relevant:
$$
\eqalign{
\cW_{D_n}(x_1,x_2,u)&={x_1}^{n-1}+\shalf x_1\, {x_2}^2
-\sum_{l=1}^{n-1}u_{2l}\,{x_1}^{n-l-1}-\tilde u_n x_2\cr
\cW_{E_6}(x_1,x_2,u)&={x_1}^3+{x_2}^4 - u_2 x_1 x_2^2-u_5 x_1 x_2-u_6
x_2^2 -u_8 x_1-u_9 x_2-u_{12} \cr
\cW_{E_7}(x_1,x_2,u)&={x_1}^3+{x_2}^4 - u_2 x_1^2 x_2-u_6 x_1^2-u_8
x_1 x_2 -u_{10} x_2^2\cr&\ \ -u_{12} x_1-u_{14} x_2-u_{18}\cr
\cW_{E_8}(x_1,x_2,u)&={x_1}^3+{x_2}^5 - u_2 x_1 x_2^3-u_8 x_1 x_2^2-
u_{12} x_2^3-u_{14} x_1 x_2\cr&\ \ -u_{18} x_2^2-u_{24} x_2-u_{30}\ ,
\cr
}\eqn\DEsing
$$
where $u_k$ are one-to-one to the Casimirs of the corresponding
algebra. The associated discriminants characterize classical
Yang-Mills theories based on the simply laced Lie algebras of type
$D$ and $E$. As for the quantum theories, we expect the underlying
curves to be of the form
$$
p(x_i)\big(\cW_{ADE}(x_i)\big)^2\ - q(x_i)\L^{2h}\ =\ 0\ ,
\eqn\guess
$$
where $h$ is the corresponding dual Coxeter number, and $p(x)$,
$q(x)$ are suitable polynomials. The key point is the quadratic
appearance of the simple singularity, which ensures that any singular
branch of the classical singularity in $\cM_0$ (describing an
unbroken $SU(2)$) splits into two quantum branches (describing
massless $SU(2)$ Seiberg-Witten monopoles). However, the genus of the
curves \guess\ does not seem to easily come out correctly, although
this
criticism may be too naive in view of the $SO(2n+1)$ curves of \DS.

We also remark that most of our considerations in \hsect{4.2}\ about
properties of BPS states in relation with vanishing cycles directly
apply or generalize to situations involving ``level'' surfaces other
than Riemann surfaces. In particular, they apply to classical $SU(n)$
\nex2 Yang-Mills theory, as was already pointed out in
\hsubsect{3.1}. They also apply to BPS states of extended
supersymmetric string compactifications, where the relevant surfaces
are $K3$ \WiVa\ and Calabi-Yau \Stro\ manifolds. We think it would be
interesting and important to develop a coherent and systematic
picture of BPS states related to vanishing cycles, especially in view
of the type II -- heterotic string duality \doubref\HT\KaVa.

%%%%%%%%%%%%%%%%%%%%%%% %%%%%%%%%%%%%%%%%%%%%%%%%%%%%%%%
\ack
%%%%%%%%%%%%%%%%%%%%%%% %%%%%%%%%%%%%%%%%%%%%%%%%%%%%%%%

We like to thank L.~Alvarez-Gaum\'e, K.\ Saito,
S.~Yankielowicz and S.T.~Yau for discussions.

%%%%%%%%%%%%%%%%%%%%%%%%%%%%% %%%%%%%%%%%%%%%%%%%%%%%%%%%%%%%
\append {A}{Asymptotic evaluation of period integrals}

The hyperelliptic period integrals $\int_{\gamma_i}\lambda$ provide a
definite basis for the solutions of the Picard-Fuchs equations
everywhere in the moduli space. In order to relate them to the
various local expansion that one gets by solving the Picard-Fuchs
equations, we have to compute some low order terms of the asymptotic
expansions in terms of those variables by which we parametrize the
vicinity of 'infinity' and the node, respectively. In this appendix
we will provide some of the details of those calculations, which
enable us to fix the physical relevant quadratic terms in the
prepotential and constitute a check of our monodromy considerations.

The integrals to be computed are of the form
$$
w_{ij}\ =\ \Coeff i{2\pi}\int_{e_i}^{e_j}\lambda\ =\
\Coeff i{2\pi}\int_{e_i}^{e_j}{x(3 x^2-u)dx\over y}
$$
where $e_i$ are the roots of the polynomial $y^2=p(x)=0$.

 The first step is to find approximate expressions for the roots
$e_i$ and then to expand the integrals such as to reduce them to
elementary integrals. We will do this for the semi-classical infinity
at $P_3$ and the node at $v=0,u=({27\over4})^{1/3}\Lambda^2$ in turn.

(i) {\it Period integrals at infinity}: Here we introduce
variables $\alpha={v\over u^{3/2}},\,\beta={\Lambda^3\over u^{3/2}}$
s.t. infinity is at the origin $(\alpha,\beta)=(0,0)$ and the roots
of $y$
are approximately $e_i\equiv\sqrt{u}\tilde e_i$, with:
$$\eqalign{
\tilde e_1\simeq-1+{1\over2}(\alpha-\beta)+{3\over 8}
(\alpha-\beta)^2&\,,\quad
\tilde e_3\simeq-\alpha-\beta\,,\quad \cr&\ \
\tilde e_5\simeq1+{1\over2}(\alpha-\beta)-{3\over 8} (\alpha-\beta)^2
\cr
\tilde e_2\simeq -1+{1\over2}(\alpha+\beta)+{3\over 8}
(\alpha+\beta)^2&\,,\quad
\tilde e_4\simeq-\alpha+\beta\,,\quad \cr&\ \
\tilde e_6\simeq1+{1\over2}(\alpha+\beta)-{3\over 8}
(\alpha+\beta)^2}
$$
with $e_{2i-1}=e_{2i},\,i=1,2,3$ at semi-classical infinity. These
roots are ordered in an obvious fashion. Their relation to the branch
points and cycles of \lfig\figxplane\ is ambiguous and depends on the
path used in the analytic continuation. This ambiguity is physically
irrelevant, and corresponds to a Weyl conjugation. By choosing a
specific path, we can make the following associations: $e_1\to e_3^-,
e_2\to e_3^+, e_3\to e_1^+, e_4\to e_1^-, e_5\to e_2^-, e_6\to
e_2^+$.
Thus, the integrals $w_{2i-1,2i}$ are related to the $a$-type periods
and will be given by pure power series, whereas $w_{2i,2i+1}$,
which are related to the $a_D$-type periods, will have logarithms.

To compute $\tilde w_{12}$, we introduce the variable
$w=x-{1\over 2}(e_2-e_1)$ and get, after expanding the integrand
in powers of small quantities:
$$
2w_{12}=-a_2\simeq{\sqrt{u}\over \pi}\int_{-\beta/2}^{\beta/2}
{dw\over\sqrt{{1\over4}\beta^2-w^2}}(-1+{1\over2}\alpha)=
-(\sqrt{u}-{1\over 2}{v\over u})
$$
Likewise we get $2w_{34}\simeq -{v\over u}$ and $2w_{56}=a_1\simeq
(\sqrt{u}+{1\over 2}{v\over u})$.

The logarithmic periods are more involved. In order to compute e.g.,
$w_{23}$, we split the range of integration into two pieces, namely
$\int_{e_2}^{e_3}\lambda=\int_{e_2}^\xi
\lambda+\int_\xi^{e_3}\lambda$, where $\xi-e_2\sim e_3-\xi$. Both
integrands can then be expanded leading to elementary integrals of
the form $\int{p(w)\over\sqrt{w^2-a^2}\,(x+b)^n}$ with $p$ a
polynomial. We find
$$
2w_{23}=\ad2\simeq-\Coeff i\pi\Big\{\sqrt{u}\Bigl(3+{1\over
2}\log({\Lambda^6\over64 u^3})\Bigr)
-{v\over u}\Bigl({3\over4}\log({\Lambda^6\over4 u^3})\Bigr)\Big\}
$$
and
$$
2w_{45}=\ad1\simeq-\Coeff i\pi\Big\{\sqrt{u}\Bigl(3+{1\over
2}\log({\Lambda^6\over64 u^3})\Bigr)
+{v\over u}\Bigl({3\over4}\log({\Lambda^6\over4 u^3})\Bigr)\Big\}
$$
(ii) {\it Period integrals at the nodes}: parametrized by
$\delta_\pm=
1-\alpha+\beta\pm2\sqrt{\beta}\,,\alpha={4 u^3\over 27\Lambda^6},\,
\beta={v^2\over \Lambda^6}$ the roots are approximately
($e_i\equiv 2^{-1/3}\Lambda\tilde e_i$):
$$\eqalign{
\tilde e_1&=-2(1-{\delta_+\over12}-{\delta_-\over36}),\quad
\tilde
e_2=-(1-i\sqrt{{\delta_+\over3}}-
{\delta_+\over36}-{\delta_-\over12}),
\cr&\quad\tilde e_3=-(1+i\sqrt{{\delta_+\over3}}-{\delta_+\over36}
-{\delta_-\over12})\cr
\tilde e_6&=2(1-{\delta_+\over36}-{\delta_-\over12}),\quad
\tilde
e_4=(1+i\sqrt{{\delta_-\over3}}-{\delta_+\over12}-{\delta_-\over36}),
\cr& \quad\tilde e_5=(1-i\sqrt{{\delta_-\over3}}-{\delta_+\over12}
-{\delta_-\over36})}
$$
We now consider the node $v=v_0=0,
u=u_0=({27\over4})^{1/3}\Lambda^2$.

Close to the node we have $v=v_0+\delta
v={\Lambda^3\over4}(\delta_+-\delta_-)$ and $u=u_0+\delta
u\simeq-{1\over6}({27\over 4})^{1/3} \Lambda^2(\delta_++\delta_-)$,
or $\delta_\pm\simeq-{2^{2/3}\over\Lambda^2}
(u-u_0)\pm{2\over\Lambda^3}v$. The computation of the $a_D$-type
periods is straightforward. Expanding the integrand in powers of
$\delta_\pm$ leads to elementary integrals, and we find
$$
2 w_{23}=\ad2\simeq-{i\Lambda\over 3^{1/2} 2^{1/3}}\delta_+
={i\Lambda\over 3^{1/2} 2^{1/3}}\Bigl({2^{2/3}\over\Lambda^2}
(u-{3\Lambda^2\over 2^{2/3}})-2{v\over\Lambda^3}\Bigr)\ ,
$$
and likewise
$$
2 w_{45}=\ad1\simeq-{i\Lambda\over 3^{1/2} 2^{1/3}}\delta_-
={i\Lambda\over 3^{1/2} 2^{1/3}}\Bigl({2^{2/3}\over\Lambda^2}
(u-{3\Lambda^2\over 2^{2/3}})+2{v\over\Lambda^3}\Bigr)\ .
$$
The computation of the logarithmic solutions is more cumbersome. We
again split the integral into two pieces in order to be able to deal
with the singularities of the integrand separately; e.g. for
$w_{34}:\,\int_{e_3}^{e_4}\lambda=\int_{e_3}^\xi\lambda+
\int_\xi^{e_4}\lambda$ such that $e_{2,3}<\xi<e_{45}$ and
$|\xi-e_3|/|e_3-e_2|,|\xi-e_4|/|e_5-e_4|\gg1$. Independence of the
choice of $\xi$ serves as a check. Expanding the integrand and the
limits of integration in powers of $\delta_\pm^{1/2}$ leads to
elementary and elliptic integrals. For instance, taking the lowest
order terms for the roots $e_i,i\neq 2,3$ and $u$ we are led to the
integral
$\int_{e_3}^\xi{x(x+1)\over\sqrt{(x-e_2)(x-e_3)}\sqrt{4-x^2}}$. To do
the integral we introduce the variable
$w=-1-x+{1\over36}\delta_++{1\over12}\delta_-$, expand
$1/\sqrt{4-x^2}$ in a power series in $x$ and then expand $x^n$ to
order $\epsilon_\pm^2$. The remaining integrals can now be performed
keeping only terms to order $\epsilon^2$ and
$\epsilon^2\log(\epsilon)$. The resulting series can then be
resummed. After some work, we finally get:
$$
2w_{12}=-a_2\simeq -\Coeff\L{\sqrt3}\Coeff{2^{2/3}3}\pi
\Big\{-3-{1\over12}\delta_+\Bigl(\log(\delta_+)
-2\log2-3\log3-1\Bigr)
+{1\over6}\delta_-\log2\Big\}
$$
and
$$
2w_{56}=a_1\simeq \Coeff\L{\sqrt3}\Coeff{2^{2/3}3}\pi
\Big\{-3-{1\over12}\delta_-\Bigl(\log(\delta_-)
-2\log2-3\log3-1\Bigr)
+{1\over6}\delta_+\log2\Big\}\ .
$$

\vfill\eject
%%%%%%%%%%%%%%%%%%%%%%%%%%%%%%%%%%%%%%%
\append{B}{Explicit expression for the period matrix}
%%%%%%%%%%%%%%%%%%%%%%%%%%%%%%%%%%%%%%%

\def\Coff{\Coeff} \def\coff{\Coeff} The period matrix, which
represents the exact $SU(3)$ quantum gauge coupling constant, is
given by $\tau(u,v)=A^{-1}B$, where
$$
A\ =\ \left(
{\del_u a_1\atop\del_v a_1}
{\del_u a_2\atop\del_v a_2}
\right)\ ,\qquad\ \
B\ =\ \left(
{\del_u \ad1 \atop\del_v \ad1 }
{\del_u \ad2 \atop\del_v \ad2 }
\right)\ .
$$
The periods are, in terms of multi-valued Appell functions $F_4$
\appeldef\
of $\a\equiv 4u^3/27\L^6$ and $\b\equiv v^2\L^6$, as follows:
$$
\eqalign{
\big(\sqrt3\L\big)\,\del_u a_1 &=\
2^{-2/3}\a^{-1/6}\,
F_4\big(\coff16,\coff56,\coff12,1;
\coff\b\a,\coff1\a\big)
\cr&-
2^{1/3}\Coff13\sqrt{\Coff\b3}\,\a^{-2/3}
F_4\big(\coff23,\coff43,\coff32,1;
\coff\b\a,\coff1\a\big)
\cr&\cr
\big(\sqrt3\L\big)\,\del_u a_2 &=\
2^{-2/3}\a^{-1/6}\,
F_4\big(\coff16,\coff56,\coff12,1;
\coff\b\a,\coff1\a\big)
\cr&+
2^{1/3}\Coff13\sqrt{\Coff\b3}\,\a^{-2/3}
F_4\big(\coff23,\coff43,\coff32,1;
\coff\b\a,\coff1\a\big)
\cr&\cr
\big(3\L^2\big)\,\del_v a_1 &=\
2^{-1/3}\a^{-1/3}\,
F_4\big(\coff13,\coff23,\coff12,1;
\coff\b\a,\coff1\a\big)
\cr&-
2^{-1/3}\sqrt{\Coff\b3}\,\a^{-5/6}
F_4\big(\coff56,\coff76,\coff32,1;
\coff\b\a,\coff1\a\big)
\cr&\cr
\big(3\L^2\big)\,\del_v a_2 &=
-2^{-1/3}\a^{-1/3}\,
F_4\big(\coff13,\coff23,\coff12,1;
\coff\b\a,\coff1\a\big)
\cr&-
2^{-1/3}\sqrt{\Coff\b3}\,\a^{-5/6}
F_4\big(\coff56,\coff76,\coff32,1;
\coff\b\a,\coff1\a\big)
}
$$

$$
\eqalign{
\big(\sqrt3\L\big)\,\del_u \ad1 &=
(-1)^{-1/3}2^{-2/3}\a^{-1/6}
F_4\big(\coff16,\coff56,\coff12,1;
\coff\b\a,\coff1\a\big)
\cr &-
\Coff23(-1)^{1/6}2^{-2/3}\sqrt\b\a^{-2/3}
F_4\big(\coff23,\coff43,\coff32,1;
\coff\b\a,\coff1\a\big)
\cr &-
(-1)^{1/6}\pi 2^{4/3}\Gamma(1/3)^{-3} \sqrt{\Coff\b3}\,
F_4\big(\coff23,\coff23,\coff32,\coff13;
\b,\a\big)
\cr &+
(-1)^{2/3}2^{-5/3}\Coff{\Gamma(1/6)^2}{\pi\Gamma(1/3)}\,
F_4\big(\coff16,\coff16,\coff12,\coff13;
\b,\a\big)
}
$$

$$
\eqalign{
\big(\sqrt3\L\big)\,\del_u \ad2 &=
(-1)^{-1/3}2^{-2/3}\a^{-1/6}
F_4\big(\coff16,\coff56,\coff12,1;
\coff\b\a,\coff1\a\big)
\cr &+
\Coff23(-1)^{1/6}2^{-2/3}\sqrt\b\a^{-2/3}
F_4\big(\coff23,\coff43,\coff32,1;
\coff\b\a,\coff1\a\big)
\cr &+
(-1)^{1/6}\pi 2^{4/3}\Gamma(1/3)^{-3} \sqrt{\Coff\b3}\,
F_4\big(\coff23,\coff23,\coff32,\coff13;
\b,\a\big)
\cr &+
(-1)^{2/3}2^{-5/3}\Coff{\Gamma(1/6)^2}{\pi\Gamma(1/3)}\,
F_4\big(\coff16,\coff16,\coff12,\coff13;
\b,\a\big)
}
$$

$$
\eqalign{
\big(3\L^2\big)\,\del_v \ad1 &=
\sqrt3(-1)^{-1/6}2^{-1/3}\a^{-1/3}
F_4\big(\coff13,\coff23,\coff12,1;
\coff\b\a,\coff1\a\big)
\cr &-
(-1)^{1/3}2^{-1/3}\a^{-5/6}\sqrt{\Coff\b3}\,
F_4\big(\coff56,\coff76,\coff32,1;
\coff\b\a,\coff1\a\big)
\cr &-
3(-1)^{1/3} 2^{2/3}\Coff{\Gamma(1/3)}{\Gamma(1/6)^2} \sqrt\b\,
F_4\big(\coff56,\coff56,\coff32,\coff23;
\b,\a\big)
\cr &+
\Coff34(-1)^{5/6}2^{-1/3}\Coff{\sqrt3}{\pi^2}\Gamma(1/3)^3\,
F_4\big(\coff13,\coff13,\coff12,\coff23;
\b,\a\big)
}
$$

$$
\eqalign{
\big(3\L^2\big)\,\del_v \ad2 &=
-\sqrt3(-1)^{-1/6}2^{-1/3}\a^{-1/3}
F_4\big(\coff13,\coff23,\coff12,1;
\coff\b\a,\coff1\a\big)
\cr &-
(-1)^{1/3}2^{-1/3}\a^{-5/6}\sqrt{\Coff\b3}\,
F_4\big(\coff56,\coff76,\coff32,1;
\coff\b\a,\coff1\a\big)
\cr &-
3(-1)^{1/3} 2^{2/3}\Coff{\Gamma(1/3)}{\Gamma(1/6)^2} \sqrt\b\,
F_4\big(\coff56,\coff56,\coff32,\coff23;
\b,\a\big)
\cr &-
\Coff34(-1)^{5/6}2^{-1/3}\Coff{\sqrt3}{\pi^2}\Gamma(1/3)^3\,
F_4\big(\coff13,\coff13,\coff12,\coff23;
\b,\a\big)
}
$$

%%%%%%%%%%%%%%%%%%%%%%%%%%%%% %%%%%%%%%%%%%%%%%%%%%%%%%%%%%%%
%\goodbreak%\vfil\eject
\refout
\end